 \documentclass[pra,longbibliography,twocolumn,showpacs,nofootinbib,
                superscriptaddress,notitlepage]{revtex4-2}
 
 \pdfoutput=1 
 
 \usepackage{mathtools}
 \usepackage{amssymb}
 \usepackage{bm}
 \usepackage{amsthm}
 \usepackage{mathrsfs}
 \usepackage{stmaryrd}
 \usepackage{latexsym}
 \usepackage{mdframed}
 
 \usepackage[table]{xcolor}   
 \usepackage{booktabs}
 \usepackage{array,makecell}
 \usepackage{boldline}
 
 \usepackage[utf8]{inputenc}  
 \usepackage[protrusion=true,expansion=true]{microtype}
 \usepackage{ragged2e}
 \usepackage{setspace}
 \usepackage{lipsum}
 \usepackage{ulem}
 
 \usepackage{graphicx}
 \usepackage{float}
 \usepackage{verbatim}
 \usepackage{listings}
 \definecolor{tomlkey}{RGB}{28,64,128}
 \definecolor{tomlstr}{RGB}{110,70,20}
 \definecolor{tomlcomment}{RGB}{110,110,110}
 \definecolor{tomlbg}{RGB}{246,246,246}
 \lstdefinelanguage{toml}{
  sensitive=true,
  comment=[l]{\#},
  string=[b]",
  morekeywords={name,distance,grid,kind,p,prepare,readout,
                id,rounds,measure,memory,xor}}
 \usepackage[caption=false]{subfig}
 
 \usepackage{enumitem} 
 
 \usepackage{dsfont}
 \usepackage{url}      
 \usepackage{keyval}
 
 \usepackage[noend]{algpseudocode} 
 \usepackage{algorithm}

 \newcommand{\apptocsection}[3]{#1 Appendix #2: \quad #3}

 \DeclareMathOperator{\supp}{supp}
 \DeclareMathOperator{\Int}{Int}
 \DeclareMathOperator{\Sweep}{Sweep}
 \DeclareMathOperator{\Term}{Term}
 \DeclareMathOperator{\Span}{Span}
 \DeclareMathOperator{\Comp}{Comp}

 \definecolor{cardinal}{rgb}{0.827,0,0}

 \usepackage[colorlinks=true,hyperindex,breaklinks,
             linkcolor=blue,urlcolor=blue,citecolor=blue]{hyperref}
 \usepackage[capitalise]{cleveref}
 
\begin{document}
 
\title{Patchwork: automated lattice surgery and space-time partitioning for real-time decoding}
 
\author{Christopher Chamberland}
\email{cchamberland@nvidia.com}
\affiliation{NVIDIA Corporation, USA}

\author{Muyuan Li}
\affiliation{NVIDIA Corporation, USA}

\author{Scott Thornton}
\affiliation{NVIDIA Corporation, USA}

\author{Justin Gage Lietz}
\affiliation{NVIDIA Corporation, USA}

\author{Melody Ren}
\affiliation{NVIDIA Corporation, USA}

\author{Antonio deMarti iOlius}
\affiliation{NVIDIA Corporation, USA}

\begin{abstract}
We introduce \texttt{patchwork}, a lattice-surgery compiler for arbitrary multi-qubit Pauli measurements on a prescribed narrow-corridor two-dimensional surface-code layout. Systematic geometric rules simplify stabilizer placement and gate scheduling, including for logical $Y$ measurements implemented using twist defects. We introduce measurement circuits and schedules that, together with protruding lobes of $Y$-extension patches, preserve the full effective code distance for all numerically tested code distances. To support parallel decoding, we develop a wavefront algorithm that partitions merged lattice-surgery patches into space-time commit, buffer, and cleanup regions. Simulations of three sequential multi-qubit Pauli measurements acting on ten logical qubits show that staged decoding achieves total failure probabilities close to those of monolithic decoding over the parameters tested. These methods provide an automated framework for fault-tolerant lattice surgery and a route toward scalable, real-time decoding.
\end{abstract}
 
\maketitle

\tableofcontents

\section{Introduction}
\label{sec:Intro}

Surface codes provide a route to fault-tolerant quantum computation based on local stabilizer measurements \cite{DennisSurface,TomitaRotatedSurface2014,fowler2012surface}. Together with magic-state preparation \cite{Bravyi12,Chamberland2019Magic1,chamberland2020very}, multi-qubit Pauli measurements support both logical Clifford and non-Clifford operations and hence universal quantum computation \cite{fowler2018low,Litinski19,Chamberland22,Chamberland22b,No_Hooks_Japan}. Lattice surgery realizes these joint logical measurements by temporarily merging surface-code patches and measuring stabilizers in the resulting geometry. It is therefore particularly well suited to implementing logical operations on two-dimensional architectures with local connectivity.

Turning a logical Pauli measurement into a physical implementation requires compatible choices of patch geometry, stabilizer placement, and measurement schedule. These choices must control fault propagation while providing the logical observables and syndrome information needed for decoding. Logical $Y$ measurements introduce additional challenges because twist defects can reduce the effective code distance. Moreover, as more logical qubits participate in a measurement, the merged patch grows, increasing the size of the classical decoding problem. A systematic approach to lattice surgery must therefore connect circuit construction with decoding.

In this work, we introduce \texttt{patchwork}, a lattice-surgery compiler that brings these tasks into a common framework for arbitrary multi-qubit Pauli measurements on a prescribed two-dimensional surface-code layout. We first develop geometric rules that make stabilizer placement systematic. Horizontal and vertical strips constrain stabilizer placement within the routing corridors, while checkpoint lines specify where elongated stabilizers are needed to resolve stabilizer-type and ancilla conflicts. These rules reduce the variability of merged-patch geometries and support automated construction of lattice-surgery operations. We also give algorithms for generating the spacelike and timelike logical observables needed to track logical errors and measurement outcomes throughout a sequence of operations.

We next develop gate-scheduling rules that adapt to the geometry of the merged patch. Cut lines and junction cells specify where stabilizer-measurement schedules must change to control fault propagation relative to the logical operators. For twist-based logical $Y$ measurements, we introduce measurement circuits and schedules together with protruding lobes of the $Y$-extension patch. The lobes occupy available routing corridors and therefore do not require a change to the underlying physical-qubit layout. For the configurations tested at distances $d \leq 11$, exhaustive numerical searches find no logical failures caused by at most $(d-1)/2$ circuit faults, supporting preservation of the full effective code distance in these cases. We also describe alternatives using wider routing corridors or an ancilla-assisted decomposition into $X$ and $Z$ measurements.

A central contribution of this work is the wavefront algorithm for parallel decoding. It partitions merged lattice-surgery patches into spatial commit, buffer, and cleanup regions that are extended into space-time decoding blocks. Commit regions are decoded in parallel using their buffers as context, followed by cleanup decoding of the residual syndrome. We construct the associated local detector error
models and specify how correction effects are combined to update the global syndrome and logical observables. This framework distributes the decoding of a merged patch across multiple classical resources and provides a route toward real-time decoding during logical computation.

We demonstrate the resulting workflow on a sequence of three multi-qubit Pauli measurements involving ten logical qubits, including measurements with logical $Y$ factors. A failure occurs when at least one deterministic logical observable is decoded incorrectly. Across the parameters tested, staged decoding achieves total failure probabilities close to, but slightly higher than, those of monolithic decoding. We also compare the areas of the largest commit and cleanup regions with the full merged-patch area for random Pauli measurements of increasing weight. Together, these studies quantify the size of the local decoding regions and the decoding-accuracy cost of partitioning, connecting automated lattice-surgery construction with parallel error correction.

The paper is organized as follows. \Cref{sec:Surface_code_stabs} reviews the required surface-code stabilizers. \Cref{sec:Layout,sec:MeasureXZ,sec:FaultTolerantY} introduce the layout and geometric rules for multi-qubit Pauli measurements. \Cref{sec:Logic_Auto_Gen,sec:Timelike_log_gen} describe logical-observable generation, and \cref{sec:Gate_Schedule} presents the gate schedules and $Y$-extension refinements. \Cref{sec:spacetime_partition,sec:spacetime_corrections} develop the wavefront construction and detector-error-model partitioning, followed by numerical results in \cref{sec:numerics} and conclusions in \cref{sec:conclusion}. Appendix~\ref{app:graph} gives the mathematical details of the wavefront algorithm, and Appendix~\ref{app:expd-spec} specifies the numerical experiment in \texttt{patchwork}.

\section{Surface code stabilizers}
\label{sec:Surface_code_stabs}

\begin{figure*}
    \centering
    \subfloat[\label{fig:Surface_Code_StandardEX}]{\includegraphics[width=.4\textwidth]{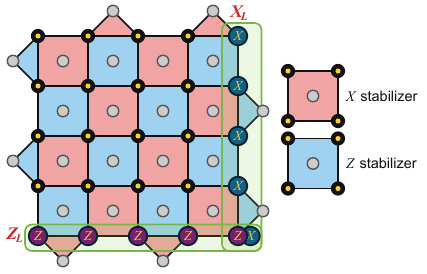}}
    \subfloat[\label{fig:LegendEX}]{\includegraphics[width=.4\textwidth]{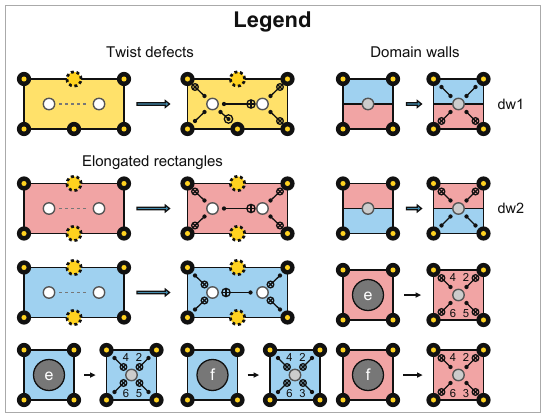}}
    \caption{(a) Distance-five surface-code patch with $X$-type (red) and $Z$-type (blue) stabilizers. Minimum-weight representatives of $X_L$ and $Z_L$ lie along the vertical and horizontal boundaries shown, respectively. Yellow vertices denote data qubits, and grey vertices denote measurement ancillas. (b) Stabilizer geometries and measurement arrangements used for lattice surgery, including standard weight-four checks with \textbf{e} and \textbf{f} gate schedules, elongated rectangles, domain walls, and a twist defect. Dashed yellow vertices denote inactive sites.}
    \label{fig:Legend}
\end{figure*}

We summarize the surface-code conventions used throughout the paper and introduce the additional stabilizer types needed for lattice surgery.

\Cref{fig:Surface_Code_StandardEX} shows a surface-code patch with $d=5$ and minimum-weight representatives of its logical operators $X_L$ and $Z_L$. A distance-$d$ patch contains a $d \times d$ array of data qubits, with weight-four bulk stabilizers arranged in a checkerboard pattern and weight-two stabilizers along the boundaries. Red and blue plaquettes represent $X$- and $Z$-type stabilizers, respectively; yellow vertices denote data qubits, and grey vertices denote ancillas used to measure the stabilizers. For the orientation shown, minimum-weight representatives of $X_L$ are vertical strings of physical $X$ operators, while those of $Z_L$ are horizontal strings of physical $Z$ operators. Other patch orientations are also used, as illustrated in \cref{fig:Layout_Example} in \cref{sec:Layout}.

These boundary representatives determine which patch boundaries must be accessed during lattice surgery. For the orientation shown, an $X_L \otimes X_L$ measurement requires access to a vertical boundary of each patch, whereas a $Z_L \otimes Z_L$ measurement requires access to a horizontal boundary. Stabilizers introduced in the routing space are chosen so that their measurement outcomes determine the desired joint logical parity. The corresponding construction rules are given in \cref{sec:MeasureXZ}.

\Cref{fig:LegendEX} summarizes the stabilizer geometries used in these constructions. In addition to standard weight-four stabilizers, shown with \textbf{e} and \textbf{f} gate schedules, lattice surgery uses elongated rectangles, domain walls, and twist defects.

An elongated weight-four stabilizer has the same Pauli factors as a standard $X$- or $Z$-type stabilizer, but its support spans a column of inactive sites for a horizontal rectangle, or a row of inactive sites for a vertical rectangle. These inactive sites are indicated by dashed yellow vertices in \cref{fig:LegendEX}. The rules governing the placement of elongated stabilizers are introduced in \cref{sec:Layout}.

Domain-wall stabilizers contain both $X$ and $Z$ factors. Ordering the active data qubits row by row from top to bottom, and from left to right within each row, the weight-four stabilizers labeled \texttt{dw1} and \texttt{dw2} are $Z \otimes Z \otimes X \otimes X$ and $X \otimes X \otimes Z \otimes Z$, respectively. In the layout of \cref{sec:Layout}, these stabilizers provide access to logical $Z_L$ operators along horizontal patch boundaries, as described in \cref{sec:MeasureXZ}.

Using the same ordering convention, the weight-five twist stabilizer shown in \cref{fig:LegendEX} is $X \otimes X \otimes Z \otimes Y \otimes X$ (the placement of the $Z$ operator depends on the orientation, see \cref{subsec:elongated_rec_gate_schedule}). Twist defects enable logical $Y_L$ measurements; their possible orientations and measurement schedules are discussed in \cref{subsec:gate_schedule_y_ext}.

Weight-two and weight-three variants of elongated rectangles and domain walls are also used in the lattice-surgery constructions, although they are not shown in the legend.

\section{Routing space layout with checkpoint lines}
\label{sec:Layout}

\begin{figure*}
    \centering
    \subfloat[\label{fig:Layout_ExampleEX}]{\includegraphics[width=.75\textwidth]{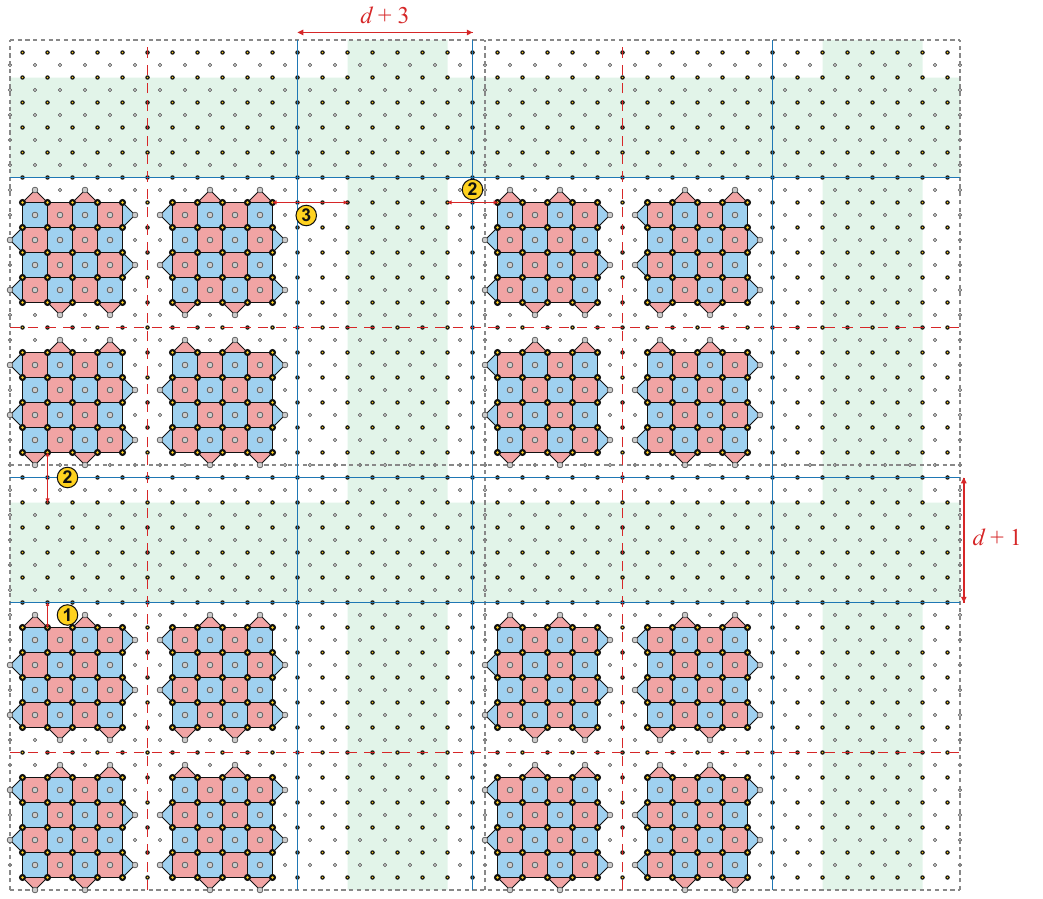}}
    \caption{Layout for multi-qubit Pauli measurements via lattice surgery, illustrated with distance-five logical patches. Yellow vertices denote data qubits, and grey vertices denote ancillas. Dashed grey lines separate unit cells, each containing four logical patches. Green strips mark the allowed stabilizer-placement regions within the routing corridors, with exceptions for patch-boundary connections and $Y$ extensions described in the text. The numbers in yellow indicate the separation between a logical patch and a green strip, measured in rows or columns of data-qubit sites. Dashed red and solid blue lines mark intra-cell and corridor checkpoint lines, respectively, which determine where elongated stabilizers are introduced.}
    \label{fig:Layout_Example}
\end{figure*}

We use a regular arrangement of logical patches and routing corridors to make the construction of lattice-surgery operations systematic. \Cref{fig:Layout_Example} illustrates the layout for $d=5$.

The layout is built from unit cells containing four logical patches, as also considered in Ref.~\cite{Chamberland22b}. Dashed grey lines in \cref{fig:Layout_Example} separate neighboring unit cells. For distance-$d$ patches, vertical routing corridors have width $d+3$, and horizontal routing corridors have height $d+1$. These dimensions accommodate the stabilizer geometries and $Y$-extension constructions developed below. The additional geometric and gate-scheduling requirements for preserving the effective code distance during $Y$ measurements are discussed in \cref{sec:FaultTolerantY,subsec:gate_schedule_y_ext,sec:protruding_lobes}.

Within each routing corridor, the green strips in \cref{fig:Layout_Example} specify where stabilizers may be placed when connecting logical patches. Stabilizers may extend beyond these strips to access patch boundaries or construct $Y$ extensions. To access a vertical boundary for a logical $X$ measurement, stabilizers may extend outside the vertical strip while remaining within the vertical extent of the patch. Similarly, access to a horizontal boundary for a logical $Z$ measurement permits stabilizers to extend outside the horizontal strip while remaining within the horizontal extent of the patch. These boundary-connection rules are developed in \cref{sec:MeasureXZ}; the additional extensions required for logical $Y$ measurements are described in \cref{sec:FaultTolerantY}.

Standard weight-four stabilizers are added along a routing corridor according to the surface-code checkerboard pattern. Continuing this pattern without modification can produce two types of conflict. First, a weight-two boundary stabilizer in the routing region may require an ancilla already used by a weight-two stabilizer of a logical patch. Second, two horizontally or vertically adjacent weight-four stabilizers may have the same Pauli type. Checkpoint lines specify where standard stabilizers must be replaced by elongated stabilizers to avoid these conflicts. We introduce the relevant rules below.

\subsection{Intra-cell checkpoint lines}
\label{subsec:Intracell}

\begin{figure}
    \centering
    \subfloat[\label{fig:Example_VICLEX}]{\includegraphics[width=.3\textwidth]{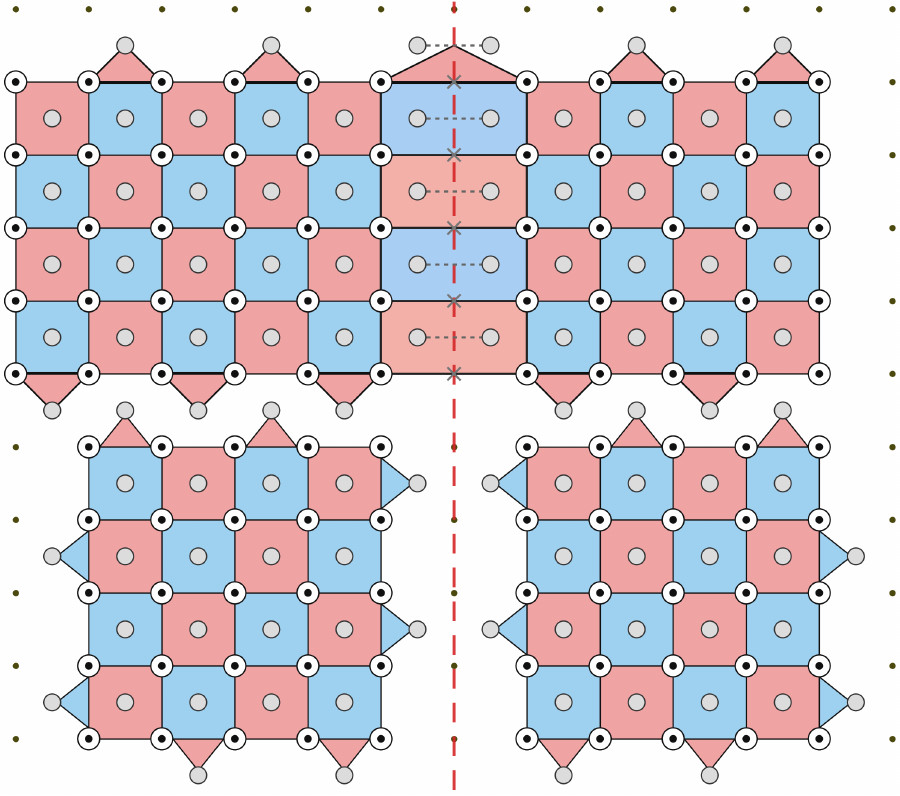}}
    \caption{Horizontal elongated stabilizers introduced where a routing path crosses a vertical intra-cell checkpoint line (VICL). Each elongated stabilizer retains the Pauli type of the corresponding standard weight-four stabilizer, with its middle inactive sites lying on the VICL. The elongated column spans the full connected transverse extent of the merged patch at the crossing.}
    \label{fig:Example_VICL}
\end{figure}

\begin{figure*}
    \centering
    \subfloat[\label{fig:VCCL_ExampleEX}]{\includegraphics[width=1.0\textwidth]{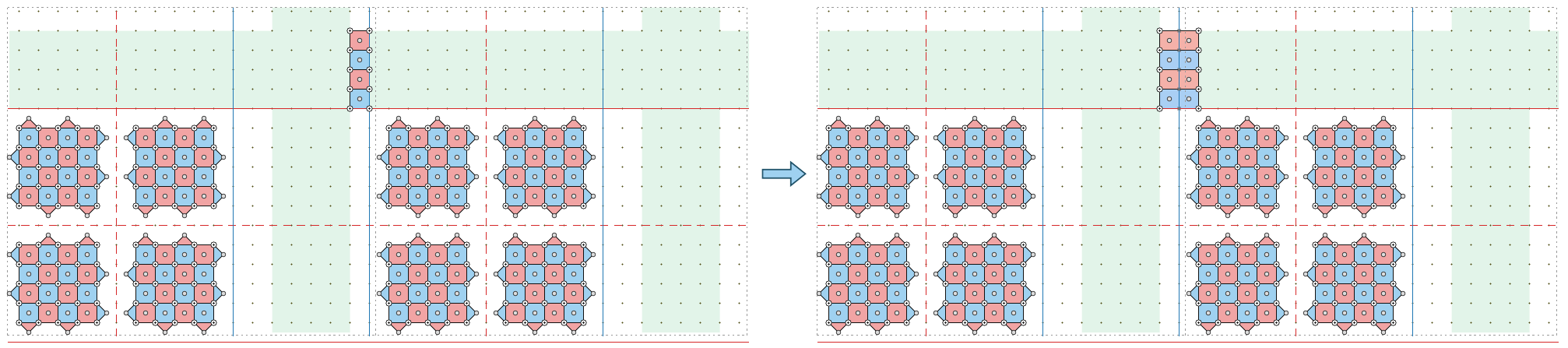}}
    \vfill
    \subfloat[\label{fig:VCCL_Example_v2EX}]{\includegraphics[width=1.0\textwidth]{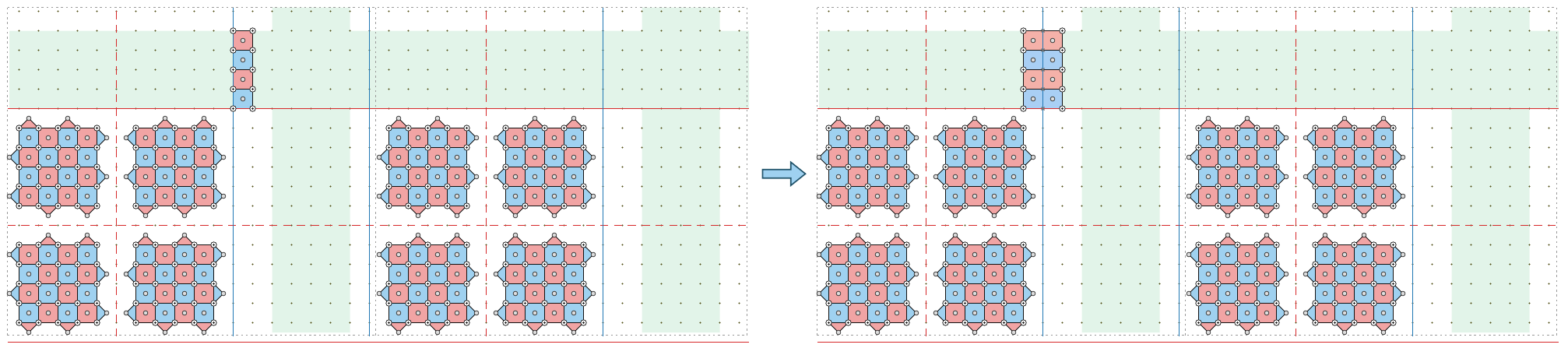}}
    \caption{Examples of the standard vertical corridor checkpoint line (VCCL) rule. (a) Left-to-right traversal of a VCCL on the right side of a vertical corridor. (b) Right-to-left traversal of a VCCL on the left side of a vertical corridor. In both cases, the bottom standard stabilizer adjoining the VCCL within the horizontal green strip is $Z$-type, triggering elongation of that stabilizer and the column of stabilizers above it within the strip. Each stabilizer retains its original Pauli type, and the middle inactive sites of the elongated stabilizers lie on the VCCL.}
    \label{fig:Example_VCCL}
\end{figure*}

Intra-cell checkpoint lines (ICLs), shown as dashed red lines in \cref{fig:Layout_Example}, lie between logical patches within a unit cell that are separated by a single row or column of data-qubit sites. We denote vertical and horizontal ICLs by VICLs and HICLs, respectively. The following rule applies when a horizontal routing path reaches a VICL.

\begin{mdframed}[frametitle={VICL Rules}]
\begin{itemize}
    \item Replace each standard weight-four stabilizer encountered at the VICL by a horizontal elongated stabilizer of the same Pauli type. The elongated column must span the full connected transverse extent of the merged patch at the crossing. This requirement is mandatory: subsequent construction steps cannot truncate the column or replace part of it by standard weight-four stabilizers. The data-qubit sites on the VICL become the middle inactive sites of the elongated stabilizer. An example is shown in \cref{fig:Example_VICL}.
\end{itemize}
\end{mdframed}

HICLs are relevant only to routing regions near logical $Y$ measurements. Their rules are given in \cref{subsec:standard_corridor_Y}.

\subsection{Vertical corridor checkpoint lines}
\label{subsec:Intercell}

Vertical corridor checkpoint lines (VCCLs) are shown as solid blue vertical lines in \cref{fig:Layout_Example}. A VCCL lies one data-qubit column to the right of the patches on the right side of a unit cell, or one column to the left of the patches on its left side. The corresponding horizontal corridor checkpoint lines (HCCLs) are treated in \cref{subsec:standard_corridor_Y}.

The standard VCCL rules depend on the traversal direction and the Pauli type of the bottom stabilizer adjoining the line. The two cases considered here are as follows.

\begin{mdframed}[frametitle={VCCL Rules}]
For a path through a horizontal routing corridor, identify the standard weight-four stabilizer at the bottom of the horizontal green strip according to the traversal direction:
\begin{itemize}
    \item \textbf{Left to right.} At a VCCL on the right side of a vertical corridor, select the stabilizer whose top-right and bottom-right data qubits lie on the VCCL (\cref{fig:VCCL_ExampleEX}).
    \item \textbf{Right to left.} At a VCCL on the left side of a vertical corridor, select the stabilizer whose top-left and bottom-left data qubits lie on the VCCL (\cref{fig:VCCL_Example_v2EX}).
\end{itemize}
If the selected stabilizer is $Z$-type, promote it and the column of stabilizers directly above it within the horizontal green strip to horizontal elongated stabilizers. Each promoted stabilizer retains its original Pauli type, and its middle inactive sites lie on the VCCL. After promotion, resume the standard checkerboard construction beyond the elongated stabilizers in the direction of traversal, continuing until the next checkpoint line is reached.
\end{mdframed}

\section{Performing multi-qubit Pauli measurements containing only $X$ and $Z$ operators}
\label{sec:MeasureXZ}

We now combine the layout and checkpoint rules of \cref{sec:Layout} to construct multi-qubit Pauli measurements containing only logical $X_L$ and $Z_L$ factors. The construction proceeds in three stages: select routing trees connecting the participating patches, place stabilizers for the required $X$ and $Z$ factors, and complete the boundaries and junctions of the merged patches.

\subsection{Multi-commodity flow algorithm for connecting logical patches}
\label{subsec:MCFAlgo}

\begin{figure*}
    \centering
    \subfloat[\label{fig:XX_Meas_dmEX}]{\includegraphics[width=0.8\textwidth]{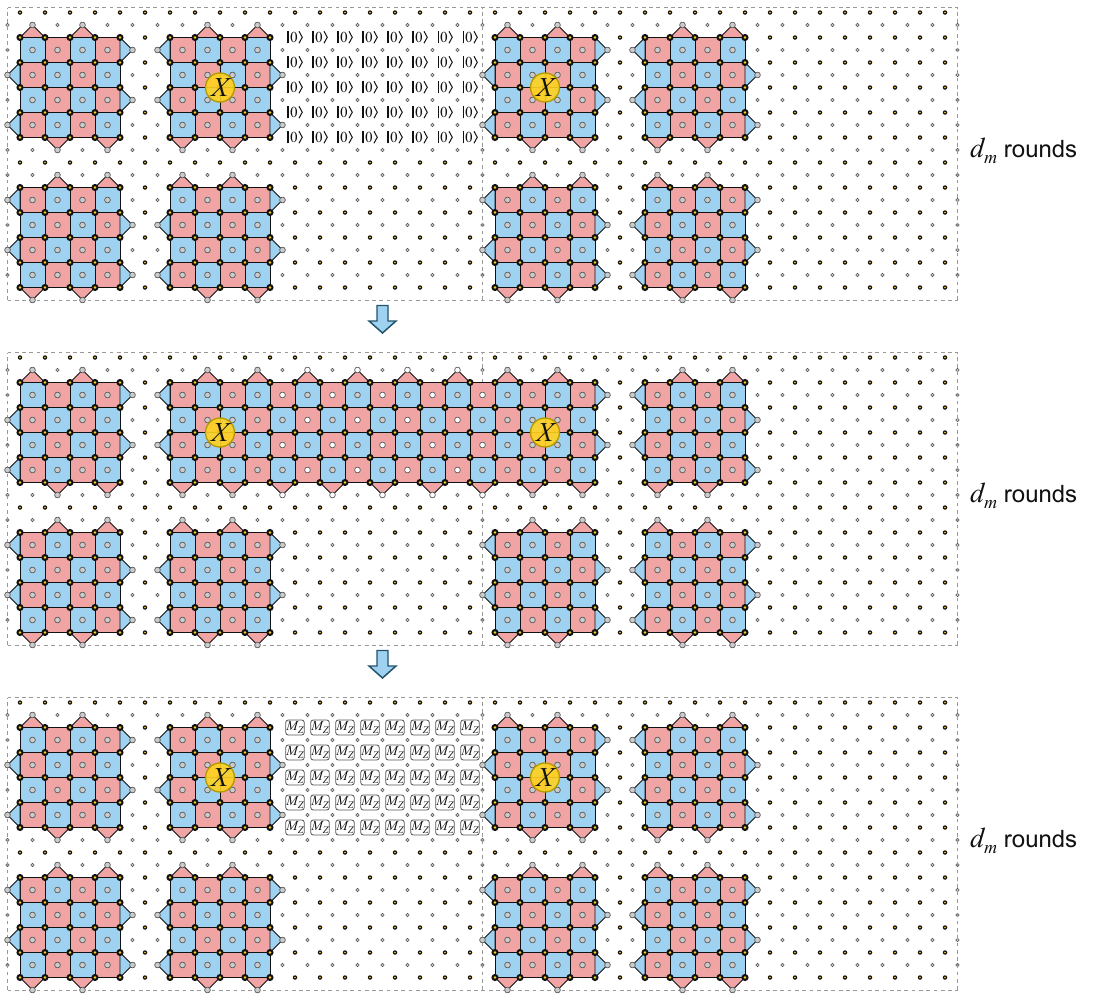}}
    \caption{Example of an $X_L \otimes X_L$ measurement. Routing-space data qubits are initialized in $|0\rangle$. After $d_m$ rounds of syndrome extraction on the separate patches, new stabilizer measurements couple their vertical logical-$X$ boundaries. White ancillas mark stabilizers with random initial outcomes; the product of the relevant outcomes encodes the joint logical parity. The merged patch is measured for $d_m$ rounds, then split by measuring the routing-space data qubits in the $Z$ basis. A final $d_m$ rounds of syndrome extraction are performed on the separated patches before either readout or the next multi-qubit Pauli measurement.}
    \label{fig:XX_Meas_dm}
\end{figure*}

\begin{figure*}
    \centering
    \subfloat[\label{fig:X_elongated_v1EX}]{\includegraphics[width=0.7\textwidth]{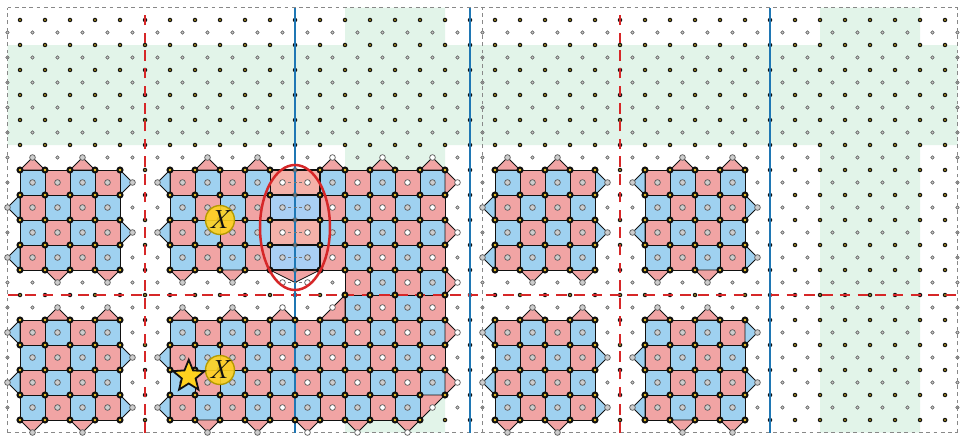}}
    \vfill
    \subfloat[\label{fig:X_elongated_v2EX}]{\includegraphics[width=0.8\textwidth]{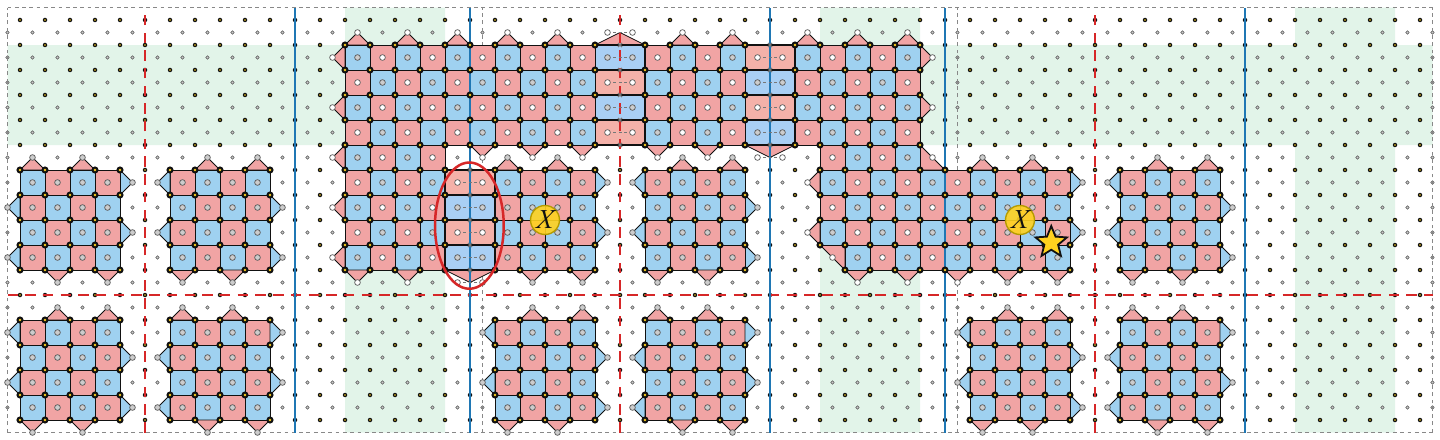}}
    \caption{(a), (b) Elongated stabilizers at vertical logical-patch boundaries in two $X_L \otimes X_L$ measurement layouts. Using standard weight-four stabilizers at the circled locations would produce adjacent stabilizers of the same Pauli type along a horizontal row. In each example, the yellow star marks the logical patch chosen as the root of the routing tree.}
    \label{fig:XX_elongated_vs}
\end{figure*}

\begin{figure*}
    \centering
    \subfloat[\label{fig:ZZ_Meas_v1EX}]{\includegraphics[width=0.7\textwidth]{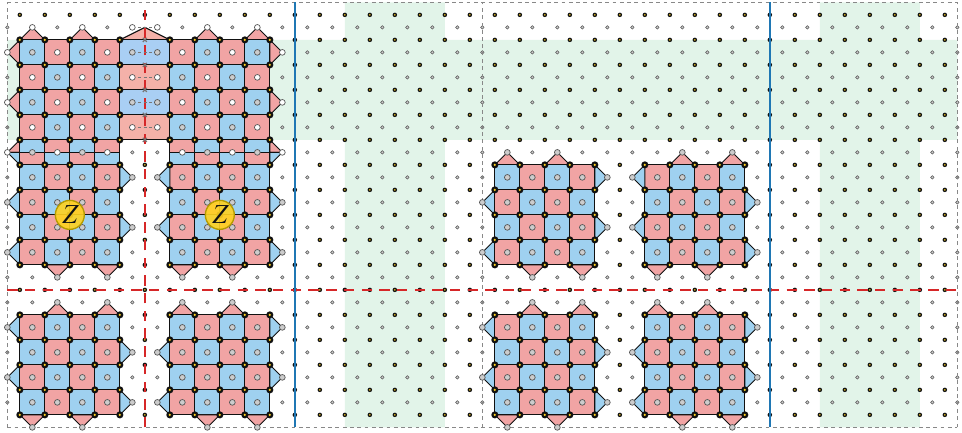}}
    \vfill
    \subfloat[\label{fig:ZZ_Meas_v2EX}]{\includegraphics[width=0.7\textwidth]{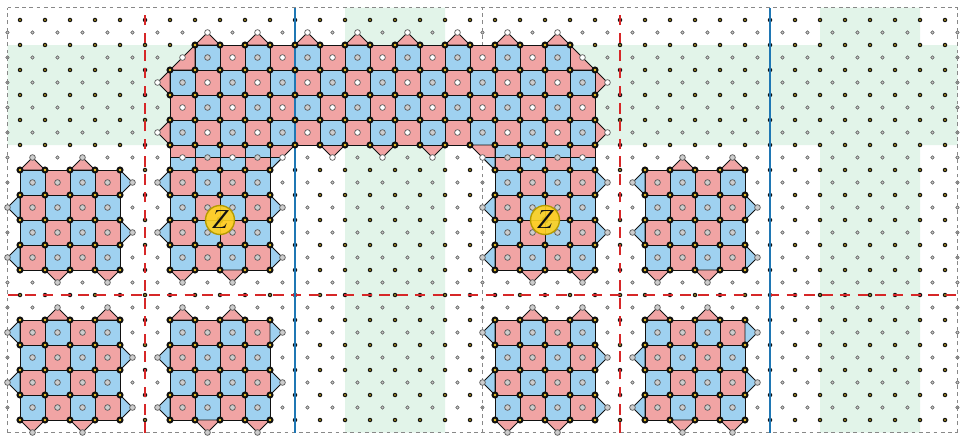}}
    \caption{Examples of $Z_L \otimes Z_L$ measurements using domain walls along the top boundaries of logical patches. (a) Weight-four and weight-two domain walls complete the boundary connections. (b) Weight-three domain walls are used where the connections meet corners of the merged patch.}
    \label{fig:XZZ_Meas_examps}
\end{figure*}

Consider a set of multi-qubit Pauli measurements (which we refer to as PPMs) $\{P_1,P_2,\ldots\}$ to be performed in parallel. Each measurement requires a connected routing region joining its participating logical patches, while the different measurements compete for the available routing space. We formulate this task as a multi-commodity routing problem \cite{Ahuja1993NetworkFlows}.

The routing network is represented by a graph $G=(V,E)$. A \textit{junction} is a region where horizontal and vertical routing corridors meet, allowing a route to turn or branch. Vertices correspond to these junctions, and edges represent allowed connections between neighboring junctions. Both edges and vertices have unit capacity: $u_e=1$ for every $e\in E$ and $u_v=1$ for every $v\in V$. Thus, each corridor segment and each junction can be occupied by at most one PPM at a time. The vertex constraint excludes routes that use different corridor segments but intersect at a shared junction.

Each Pauli measurement defines a commodity. For a two-qubit measurement between patches $A$ and $B$, this commodity requires one unit of flow from $A$ to $B$. For example, the simultaneous measurements $P_1=X_1\otimes X_2$ and $P_2=Z_5\otimes Z_7$ require paths connecting patches 1 and 2, and patches 5 and 7, respectively. The paths are selected jointly subject to the corridor and junction capacities, allowing one measurement to be rerouted when the individually preferred routes overlap.

A measurement involving more than two logical qubits generally requires a branching connection. For example, $P=X_1\otimes X_3\otimes X_8\otimes X_{10}$ requires a connected stabilizer region spanning four terminals. We describe such connections using Steiner trees \cite{Dreyfus1971Steiner}. Each measurement is assigned a root patch, and connectivity is enforced from that root to every other participating patch. The resulting multi-commodity Steiner routing problem seeks one tree per measurement, with no corridor edge or junction vertex shared between different trees. Branches of the same tree may share corridor segments and junctions, since capacities count distinct measurements rather than individual root-to-terminal paths.

The routing procedure used throughout this work is therefore:
\begin{enumerate}
    \item For each $P_i$, identify the logical patches in its support as terminals and choose one as the root.
    \item Form a rooted Steiner-tree routing request for each measurement.
    \item Assign unit capacity to every corridor edge and junction vertex.
    \item Solve the routing requests jointly to obtain the measurement trees.
\end{enumerate}
Stabilizers along these trees are placed within the green corridor strips, with the boundary-access exceptions described in \cref{sec:Layout}. After placing all boundary and junction stabilizers, we verify that each merged patch is connected and contains no \textit{holes}, defined as unmerged regions completely surrounded by the patch. We also verify that the completed construction implements the requested multi-qubit Pauli measurement.

\subsection{Elongated stabilizers for logical $X_L$ measurements}
\label{subsec:ElongatedXMeas}

We first describe stabilizer placement for the logical $X$ factors. Throughout this work, routing-space data qubits are initialized in $|0\rangle$ before a multi-qubit Pauli measurement, regardless of its Pauli factors. With this convention, logical $X$ factors are coupled to the routing region using standard surface-code stabilizers, while logical $Z$ factors use domain walls, as described in \cref{subsec:ZMeasRules}. Newly measured checks that anticommute with a $Z$ operator stabilizing an initialized routing qubit have random individual outcomes in the first merge round. Nevertheless, the product of the relevant outcomes encodes the parity of the measured Pauli operator \cite{fowler2018low}. White ancillas distinguish checks with random initial outcomes from those shown with grey ancillas; \cref{fig:XX_Meas_dm} illustrates an $X_L\otimes X_L$ measurement.

For the patch orientations used here, minimum-weight representatives of $X_L$ are vertical strings of physical $X$ operators. For each logical $X$ factor, the gauge-fixing step therefore measures $X$-type checks whose product acts as $X$ on every data qubit along the relevant vertical patch boundary. Along the routing tree, stabilizers follow the standard checkerboard pattern, with elongated stabilizers inserted according to the VICL and VCCL rules of \cref{sec:Layout}. Boundary and corner checks are completed using the rules in \cref{sec:MBoundaryStabs}.

Stabilizer placement begins at the root patch selected in \cref{subsec:MCFAlgo}. If the measurement acts as $X$ on this patch, we choose a vertical boundary adjacent to the route and extend standard weight-four stabilizers along the route. If the measurement acts as $Z$, we instead choose a horizontal boundary, add a row of domain walls according to \cref{subsec:ZMeasRules}, and then continue with standard weight-four stabilizers along the route.

Elongated stabilizers can also be required away from checkpoint lines. When connecting a vertical green strip to the vertical boundary of a patch undergoing an $X$ measurement, standard weight-four stabilizers may produce a Pauli-type conflict: two horizontally adjacent stabilizers would have the same type. In this case, the weight-four stabilizers at the patch boundary are promoted to elongated stabilizers of the same type. \Cref{fig:XX_elongated_vs} shows two examples.

\subsection{Logical $Z$ measurement rules}
\label{subsec:ZMeasRules}

\begin{figure*}
    \centering
    \subfloat[\label{fig:Corner_v1EX}]{\includegraphics[width=0.1\textwidth]{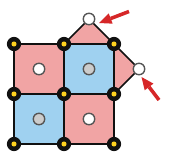}}
    \subfloat[\label{fig:Corner_v2EX}]{\includegraphics[width=0.2\textwidth]{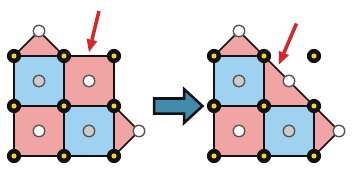}}
    \subfloat[\label{fig:Corner_v3EX}]{\includegraphics[width=0.15\textwidth]{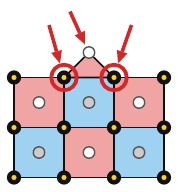}}
    \subfloat[\label{fig:Corner_v4EX}]{\includegraphics[width=0.15\textwidth]{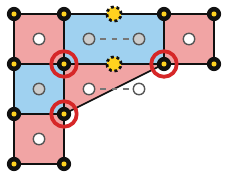}}
    \subfloat[\label{fig:Corner_v5EX}]{\includegraphics[width=0.15\textwidth]{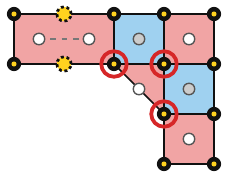}}
    \subfloat[\label{fig:Corner_v6EX}]{\includegraphics[width=0.125\textwidth]{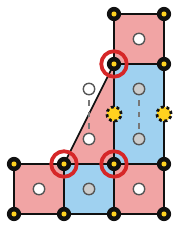}}
    \vfill
    \subfloat[\label{fig:Corner_v7EX}]{\includegraphics[width=0.15\textwidth]{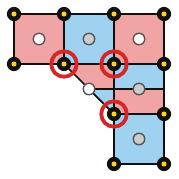}}
    \subfloat[\label{fig:Corner_v8EX}]{\includegraphics[width=0.125\textwidth]{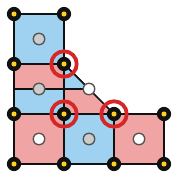}}
    \subfloat[\label{fig:Corner_v9EX}]{\includegraphics[width=0.4\textwidth]{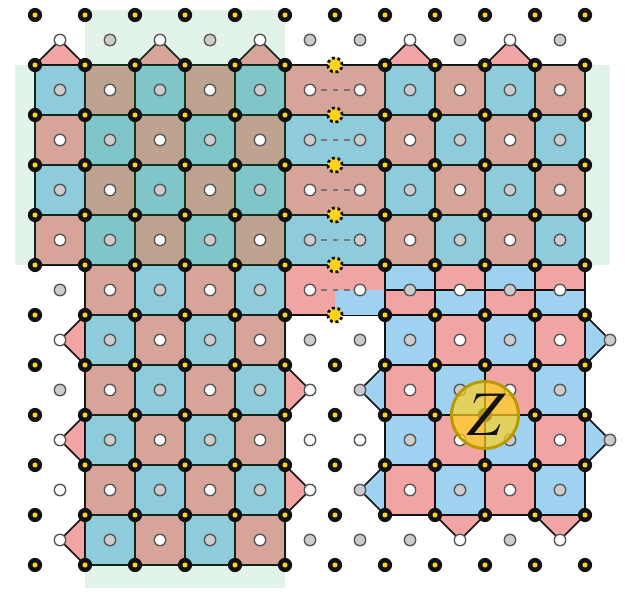}}
    \vfill
    \subfloat[\label{fig:Weight_4_X_JunctionEX}]{\includegraphics[width=0.4\textwidth]{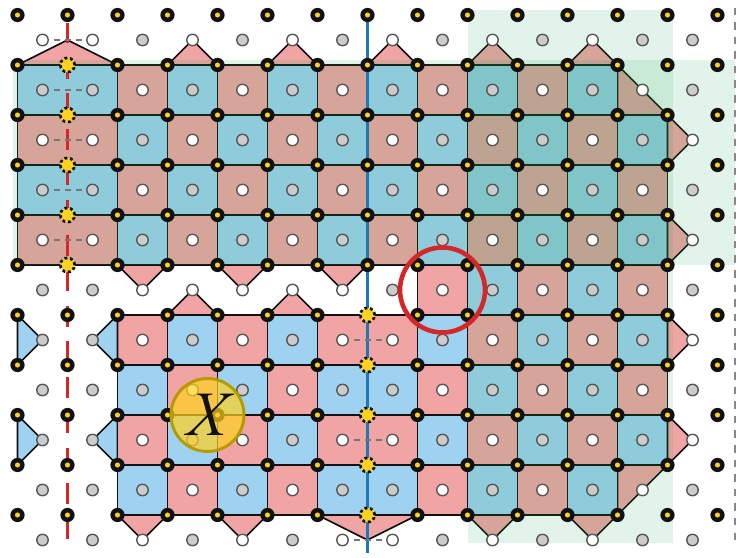}}
    \caption{Boundary and junction stabilizers. (a) Two weight-two $X$ checks complete the exposed edges of a corner weight-four $Z$ check. (b) A corner weight-four $X$ check is reduced to weight three by omitting its exposed corner qubit. (c) A weight-two $X$ check completes an exposed edge of a weight-four $Z$ check. (d)--(f) Weight-three $X$ checks complete corners formed by pairs of weight-four $Z$ checks, including elongated variants. (g), (h) Weight-three domain walls complete corners formed by a domain wall and a weight-four $Z$ check. (i) An elongated domain wall with three $X$ factors and one $Z$ factor completes a junction between two vertical strips. (j) A weight-four $X$ check (circled in red) completes a junction with weight-four $Z$ checks to its north, east, and south.}
    \label{fig:Corner_examps}
\end{figure*}

In this construction, logical $Z$ factors couple to horizontal patch boundaries through the \texttt{dw1} and \texttt{dw2} domain-wall stabilizers shown in \cref{fig:Legend}. At a bottom boundary, place weight-four domain walls according to the stabilizer geometry before the merge:
\begin{itemize}
    \item The $Z$ component acts on the bottom-left and bottom-right data qubits of a weight-four $X$-type stabilizer.
    \item The $X$ component acts on the bottom-left and bottom-right data qubits of a weight-four $Z$-type stabilizer.
\end{itemize}
At a top boundary, use the corresponding top-left and top-right data qubits instead.

The $Z$ components of these weight-four domain walls cover all but one boundary data qubit. Denote the remaining qubit by $q_b$. The final domain wall depends on the local geometry:
\begin{itemize}
    \item If $q_b$ is not at a corner where the route changes between horizontal and vertical directions, add a weight-two domain wall whose $Z$ component acts on $q_b$.
    \item If $q_b$ is at such a corner, add a weight-three domain wall whose unique $Z$ component acts on $q_b$.
\end{itemize}
Together, these checks supply the required logical-$Z$ boundary support for the joint Pauli measurement. Examples are shown in \cref{fig:XZZ_Meas_examps}.

\subsection{Boundary stabilizer rules}
\label{sec:MBoundaryStabs}

We complete the merged-patch boundaries by adding or modifying checks around the stabilizers constructed in \cref{subsec:ElongatedXMeas,subsec:ZMeasRules}. The local rules are:
\begin{itemize}
    \item \textbf{Corner $Z$ check.} If a weight-four $Z$-type stabilizer lies at a corner, add a weight-two $X$-type stabilizer on each of its two exposed edges (\cref{fig:Corner_v1EX}).

    \item \textbf{Corner $X$ check.} If a corner weight-four stabilizer is $X$-type, reduce it to weight three by removing from its support the corner data qubit that belongs to no $Z$-type stabilizer (\cref{fig:Corner_v2EX}).

    \item \textbf{Exposed edge of a $Z$ check.} If two data qubits $q_1$ and $q_2$ of a weight-four $Z$-type stabilizer form an exposed boundary edge, add the weight-two check $X_{q_1}X_{q_2}$. \Cref{fig:Corner_v3EX} illustrates the case in which $q_1$ and $q_2$ are the top-right and top-left data qubits.

    \item \textbf{Corner between two $Z$ checks.} If two weight-four $Z$-type stabilizers form a $90^\circ$ corner at the boundary, add a weight-three $X$-type stabilizer on the three boundary data qubits in the union of their supports (\cref{fig:Corner_v4EX,fig:Corner_v5EX,fig:Corner_v6EX}).

    \item \textbf{Corner between a domain wall and a $Z$ check.} Suppose a domain wall and a weight-four $Z$-type stabilizer form a $90^\circ$ boundary corner, with boundary data qubits $q_1$, $q_2$, and $q_3$ in the union of their supports. If $q_1$ lies in the support of the domain wall's $X$ component, add the weight-three domain-wall stabilizer $Z_{q_1}X_{q_2}X_{q_3}$ (\cref{fig:Corner_v7EX,fig:Corner_v8EX}).
\end{itemize}

\begin{figure*}
    \centering
    \subfloat[\label{fig:YX_MeasEX}]{\includegraphics[width=0.7\textwidth]{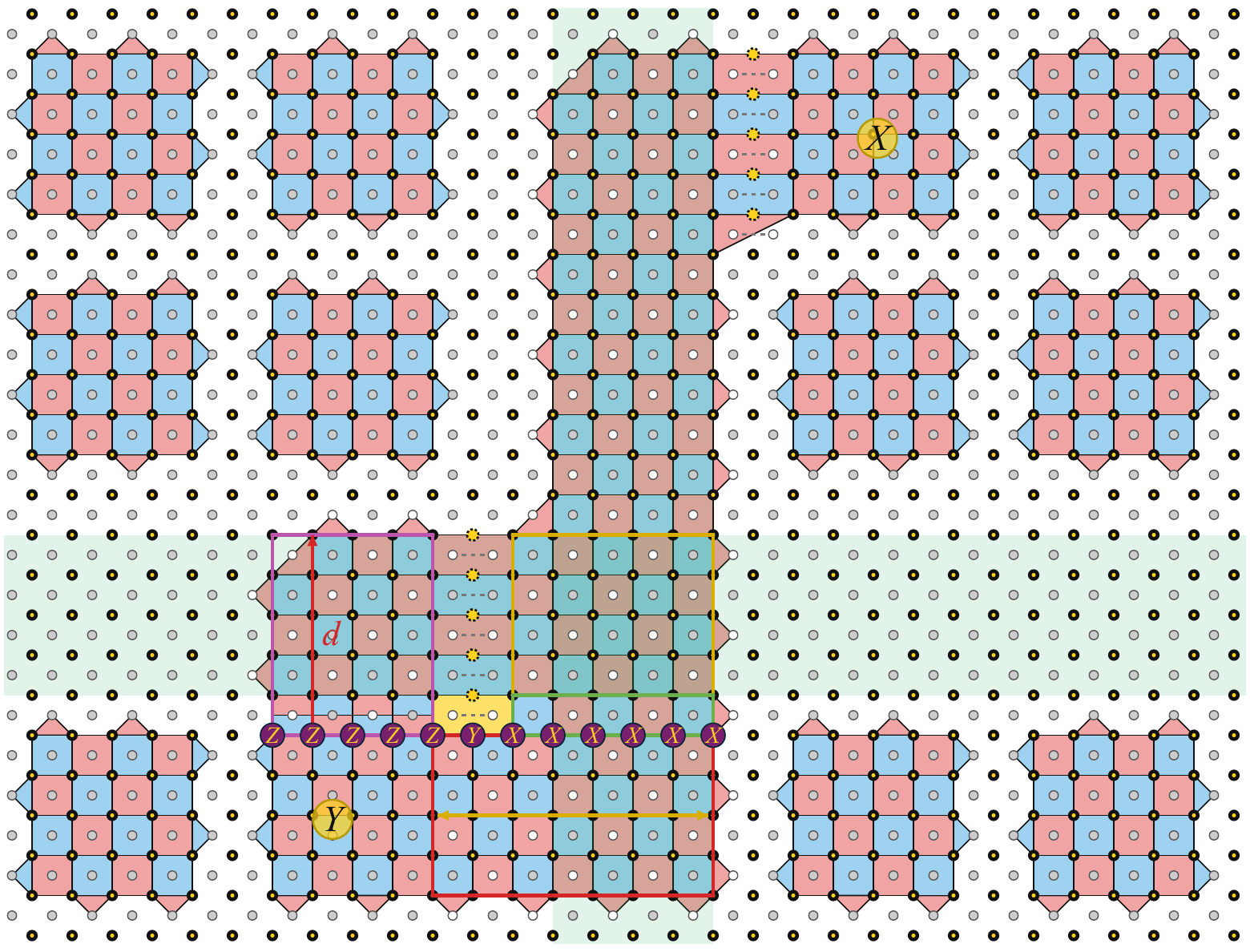}}
    \vfill
    \subfloat[\label{fig:Y_measure_mergeEX}]{\includegraphics[width=0.7\textwidth]{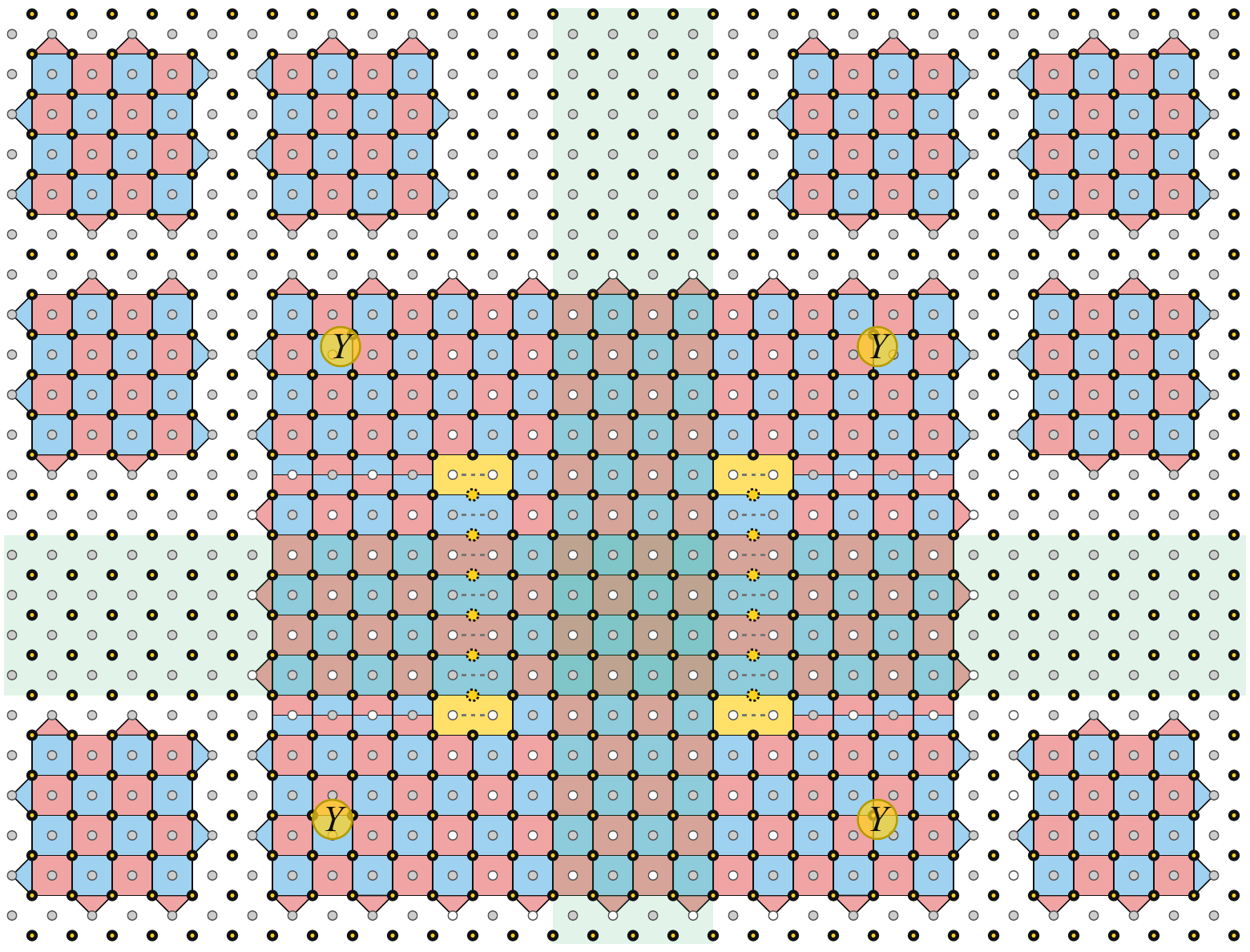}}
    \caption{(a) Example of a $Y_L \otimes X_L$ measurement. The Pauli operators in purple circles form a representative of the extended logical $Y_L$ operator. Colored boxes mark the regions introduced by the $Y$-extension procedure of \cref{subsec:standard_corridor_Y}. (b) Neighboring logical patches undergoing $Y$ extensions share horizontal and vertical routing corridors. Stabilizers shared by overlapping extensions are merged into single checks.}
    \label{fig:YX_Meas}
\end{figure*}

In the corner rule for two $Z$-type checks, one check may be an elongated rectangle. The added weight-three $X$ check is then an elongated triangle measured using two ancillas, as illustrated in \cref{fig:Corner_v4EX,fig:Corner_v6EX}.

\textbf{Facing weight-two boundary checks.}
If weight-two $X$ checks on facing horizontal corridor boundaries require the same ancilla, we resolve the conflict by introducing elongated columns at both VCCLs bounding the intervening vertical corridor, within one adjoining horizontal strip. The pair changes the checkerboard pattern between the two lines and restores it beyond the second line. Both columns must be compatible with the existing geometry and fixed checks, and the completed construction is checked again. The same procedure applies to either horizontal traversal direction.

\textbf{Oversized elongated corners.}
The elongated triangles described above remain permitted when their support can be reached by the prescribed measurement ancillas. If boundary completion instead produces a weight-three $X$ check whose support cannot be reached by its two measurement ancillas, as can occur where horizontal and vertical elongated $Z$ checks meet, we retry the construction using compatible alternative VCCL placements along the same routing tree. This additional search is used only when such an oversized corner is present. Every mandatory VICL column is retained in full, and the revised construction must satisfy the stabilizer and measurement-circuit constraints.

\textbf{Weight-four elongated domain wall.}
An additional rule applies where two vertical strips of the merged patch, separated by a single column of data-qubit sites, are joined by a horizontal strip. Let $S_e(Z)$ be a weight-four horizontal elongated $Z$ stabilizer connecting the vertical strips at this junction. Suppose a \texttt{dw1} stabilizer $S_d$ shares exactly one data qubit, $q_{j_1}$, with $S_e(Z)$ and acts as $Z$ on that qubit. Let $q_{j_2}$ be the data qubit in the support of $S_e(Z)$ across the inactive column from $q_{j_1}$ on the same horizontal row. Suppose a standard weight-four $Z$ stabilizer $S_{\text{st}}(Z)$ shares exactly this qubit with $S_e(Z)$.

Add an elongated weight-four domain-wall stabilizer $S_{\text{edw}}$ containing three $X$ factors and one $Z$ factor. Its two horizontally aligned $X$ factors act on $q_{j_1}$ and $q_{j_2}$, and its $Z$ factor acts on another data qubit shared with $S_d$. The remaining $X$ factor completes the elongated rectangle illustrated in \cref{fig:Corner_v9EX}.

\textbf{Weight-four $X$ check at a junction.}
Consider two horizontal strips of a merged patch separated by a single row of \textit{ancilla} sites and joined by a vertical strip. Let $S_1(Z)$ be a standard weight-four $Z$ stabilizer connecting the horizontal strips at the junction. Choose two data qubits $q_1$ and $q_2$ in its support that lie on the same column and form an edge facing the junction boundary. No additional stabilizers extend beyond this edge on its exposed eastward or westward side.

Suppose $q_1$ also belongs to a standard weight-four $Z$ stabilizer $S_2(Z)$, and $q_2$ belongs to a standard weight-four $Z$ stabilizer $S_3(Z)$. Let $q_3$ be the other data qubit of $S_2(Z)$ on the same row as $q_1$, and let $q_4$ be the other data qubit of $S_3(Z)$ on the same row as $q_2$. Complete the junction by adding the weight-four stabilizer $X_{q_1}X_{q_2}X_{q_3}X_{q_4}$, as shown in \cref{fig:Weight_4_X_JunctionEX}.

\section{Fault-tolerant implementation of logical $Y$ measurements}
\label{sec:FaultTolerantY}

We now describe the stabilizer constructions used for logical $Y$ measurements. The objective is to preserve the full effective code distance $d$, so that no set of at most $(d-1)/2$ circuit faults causes a logical failure. We first introduce the underlying geometry; the gate schedules and protruding-lobe refinements used to protect against circuit faults are presented in \cref{subsec:gate_schedule_y_ext,sec:protruding_lobes}. We refer to the standard-corridor stabilizer layout defined in \cref{subsec:standard_corridor_Y} as the ``legacy $Y$-extension patch.'' The numerical evidence for distance preservation and the alternatives for overlapping extensions are discussed below.

\subsection{Standard corridor $Y$ measurements}
\label{subsec:standard_corridor_Y}

For the layout in \cref{fig:Layout_Example}, a logical $Y$ measurement is implemented by extending the patch so that a representative of $Y_L$ lies along a single accessible boundary. We extend the patch horizontally to a width of $2d$, following the approach also considered in Ref.~\cite{Chamberland22b}. \Cref{fig:YX_MeasEX} illustrates a $Y_L\otimes X_L$ measurement: the Pauli operators marked in purple form a minimum-weight representative of the extended logical $Y_L$ operator. The elongated yellow rectangle denotes a weight-five twist stabilizer measuring $X\otimes X\otimes Z\otimes Y\otimes X$, with active data qubits ordered row by row from top to bottom and from left to right within each row. \Cref{fig:LegendEX} illustrates a twist defect and its physical measurement circuit.

We construct the local stabilizer regions for all logical $Y$ factors before completing the routing trees and placing the remaining routing-space stabilizers. We call this preprocessing the \textit{$Y$-extension step}. For each participating patch, the route first leaves horizontally toward an allowed vertical routing strip. Define $s=+1$ for a patch at the top of its unit cell and $s=-1$ for a patch at the bottom. Thus, $s$ specifies the outward vertical direction toward the adjacent horizontal routing strip: north for a top patch and south for a bottom patch.

\begin{enumerate}
    \item \textbf{Horizontal extension.} Extend standard weight-four stabilizers from the logical patch in the direction of the route until the measured region reaches the allowed vertical routing strip. This region is marked by the red box in \cref{fig:YX_MeasEX}.

    \item \textbf{Domain-wall row and vertical extension.} Place a horizontal row of domain walls along the outward-facing horizontal boundary of the original patch, using the logical-$Z$ construction. For $s=+1$, use the top boundary and extend standard weight-four stabilizers northward above the domain walls. For $s=-1$, use the bottom boundary and extend them southward below the domain walls. Continue until the full vertical extent of the allowed horizontal strip is covered. In particular, the northward extension must stop at the strip boundary; no additional rows are added beyond it. This stopping condition produces an asymmetry between extensions from top and bottom patches. The domain-wall region is marked by the purple box in \cref{fig:YX_MeasEX}.

    \item \textbf{Twist and elongated column.} Place a weight-five twist stabilizer adjacent to the domain wall nearest the vertical routing corridor. Immediately beyond the twist in direction $s$, place an elongated $Z$-type stabilizer, followed by an elongated $X$-type stabilizer. Continue the sequence $Z,X,Z,X,\ldots$ northward for $s=+1$ or southward for $s=-1$ until the full vertical extent of the allowed horizontal strip is covered. The yellow rectangle in \cref{fig:YX_MeasEX} denotes the twist defect.

    \item \textbf{Rows adjoining elongated stabilizers.} From each elongated stabilizer introduced in step 3, extend a horizontal row of standard weight-four stabilizers toward the vertical routing corridor, spanning the full width of its allowed strip. The first standard stabilizer has the opposite Pauli type from the adjoining elongated stabilizer; subsequent types alternate along the row. These rows are marked by the yellow box in \cref{fig:YX_MeasEX}.

    \item \textbf{Row adjoining the twist.} Add a horizontal row of standard weight-four stabilizers adjacent to the twist and extending toward the vertical routing corridor. This row begins with a $Z$-type stabilizer and alternates between $Z$ and $X$ until it spans the full width of the allowed vertical strip. The row is marked by the green box in \cref{fig:YX_MeasEX}.

    \item \textbf{Complete the routing construction.} Once all local $Y$ extensions have been generated and combined as described below, continue placing standard weight-four stabilizers along the horizontal and vertical routing corridors. Apply the $X$- and $Z$-measurement routing rules, together with the mandatory VICL rule of \cref{subsec:Intracell} and the VCCL, HICL, and HCCL rules introduced below.
\end{enumerate}

\begin{figure*}
    \centering
    \subfloat[\label{fig:Update_HCCL_YmeasEX}]{\includegraphics[width=0.6\textwidth]{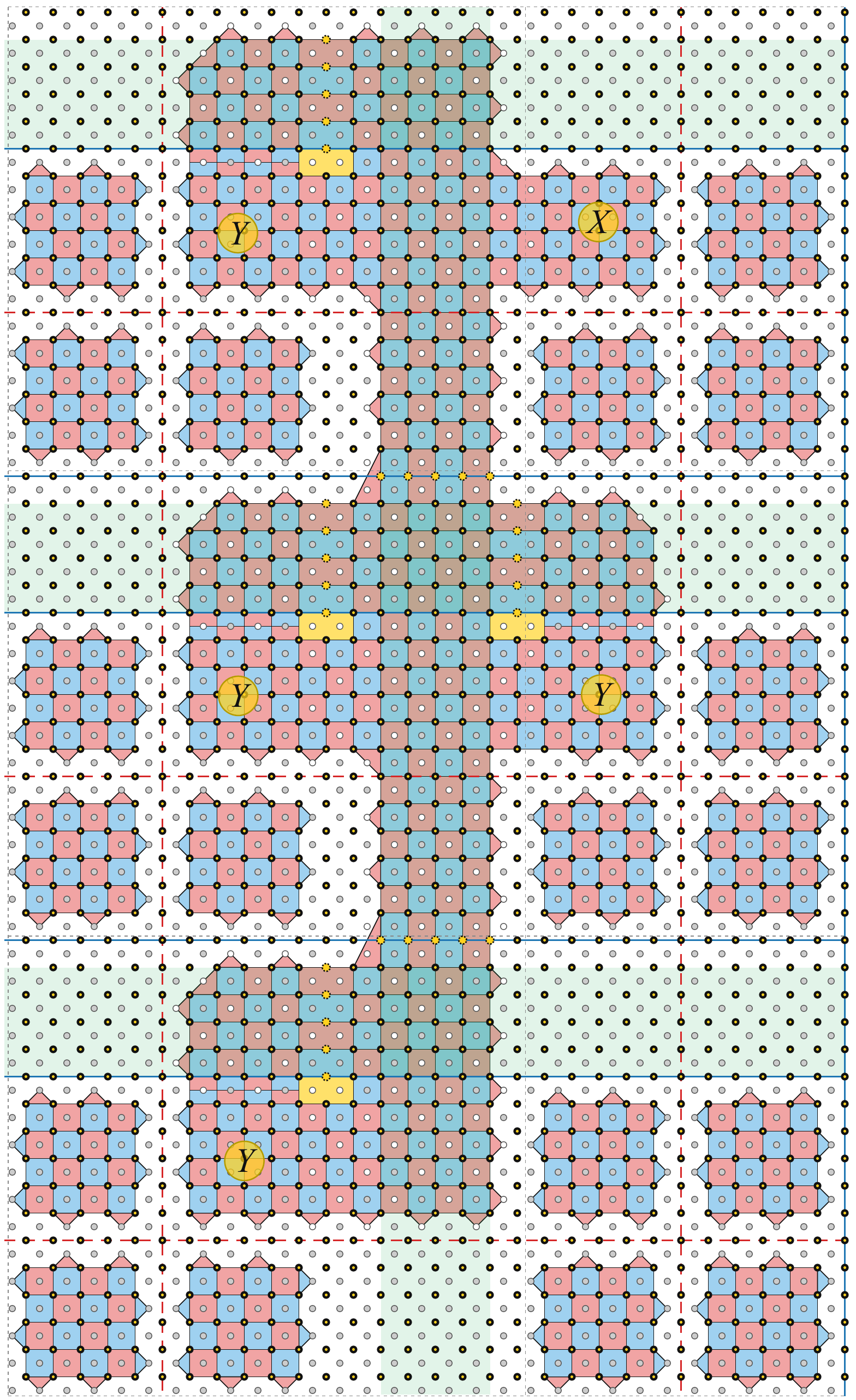}}
    \caption{Elongated checks introduced by the HICL and HCCL look-ahead rules when routing through regions containing fixed $Y$ extensions. The checkpoint rules reconcile the routing stabilizers with the stabilizers already present in these regions.}
    \label{fig:YUpdate_HCCL_Ymeas}
\end{figure*}

When several logical $Y$ factors of the same multi-qubit Pauli measurement have extensions entering a common routing corridor, generate their local constructions independently and then combine the measured regions. Stabilizers with identical support and Pauli action are identified and measured only once. For example, if extensions from opposite sides of a vertical corridor place the same $X$-type stabilizer on the same data qubits, the combined region contains a single $X$ check there. The same rule applies to extensions meeting across a horizontal corridor. \Cref{fig:Y_measure_mergeEX} shows an example.

The data-qubit supports and Pauli assignments of the legacy $Y$-extension stabilizers remain fixed during subsequent routing. This applies to standard weight-four checks, elongated checks, domain walls, and twist defects. If a routing strip overlaps an extension, identical stabilizers are included only once in the merged patch. The remaining routing stabilizers are matched to the fixed extension checks using the checkpoint rules introduced below. Where required, a standard weight-four routing stabilizer is replaced by an elongated stabilizer of the same Pauli type, allowing the checkerboard pattern beyond it to match the extension. Every mandatory VICL column remains fully elongated throughout this process. These rules also apply when a patch undergoing a $Y$ measurement is chosen as the root of the routing tree. The measurement circuits are assigned using the gate-scheduling rules described later.

The additional protruding lobes may be adjusted to prevent holes, as described in \cref{sec:protruding_lobes}. These adjustments leave the legacy $Y$ extension unchanged and retain access to the logical boundaries required by the Pauli measurement. Every connection in the final merged patch must be completed by a compatible set of stabilizers across the entire connecting strip, including its boundaries.

The following look-ahead rule applies initially to both HICLs and HCCLs when routing through regions containing fixed extensions.

\begin{mdframed}[frametitle={Look-ahead HICL and HCCL Rules}]
When extending standard weight-four stabilizers vertically across an HICL or HCCL, inspect the stabilizer location two steps beyond the checkpoint line in the traversal direction. Compare any existing stabilizer there with the newly added standard stabilizer immediately before the crossing.
\begin{itemize}
    \item \textbf{The inspected location is empty.} Ignore the checkpoint line and continue the standard checkerboard pattern.
    \item \textbf{The inspected stabilizer has the same type.} Continue the routing construction normally.
    \item \textbf{The inspected stabilizer has the opposite type.} Replace the standard weight-four stabilizer immediately before the checkpoint line by a vertical elongated stabilizer of the same type as the replaced check. Continue the construction using this elongated stabilizer.
\end{itemize}
\end{mdframed}

An example is shown in \cref{fig:YUpdate_HCCL_Ymeas}. For VCCLs, apply the same look-ahead rule in the horizontal traversal direction and use horizontal elongated stabilizers. These VCCL rules apply only outside the fixed $Y$-extension patches, whose stabilizer construction takes precedence over all checkpoint rules. If look-ahead finds a stabilizer at any inspected location, it determines the checkpoint choice for the entire connected crossing, including rows whose inspected locations are empty. When elongation is required, the elongated checks form a complete column spanning the full connected transverse extent of the merged patch, with alternating $X$ and $Z$ types. Only when all inspected locations are empty do we apply the standard VCCL rule of \cref{subsec:Intercell} to the entire crossing. A single connected crossing must not contain a mixture of elongated and standard weight-four checks along the same VCCL.

If the preceding rules leave anticommuting stabilizers at a fixed-extension boundary, or require an oversized weight-three $X$ stabilizer whose data qubits cannot all be reached by its measurement ancillas, we retry the construction using an earlier HICL along the same routing strip. Outside the fixed extensions, we replace a complete row of standard weight-four checks by vertical elongated checks of the same Pauli types, alternating between $X$ and $Z$ across the full connected width. This changes the checkerboard pattern beyond the HICL, allowing it to match the fixed extension. This fallback preserves the routing path, the legacy $Y$-extension checks, and the mandatory VICL columns. The resulting construction is accepted only after checking the complete stabilizer placement and gate schedule.

VICLs always obey the mandatory rule of \cref{subsec:Intracell}: the elongated column spans the full connected transverse extent of the merged patch at the crossing. Existing stabilizers and look-ahead constraints must be reconciled with this complete column; they cannot truncate it or replace part of it by standard weight-four checks. A construction that cannot satisfy these requirements is not accepted.

A related preprocessing step is used for top-left and top-right patches of a unit cell undergoing a logical $X$ measurement when the horizontal corridor above them will be occupied. Before completing the inter-patch routing, construct the required boundary extension to place the appropriate elongated stabilizers. We call this the \textit{$X$-extension step}. The supports and Pauli assignments of these extension checks are also fixed, and the HICL/HCCL look-ahead rules apply when subsequent routing encounters them.

\subsection{Doubled corridor $Y$ measurements}
\label{subsec:doubled_corridor_Y}

\begin{figure*}
    \centering
    \subfloat[\label{fig:double_gridEX}]{\includegraphics[width=.5\textwidth]{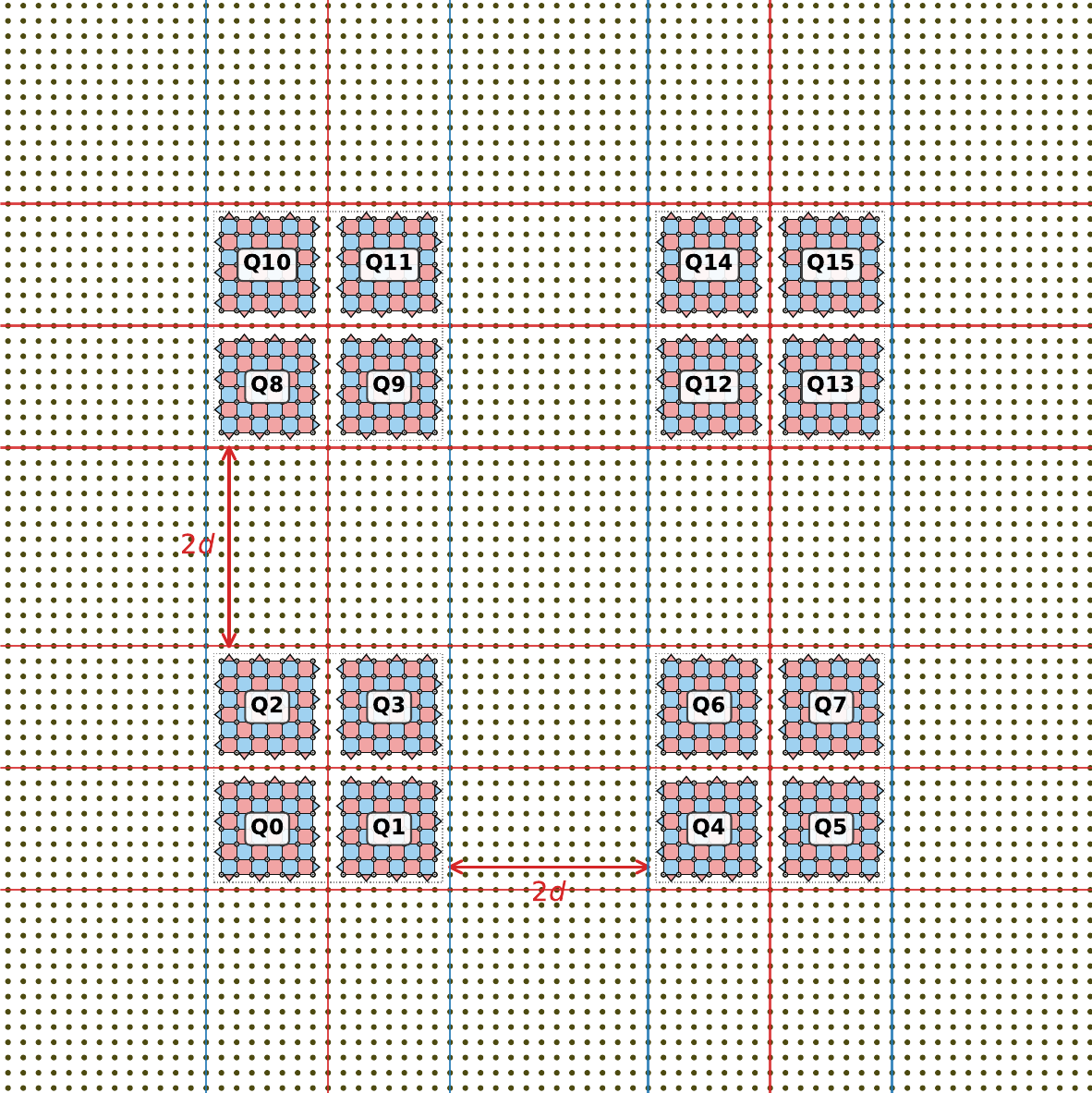}}
    \vfill
    \subfloat[\label{fig:Y_double_exampEX}]{\includegraphics[width=.5\textwidth]{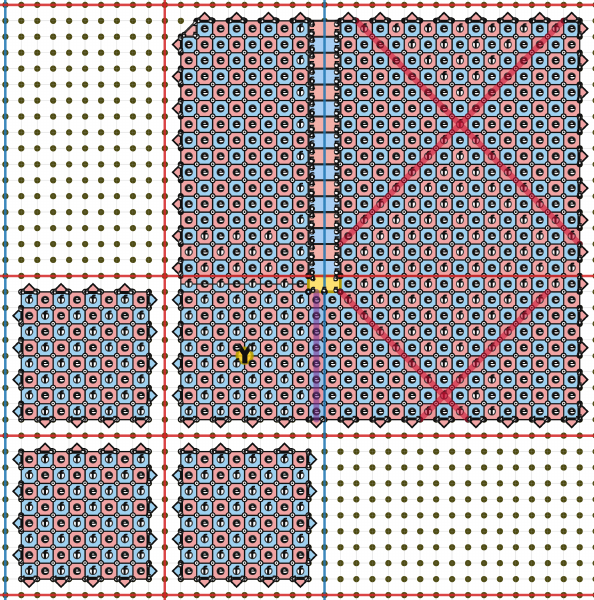}}
    \caption{(a) Enlarged layout with horizontal routing corridors of height $2d$ and vertical routing corridors of width $2d$. (b) Example for $d=9$, with the $Y$-extension region extending beyond the original logical patch by $2d-1$ rows and columns of data qubits. The additional space provides padding for the logical-$Y$ measurement construction.}
    \label{fig:double_grid}
\end{figure*}

The gate schedules and protruding lobes described in \cref{subsec:gate_schedule_y_ext,sec:protruding_lobes} are designed to preserve the full effective code distance in the presence of twist defects. Overlapping $Y$ extensions can require partitioning the gate schedules as described in \cref{sec:neighbor_Y} and adapting the protruding lobes as described in \cref{sec:protruding_lobes}. The effective distance must be checked for the resulting complete circuit. We consider two alternatives for these configurations.

\textbf{Option 1: Twist-free treatment of overlapping $Y$ factors.}
For nonoverlapping $Y$ extensions, retain the direct measurement construction with the gate schedules of \cref{subsec:gate_schedule_y_ext}. For overlapping extensions, use the twist-free scheme of Ref.~\cite{Chamberland22}, which introduces logical ancillas to decompose the required measurement into consecutive $X$- and $Z$-type measurements. If the overlapping $Y$ factors treated by this decomposition are even in number, only a $|0\rangle$ ancilla is required. If their number is odd, both $|0\rangle$ and $|Y\rangle$ ancillas are required. This approach requires additional ancilla preparation and approximately doubles the measurement time because the $X$- and $Z$-type measurements are performed consecutively. Implementation details are given in Ref.~\cite{Chamberland22}.

\textbf{Option 2: Enlarged routing corridors.}
Replace the horizontal corridors of height $d+1$ and vertical corridors of width $d+3$ by corridors of height and width $2d$, respectively, as shown in \cref{fig:double_gridEX}. The additional space allows the $Y$-extension region to be padded in both directions, extending beyond the original logical patch by $2d-1$ rows and columns of data qubits (\cref{fig:Y_double_exampEX}). This enlarged geometry provides padding for the $Y$-extension construction without the protruding lobes of \cref{sec:protruding_lobes}. We apply the gate-scheduling rules of \cref{subsec:gate_schedule_y_ext} to the enlarged legacy $Y$-extension patch, although other schedules may also be used.

The enlarged corridors also simplify stabilizer placement. The allowed strips remain $d$ columns wide in vertical corridors and $d$ rows high in horizontal corridors, but are centered within the larger corridors. This separation prevents boundary checks in the routing region from conflicting with those of the logical patches. The standard VCCL rule of \cref{subsec:Intercell} and the VICL rule of \cref{subsec:Intracell} can therefore both be omitted. Only the look-ahead rules of \cref{subsec:standard_corridor_Y} are retained to match the routing stabilizers to existing fixed $Y$-extension patches, inserting elongated checks where required.

In \texttt{patchwork}, the enlarged layout is selected by passing \texttt{--grid\_double "True"} to \texttt{ls\_driver.py}, with \texttt{--ancilla\_twist\_free} left at its default value of \texttt{"False"}. This sets the horizontal corridor height and vertical corridor width to $2d$, centers the $d$-high and $d$-wide stabilizer strips, and disables \texttt{--Y\_extra\_pad}, since the enlarged extension is used in place of protruding lobes.

Alternatively, \texttt{--ancilla\_twist\_free "True"} selects a fully twist-free implementation. This option takes precedence over \texttt{--grid\_double}, restores horizontal corridor height $d+1$ and vertical corridor width $d+3$, and bypasses all direct $Y$-extension and twist-defect constructions in favor of the ancilla-assisted decomposition into consecutive $X$- and $Z$-type measurements.

\section{Generation of spacelike logical observables}
\label{sec:Logic_Auto_Gen}

In this section, we describe the protocol used to automatically construct logical observable representatives for a multi-qubit Pauli measurement. Consider the measurement $P=P_1\otimes P_2\otimes\cdots\otimes P_N$, where $Q_j$ denotes the logical patch associated with $P_j$. The construction follows the boundary components of the merged patch. When several multi-qubit Pauli measurements are performed in parallel, the boundary components belonging to different merged patches are treated separately.

Let $\mathcal{D}$ denote the set of active physical data qubits in the merged patch, and let $\mathcal{S}(P)$ denote the complete set of active stabilizers after the merge. This includes stabilizers retained from the original logical patches, routing-space stabilizers, boundary stabilizers, domain walls, twist defects, elongated stabilizers, triangular stabilizers, and trapezoidal stabilizers. Stabilizers that are suppressed or replaced during the merge are excluded from $\mathcal{S}(P)$. For a Pauli operator $S$, define
\begin{align*}
    &\operatorname{Commutes}(S,\mathcal{S}(P))\\
    &\qquad=
    \begin{cases}
        \mathrm{true}, & [S,g]=0\quad\forall g\in\mathcal{S}(P),\\
        \mathrm{false}, & \text{otherwise}.
    \end{cases}
\end{align*}

\subsection{Boundary representation of the merged patch}

Each connected boundary component is represented by a cyclically ordered sequence of lattice coordinates
\begin{align*}
    \partial\mathcal{D}_{\alpha}
    &= (b_0,b_1,\ldots,b_{M_{\alpha}-1}),
\end{align*}
with indices understood modulo $M_{\alpha}$. The reference ordering keeps the merged-patch region on the left of the directed boundary. Consequently, exterior boundaries are ordered counterclockwise, while boundaries surrounding holes are ordered clockwise.

We specify the direction of a walk relative to this reference ordering using $d_{\mathrm{walk}}\in\{-1,+1\}$. For an ordinary boundary step,
\begin{align*}
    \operatorname{Next}_{\partial}(b_i,d_{\mathrm{walk}})
    &= b_{(i+d_{\mathrm{walk}})\bmod M_{\alpha}}.
\end{align*}
Thus, on an exterior boundary, $d_{\mathrm{walk}}=+1$ gives counterclockwise motion and $d_{\mathrm{walk}}=-1$ gives clockwise motion. The corresponding geometric directions are reversed on a hole boundary.

Boundary coordinates need not all correspond to active data qubits. For example, the geometric perimeter of an elongated stabilizer can pass an inactive data-qubit position. Such a position is traversed geometrically but is not added to the Pauli string. Supported triangular and weight-three elongated checks also provide outer-edge continuations that can bypass intermediate coordinates of the geometric contour. These continuations follow the selected outer support edge rather than adding qubits along an artificial rectangular outline.

For a weight-four trapezoidal stabilizer, the boundary is constructed from its physical support polygon. When the walk follows an exposed slanted edge, it includes the two support qubits at the ends of that edge. It does not replace the slanted edge by the rectangular bounding contour of the trapezoid.

Define the complementary Pauli operation by $\overline{X}=Z$ and $\overline{Z}=X$, and let $\mathcal{E}_{\mathrm{DW}}$ denote the boundary edges across which the walk crosses a domain wall. When the walk crosses a domain wall, the running Pauli label is exchanged according to $X\leftrightarrow Z$. This applies to both standard and elongated domain walls. Inactive positions are not included in the Pauli string, but domain-wall crossings encountered while traversing those positions are still taken into account.

\subsection{Common boundary-walk procedure}

A boundary-walk seed specifies a starting qubit $q_{\mathrm{start}}$, a coordinate $q_{\mathrm{join}}$ at which an initial straight segment joins the cyclic boundary, a traversal direction $d_{\mathrm{walk}}$, and an initial running Pauli label $\tau_0\in\{X,Z\}$. It also specifies the physical Pauli $\sigma_0$ placed at the starting qubit. Normally $\sigma_0=\tau_0$; the $Y$-corner construction described below instead uses $\sigma_0=Y$ and $\tau_0=X$.

The initial segment is horizontal or vertical. Its direction is determined by $q_{\mathrm{start}}$ and $q_{\mathrm{join}}$, whereas $d_{\mathrm{walk}}$ determines the subsequent direction around the boundary. The coordinate $q_{\mathrm{join}}$ is not a prescribed final endpoint. The walk can terminate earlier, or continue beyond it, according to the commutation test.

We describe the accumulated operator by its local Pauli $S(q)$ at each data qubit. Adding a qubit to the candidate means assigning its local Pauli; all other local Paulis remain unchanged. The algorithm is as follows.
\begin{enumerate}
    \item Initialize $S\leftarrow I$ and $\tau\leftarrow\tau_0$. Set
    \begin{align*}
        S(q_{\mathrm{start}})&\leftarrow\sigma_0.
    \end{align*}
    If $\operatorname{Commutes}(S,\mathcal{S}(P))=\mathrm{true}$, output $S$ and terminate.

    \item Follow the initial straight segment towards $q_{\mathrm{join}}$. Before adding each new site, toggle $\tau$ whenever the traversed edge belongs to $\mathcal{E}_{\mathrm{DW}}$. At each retained data qubit $q$, set
    \begin{align*}
        S(q)&\leftarrow\tau.
    \end{align*}
    Recognized inactive positions associated with elongated checks are omitted from the Pauli string. After every addition, output $S$ and terminate if it commutes with every stabilizer in $\mathcal{S}(P)$.

    \item If no commuting representative has been found upon reaching $q_{\mathrm{join}}$, locate the boundary cycle containing that coordinate. If no such cycle exists, return a failure flag for the seed.

    \item Continue along that cycle in direction $d_{\mathrm{walk}}$, using the supported outer-edge continuations at triangular or elongated checks when applicable. Apply the same domain-wall toggle and local-Pauli assignment rules. After every added data qubit, output $S$ and terminate if it commutes with every stabilizer in $\mathcal{S}(P)$.

    \item Do not add the starting qubit a second time. If the continuation returns to $q_{\mathrm{start}}$ before finding a commuting string, return a failure flag. The continuation is also bounded by $M_{\alpha}-1$ iterations, where an outer-edge shortcut can advance past several stored boundary coordinates in one iteration. Exhausting this bound without finding a commuting string also returns a failure flag.
\end{enumerate}

The walk is not required to terminate at a predetermined logical patch. It retains the first candidate along its prescribed traversal that commutes with every active stabilizer. A failed seed does not establish that no logical representative exists; it only means that this particular construction did not produce one.

\subsection{Logical $X$ representatives}

For a logical patch not undergoing a $Y$ measurement, a logical $X$ representative begins with a vertical $X$ string. The routing-space-facing vertical boundary is used as the primary seed, while the opposite vertical boundary is used as an independent alternate seed. Both walks are performed.

The routing-space-facing sides of the four patches in a unit cell are
\begin{center}
\small
\setlength{\tabcolsep}{4pt}
\begin{tabular}{c|cc}
    Patch position & $X_L$ side & $Z_L$ side\\
    \hline
    Top-left     & West & North\\
    Top-right    & East & North\\
    Bottom-left  & West & South\\
    Bottom-right & East & South
\end{tabular}
\end{center}
The alternate vertical side is obtained by interchanging east and west.

For a selected vertical side, let $q^{\mathrm{S}}$ and $q^{\mathrm{N}}$ denote its southern and northern corner data qubits. If an active stabilizer face lies immediately below $q^{\mathrm{S}}$, the initial segment is traversed from $q^{\mathrm{N}}$ to $q^{\mathrm{S}}$. Otherwise, it is traversed from $q^{\mathrm{S}}$ to $q^{\mathrm{N}}$. The corresponding traversal directions are
\begin{center}
\begin{tabular}{c|c|c}
    Selected side & Initial traversal & $d_{\mathrm{walk}}$\\
    \hline
    West & South to north & $-1$\\
    West & North to south & $+1$\\
    East & South to north & $+1$\\
    East & North to south & $-1$
\end{tabular}
\end{center}
The starting and running Pauli labels are $\sigma_0=\tau_0=X$.

Using both vertical boundaries is important because the two walks can terminate on different logical patches and produce distinct logical products. The two physical strings are treated separately rather than combined. Both successful representatives are retained when their logical products differ. If one seed fails, this does not invalidate a successful representative obtained from the other seed.

\subsection{Logical $Z$ representatives}

For a logical patch not undergoing a $Y$ measurement, the logical $Z$ construction is the horizontal analogue of the logical $X$ construction. The routing-space-facing horizontal boundary is used as the primary seed, while the opposite horizontal boundary is used as an independent alternate seed. Both walks are evaluated, with $\sigma_0=\tau_0=Z$.

For a selected horizontal side, let $q^{\mathrm{W}}$ and $q^{\mathrm{E}}$ denote its western and eastern corner data qubits. If an active stabilizer face lies immediately to the west of $q^{\mathrm{W}}$, the initial segment is traversed from $q^{\mathrm{E}}$ to $q^{\mathrm{W}}$. Otherwise, it is traversed from $q^{\mathrm{W}}$ to $q^{\mathrm{E}}$. The corresponding traversal directions are
\begin{center}
\begin{tabular}{c|c|c}
    Selected side & Initial traversal & $d_{\mathrm{walk}}$\\
    \hline
    South & West to east & $+1$\\
    South & East to west & $-1$\\
    North & West to east & $-1$\\
    North & East to west & $+1$
\end{tabular}
\end{center}

As in the logical $X$ construction, the two horizontal seeds are treated as separate candidates. Successful walks beginning on opposite horizontal boundaries are both retained whenever they produce distinct logical products.

\subsection{Logical representatives for $Y$-extension patches}

A logical patch undergoing a $Y$ measurement uses a separate vertical-walk construction. Its designated starting qubit is the corner of the original logical patch diagonally opposite the two routing-space-facing sides occupied by the $Y$ extension. The four choices are
\begin{center}
\begin{tabular}{c|c}
    Patch position & Starting corner\\
    \hline
    Top-left & Southeast\\
    Top-right & Southwest\\
    Bottom-left & Northeast\\
    Bottom-right & Northwest
\end{tabular}
\end{center}

Let $v_j\in\{\mathrm{E},\mathrm{W}\}$ and $h_j\in\{\mathrm{N},\mathrm{S}\}$ denote the routing-space-facing vertical and horizontal sides of the original patch. The starting corner lies on the vertical side opposite $v_j$ and the horizontal side opposite $h_j$.

The preferred seed starts at this corner with $\sigma_0=\tau_0=X$ and proceeds vertically in direction $h_j$. Its proposed joining coordinate is the nearest coordinate beyond the original patch on that vertical ray that belongs to the boundary cycle containing the starting corner. Unless the commutation test terminates the walk earlier, the initial segment reaches this coordinate and the walk then continues around the selected boundary. The direction is chosen using the same vertical-side convention as above.

This construction uses the boundary of the actual merged geometry, including protruding lobes when present (see \cref{sec:protruding_lobes}), rather than the outline of the original $Y$-extension rectangle. The same boundary representation is used when the geometry includes the wider routing-space corridors discussed in \cref{subsec:standard_corridor_Y}. 

If the preferred construction does not produce an accepted representative, a second construction is tried on the same vertical side of the original patch. It uses the ordinary vertical seed rule, but places a physical $Y$ at the starting corner:
\begin{align*}
    \sigma_0&=Y, & \tau_0&=X.
\end{align*}
The remaining sites follow the running $X/Z$ rule and the same commutation test. This candidate is accepted only if its starting qubit is the designated corner given above. In particular, this corner construction can produce a logical $Y$ representative for an isolated $Y$-extension patch.

For either successful vertical construction, the logical factor assigned to the origin patch depends on the final endpoint. If the endpoint lies within the original extent of $Q_j$, the origin factor is $Y_L^{(j)}$. Otherwise, it is $X_L^{(j)}$. Thus, initializing the running Pauli as $X$ does not imply that the resulting representative must belong to the logical $X$ family. Additional logical factors are assigned from the other participating patches encountered by the walk, as described below.

The logical $Z$ construction for a $Y$-extension patch evaluates both horizontal boundaries of the original patch, with $\sigma_0=\tau_0=Z$. The routing-space-facing boundary is initially preferred. If its initial horizontal segment intersects the support of a twist defect, the opposite horizontal boundary is tried first instead. This changes the order of the candidates; it does not remove either candidate. Each walk uses the common boundary-walk procedure, and every successful distinct logical product is retained. The origin factor for these horizontal walks is $Z_L^{(j)}$.

\subsection{Consistent endpoint boundaries}
\label{subsec:logical_endpoint_boundaries}

\begin{figure*}[t]
    \centering
    \includegraphics[width=0.85\textwidth]{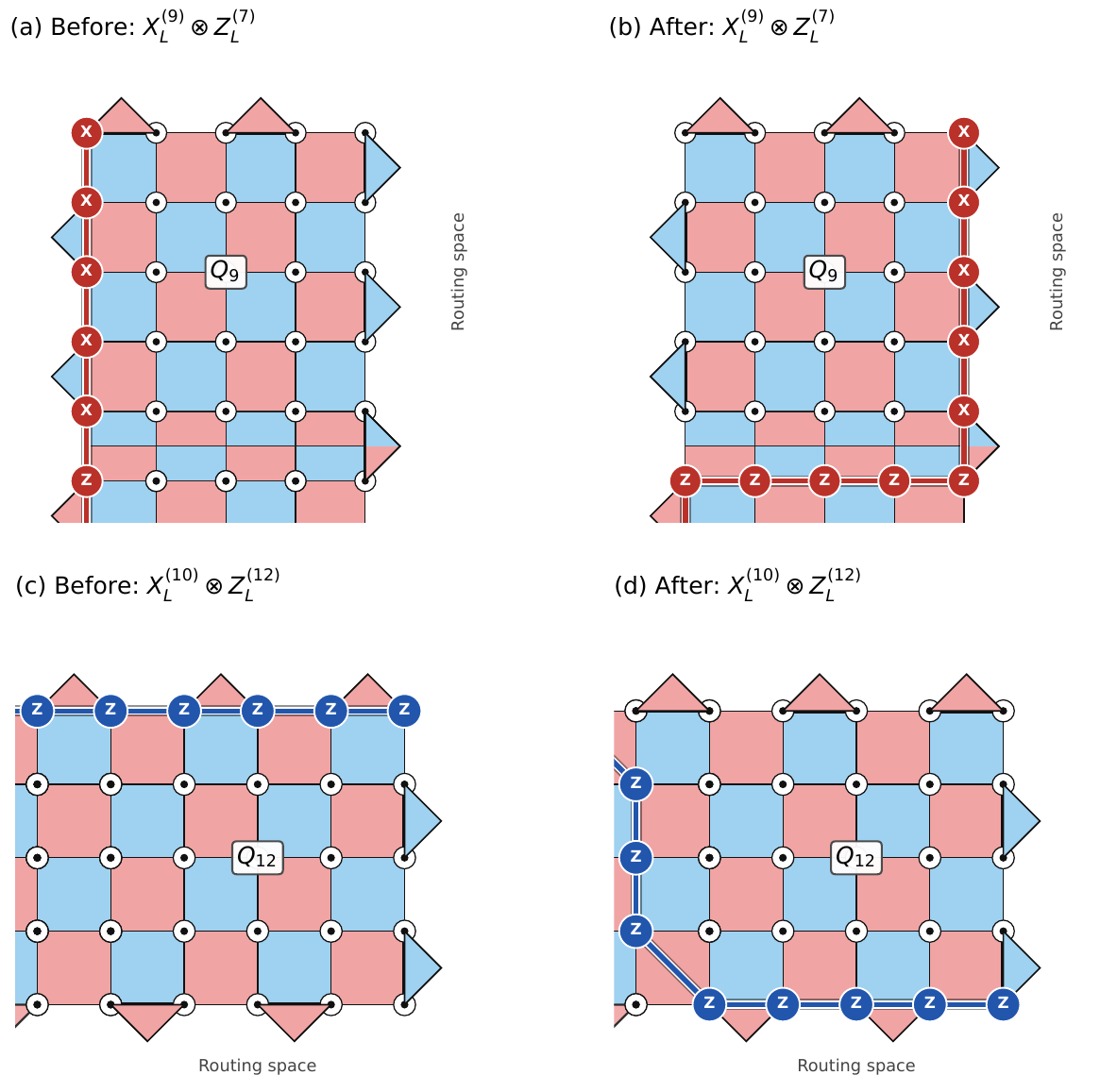}
    \caption{Shifting spacelike logical strings to the patch boundaries facing routing space. Panels (a,c) show the original representatives, and (b,d) show the corrected representatives. (a,b) The vertical $X$ segment on $Q_9$ moves from the left boundary to the right boundary by multiplying the full representative by retained patch $X$ stabilizers and adjoining weight-four domain-wall stabilizers.
(c,d) The horizontal $Z$ segment on $Q_{12}$ moves from the top boundary to the bottom boundary using retained patch $Z$ stabilizers. Both transformations preserve the logical action of the full representative.}
    \label{fig:logical_endpoint_shifts}
\end{figure*}

After a successful walk, each string connecting distinct original logical patches is corrected independently at both ends. This includes the starting patch when an alternate seed is used. At an endpoint patch $Q_j$, a complete horizontal $Z$ chain on the boundary opposite the routing-space-facing horizontal side is corrected by first multiplying the full representative by all retained pure-$Z$ stabilizers supported entirely within the original patch $Q_j$. The analogous correction for a complete vertical $X$ chain uses the retained pure-$X$ stabilizers supported entirely within $Q_j$. Suppressed or replaced checks are excluded.

For a horizontal-$Z$ correction, we additionally multiply by all active pure-$Z$ stabilizers with support both within $Q_j$ and in routing space, but none in any other original logical patch. For a vertical-$X$ correction, we additionally multiply by all active weight-four $X/Z$ domain-wall stabilizers whose $X$ support lies within the original patch $Q_j$ and whose $Z$ support lies in routing space, outside every original logical patch. The products are evaluated up to overall phase. Support outside the multiplied stabilizers is unchanged, and the resulting representative remains in the same logical class. Strings whose endpoints lie in the same original patch, including returning logical-$Y$ walks, retain their seed convention.

\cref{fig:logical_endpoint_shifts} illustrates these corrections: the vertical $X$ segment on $Q_9$ moves from left to right, and the horizontal $Z$ segment on $Q_{12}$ moves from top to bottom, placing both segments on their respective routing-space-facing boundaries. The additional $Z$ stabilizers move the adjoining vertical segment off $Q_{12}$ and into routing space.

Logical labels are assigned from the original walk before these corrections, and duplicate removal is performed afterwards.

\subsection{Logical labels and removal of duplicates}

\begin{figure*}[t]
    \centering
    \includegraphics[width=0.85\textwidth]{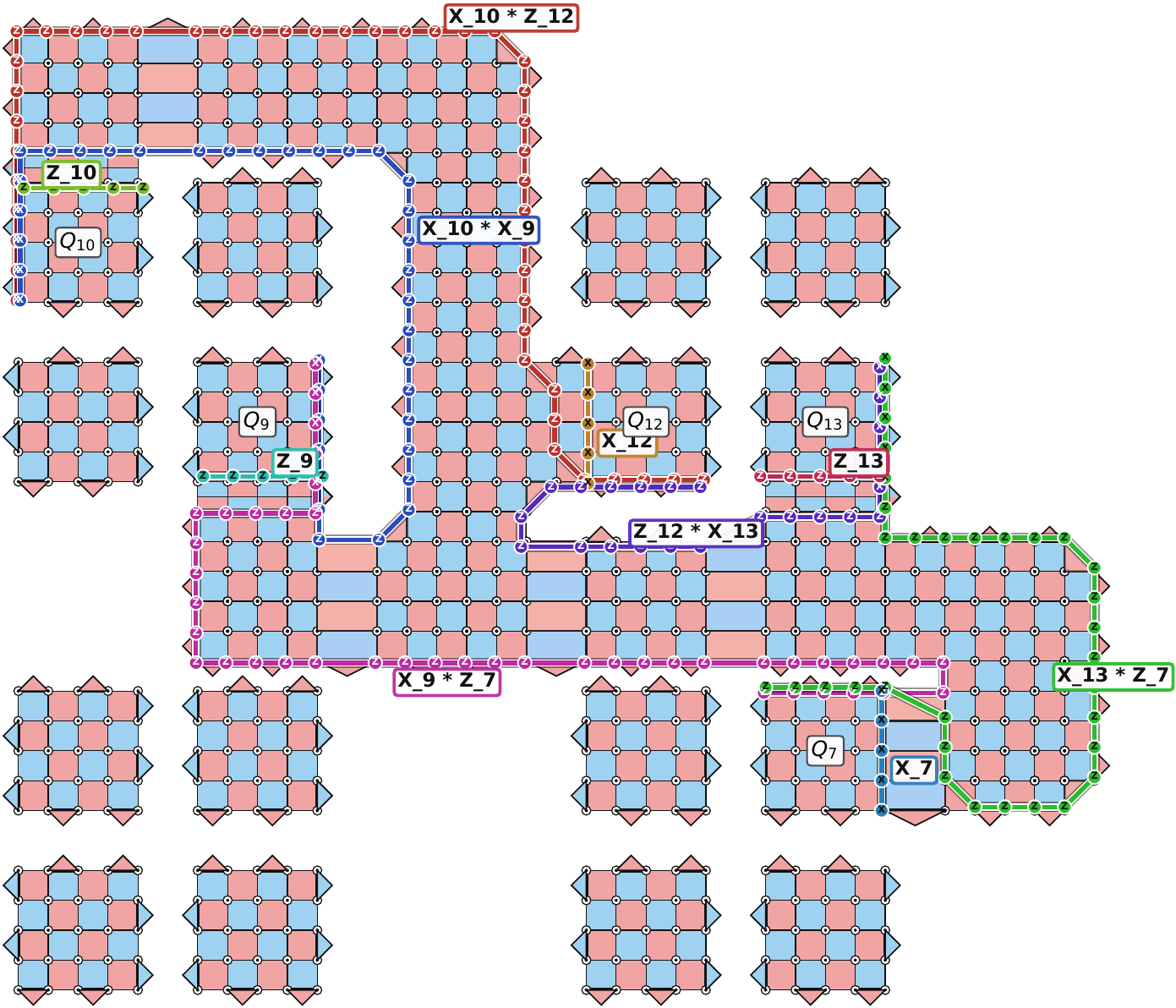}
    \caption{Example construction of spacelike logical observables on the full merged patch of a $P = Z_{Q_{10}} \otimes Z_{Q_{9}} \otimes X_{Q_{12}} \otimes Z_{Q_{13}} \otimes X_{Q_{7}}$ measurement.}
    \label{fig:v18_logicals}
\end{figure*}

A physical walk can pass through or terminate on logical patches other than its origin. Its logical product is therefore assigned using both the origin construction and the complete traversal sequence. Only logical patches participating in the requested Pauli measurement are considered.

The origin patch is first assigned its logical factor. This is $X$ for an ordinary vertical construction and $Z$ for a horizontal construction. For the vertical construction of a $Y$-extension patch, it is $X$ or $Y$ according to the endpoint rule above.

The original physical string, before endpoint correction, is then scanned in traversal order. Whenever it first enters the original extent of another participating logical patch $Q_k$, that patch is assigned the physical Pauli carried by the walk at the first encountered data qubit. Each logical patch is included at most once. Revisiting a patch does not change the factor already assigned to it.

Let $\mathcal{I}_X(S)$, $\mathcal{I}_Y(S)$, and $\mathcal{I}_Z(S)$ denote the sets of logical patches assigned $X$, $Y$, and $Z$, respectively. The logical label assigned to the physical string is
\begin{align}
    \mathcal{L}(S)
    &= \bigotimes_{k\in\mathcal{I}_X(S)}X_L^{(k)}
       \otimes\bigotimes_{k\in\mathcal{I}_Y(S)}Y_L^{(k)}
       \nonumber\\
    &\qquad\otimes\bigotimes_{k\in\mathcal{I}_Z(S)}Z_L^{(k)}.
    \label{eq:XL_True_Label}
\end{align}
For example, an $X$-family walk originating on $Q_j$ and entering $Q_k$ while carrying a $Z$ string is assigned the factors $X_L^{(j)}\otimes Z_L^{(k)}$. If it also enters $Q_t$ while carrying an $X$ string, the product additionally contains $X_L^{(t)}$.

The order of factors on different logical patches is immaterial. Consequently, $X_L^{(j)}\otimes Z_L^{(k)}$ and $Z_L^{(k)}\otimes X_L^{(j)}$ describe the same logical product. The implementation compares the order-independent collections of patch--Pauli pairs and retains the first successful representative of each distinct product. This removes duplicates obtained from opposite sides of a patch or from walks originating on different patches. Representatives with different logical products remain separate, even when their physical paths overlap.

This procedure is a deterministic constructor based on boundary walks and local stabilizer deformations. It is not an exhaustive search over all homology classes and does not attempt to minimize the weight of a representative. The commutation test checks compatibility with the active stabilizers; the logical-label rule describes how the implementation assigns the product on the participating patches. Duplicate removal compares these assigned products and does not perform a general classification modulo the stabilizer group.

An illustration of the generation of spacelike logical observables using the methods presented in this section on a merged patch resulting from a $P = Z_{Q_{10}} \otimes Z_{Q_{9}} \otimes X_{Q_{12}} \otimes Z_{Q_{13}} \otimes X_{Q_{7}}$ measurement is shown in \cref{fig:v18_logicals}.

In Patchwork, spacelike logical-observable construction and visualization are enabled by setting \texttt{--logical "True"}. Each retained logical representative is displayed using a distinct color so that overlapping physical paths can be distinguished.

\begin{figure}
     \centering
 \subfloat[\label{fig:timelike_log_exampEX} ]{\includegraphics[width=.4\textwidth]{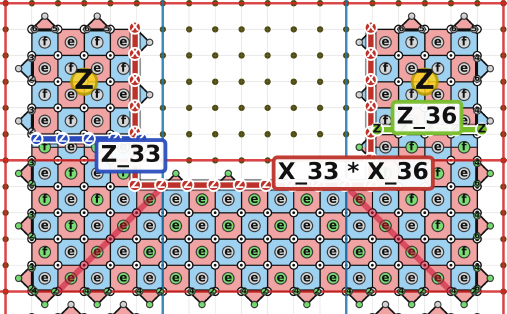}}
 \vfill
 \subfloat[\label{fig:timelike_log_examp2EX} ]{\includegraphics[width=.4\textwidth]{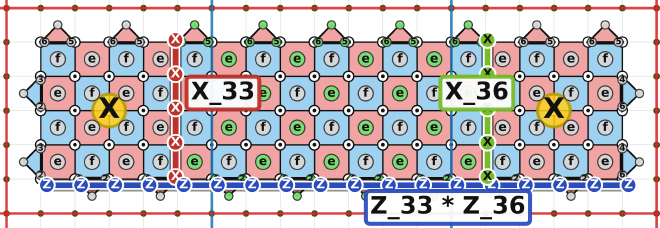}}
     \caption{ (a) Example of a $Z \otimes Z$ measurement performed between two logical qubits done in round $n_t$. Although their measurements are random, the product of all stabilizers with green ancillas encode the parity of $Z \otimes Z$ in the first round of the merged patch. The spacelike logical observables, and their corresponding labels, are shown by the red, green and blue strings of Pauli operators. (b) Example of an $X \otimes X$ measurement performed on the same logical qubits as in (a) but in a syndrome measurement round $n_s < n_t$. } 
     \label{fig:timelike_log_examp}
\end{figure}

\section{Generation of timelike logical observables}
\label{sec:Timelike_log_gen}

When performing a multi-qubit Pauli measurement via lattice surgery, errors can flip the parity of the measurement resulting in the wrong measured parity. Such a failure corresponds to a logical timelike failure. To keep track of the correct parity of a multi-qubit Pauli measurement, we need to properly define representatives for timelike logical observables. Such operators contain both a spatial and temporal component which we now define.

Let $P = P_{Q_1} \otimes \cdots \otimes P_{Q_N}$ be the multi-qubit Pauli being measured in round $n_t$. Let $T_P$ be the timelike logical observable for the $P$ measurement. The temporal component of $T_P$, labeled $T_P(t)$ corresponds to the product of all stabilizers applied in the routing space region which anti-commute with a subset of the stabilizers defining the logical qubits $\{Q_1, \cdots, Q_N\}$ prior to the $P$ measurements. The reason is that the product of all such stabilizers when $P$ is measured encode the parity of the $P$ measurement. We can then write
\begin{align}
    T_P(t) = \prod_{s \in S_{\text{rout}}} s(n_t),
    \label{eq:temporal_timelike}
\end{align}
where $s(n_t)$ are stabilizers in the routing space region in the syndrome measurement round $n_t$ whose product gives the parity of $P$. 

We label the spacelike component of $T_P$ as $T_P(s)$. Consider all syndrome measurement rounds from the moment a logical qubit is initialized and until $P$ is measured (i.e., rounds 1 to $n_t$). In this full history, we need to keep track of strings of spacelike errors which flip the parity of the $P$ measurement in round $n_t$. Such bookkeeping is done by storing the spacelike logical observables of each $P_{Q_j}$ term in $P$ for the full syndrome history. In other words, we write
\begin{align}
    T_P(s) = \overline{P}_{Q_1}(1{:}n_t) \otimes \cdots \otimes \overline{P}_{Q_N}(1{:}n_t),
    \label{eq:spatial_timelike}
\end{align}
where $\overline{P}_{Q_j}(1{:}n_t)$ corresponds to representatives of the spacelike logical observable of $P_{Q_j}$ in rounds 1 to $n_t$. Note that $\overline{P}_{Q_j}(t_1)$ is not necessarily equal to $\overline{P}_{Q_j}(t_2)$ when $t_1 \neq t_2$ since $Q_j$ can participate in multiple multi-qubit Pauli measurements \textit{prior} to measuring $P$. We also note that multiple representatives can be chosen for $\overline{P}_{Q_j}(t_j)$. As described in \cref{sec:Logic_Auto_Gen}, we always choose the representative that has support along the boundary in which $P_{Q_j}$ is measured (i.e., the boundary facing the routing space corridor used to measure $P_{Q_j}$). Putting everything together, we have that
\begin{align}
T(P) = T_P(s) \cdot T_P(t) 
\end{align}

In Patchwork, the stabilizers corresponding to $T_P(t)$ are obtained by setting \texttt{--stabilizers\_track "True"}. They can be visualized and are shown by stabilizers with green vertices for the ancilla qubits. 

An example of the above formalism is given in \cref{fig:timelike_log_examp}. In \cref{fig:timelike_log_exampEX}, we consider a $P_1 = Z \otimes Z$ measurement performed over $d$ rounds starting in round $n_t$. In this case, $\overline{Z}_{Q_1}(n_t)$ ($\overline{Z}_{Q_2}(n_t)$) is be given by the blue (green) string of $Z$ operators. Note also that these strings are chosen along the horizontal boundary facing the routing space corridor (following the convention mentioned above). In \cref{fig:timelike_log_examp2EX}, we consider a $P_2 = X \otimes X$ measurement performed on the same two logical qubits and over $d$ rounds but where the first round of the measurement occurs in round $n_s < n_t$. Notice that for this measurement, we have that both $\overline{Z}_{Q_1}(n_s{:}(n_s+d-1))$ and $\overline{Z}_{Q_2}(n_s{:}(n_s+d-1))$ are given by the wider blue string of $Z$ Paulis which contain additional Paulis compared to those in $\overline{Z}_{Q_1}(n_t)$ and $\overline{Z}_{Q_2}(n_t)$.

\section{Gate scheduling for distance preservation}
\label{sec:Gate_Schedule}

\begin{figure}
     \centering
 \subfloat[\label{fig:abANDefEX} ]{\includegraphics[width=.3\textwidth]{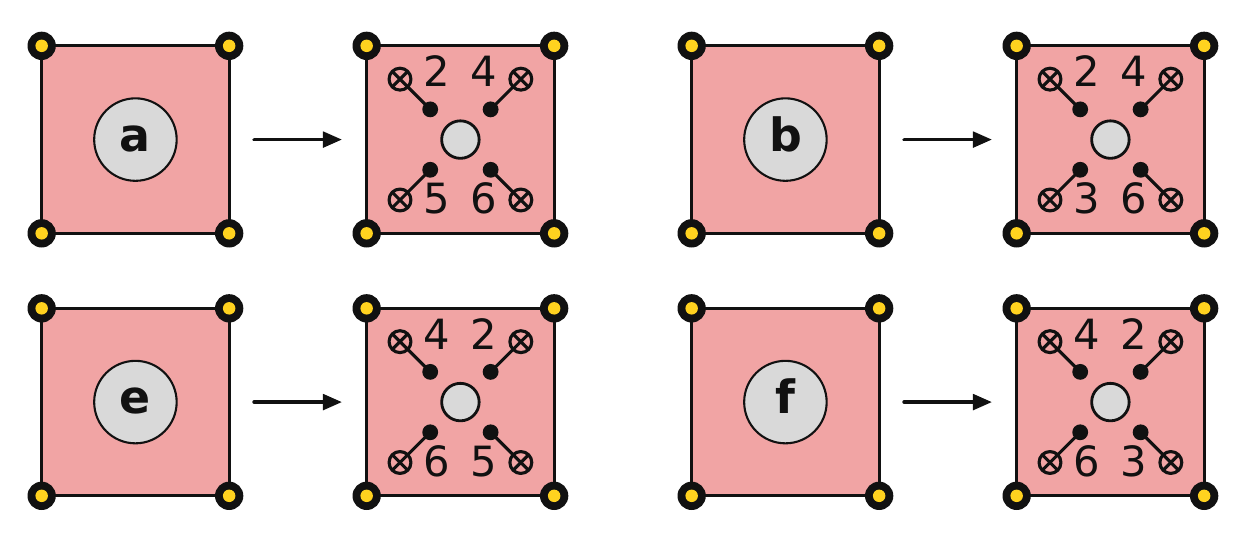}}
     \caption{ Examples of valid gate schedules for the surface code stabilizers when performing lattice surgery. We use five time steps to implement the two-qubit gates rather than the four time steps to avoid gate scheduling clashes. Several gate schedules are possible. Here we illustrate the \textbf{a} and \textbf{b} as well as \textbf{e} and \textbf{f} gate schedules. In this work we use an \textbf{e} and \textbf{f} gate schedules. } 
     \label{fig:abANDef}
\end{figure}

\begin{figure*}
     \centering
 \subfloat[\label{fig:CuttingCornersEX} ]{\includegraphics[width=.6\textwidth]{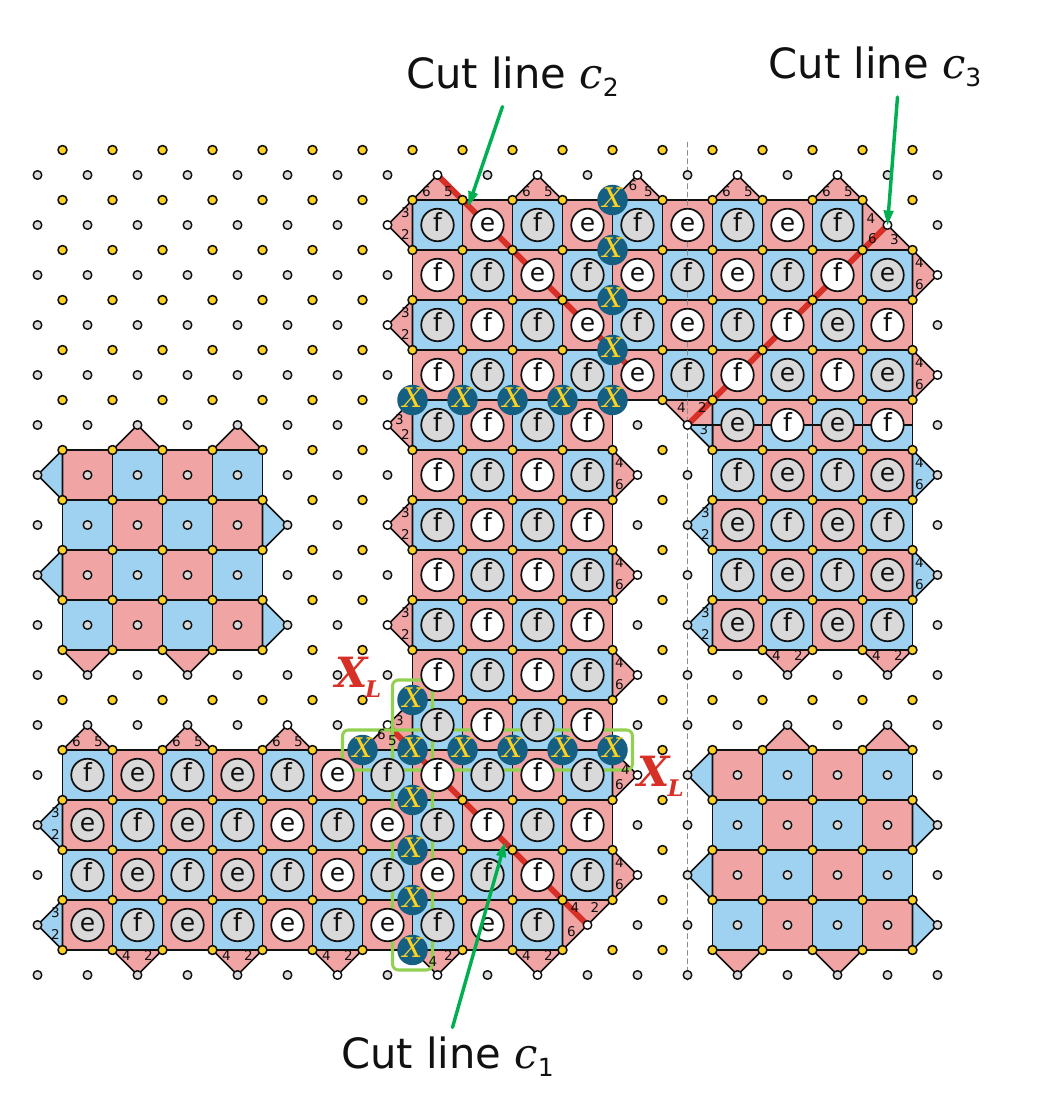}}
     \caption{ Multi-qubit Pauli measurement $X_L \otimes Z_L$ between two surface code patches. We show explicitly minimum-weight representatives of a logical observable $X^{(1)}_L$. Starting from the surface code patch at the bottom left of the merged patch, minimum-weight $X_L$ representatives form vertical strings. However, when transitioning to the vertical portion of the merge patch, they shift to horizontal strings and remain horizontal until encountering the next corner of the merged patch. Cut lines (shown by red lines) indicate where a change in gate schedule takes place. $X$-cut lines are shown following the construction in the main text.  } 
     \label{fig:CuttingCorners}
\end{figure*}

In this section we describe the gate scheduling protocol used in patchwork to perform lattice-surgery. In particular, we introduce the notion of a \emph{cut line}, which marks a transition between regions assigned different stabilizer-measurement schedules. Cut lines are introduced to ensure that errors arising from single faults propagate predominantly perpendicular to the minimum-weight representatives of the relevant logical observables.

Throughout this work, we use the \textbf{e} and \textbf{f} gate schedules shown in \cref{fig:abANDef}. Other gate schedules are also possible. For instance, the \textbf{a} and \textbf{b} schedules shown in \cref{fig:abANDef} also yield a valid surface-code memory circuit. Unlike the standard four-step surface-code schedule \cite{TomitaRotatedSurface2014}, an \textbf{e} and \textbf{f} schedules use five CNOT time steps. The additional time step allows adjacent \textbf{e} and \textbf{f} schedules to be stacked without clashes, i.e., without assigning multiple CNOT gates to the same data qubit during a single time step or violating the commutation relations required for stabilizer measurement.

This property is essential for lattice surgery because minimum-weight representatives of logical observables may change orientation within a merged patch. In particular, a logical observable that has a minimum weight representative represented by horizontal strings in one region may have a minimum weight representative represented by vertical strings in another region after traversing a corner of a routing corridor. To preserve the effective code distance, the stabilizer-measurement schedule must adapt to such changes.

The ZX-interleaving construction of Ref.~\cite{No_Hooks_Japan} preserves fault distance for regular stabilizer layouts using four CNOT layers. Our approach instead uses five layers and spatially varying schedules separated by cut lines to control hook-error propagation. We additionally address elongated stabilizers and twist-based $Y$ measurements through dedicated measurement circuits and protruding lobes, which extend beyond the scope of the regular-layout construction of Ref.~\cite{No_Hooks_Japan}.

Throughout this section, we label the top-left, top-right, bottom-left, and bottom-right data qubits in the support of a stabilizer by $q_{\text{tl}}$, $q_{\text{tr}}$, $q_{\text{bl}}$ and $q_{\text{br}}$, respectively.

\subsection{Gate schedules for standard weight-four stabilizers}
\label{subsec:standard_weight_4_gate_schedule}

For an $X$-type stabilizer measured using an \textbf{e} schedule, a single fault can produce a horizontal weight-two $X$ error. For an \textbf{f} schedule, a single fault can produce a vertical weight-two $X$ error. The same propagation rules apply to $Z$-type stabilizers, with $X$ errors replaced by $Z$ errors.

Since minimum-weight representatives of logical $X_L$ ($Z_L$) observables are vertical (horizontal) strings of $X$ ($Z$) errors, standard weight-four $X$-type stabilizers of logical patches are assigned an \textbf{e} schedule, while standard weight-four $Z$-type stabilizers are assigned an \textbf{f} schedule. This ensures that $X$ ($Z$) errors propagate perpendicular to their $X_L$ ($Z_L$) minimum-weight representatives. Weight-two boundary stabilizers are assigned schedules consistent with these conventions. Specifically, for weight-two $Z$-type stabilizers on the left boundary of a logical patch, the CNOT gates acting on $q_{\mathrm{tr}}$ and $q_{\mathrm{br}}$ occur during time steps $3$ and $2$, respectively. For weight-two $Z$-type stabilizers on the right boundary, the CNOT gates acting on $q_{\mathrm{tl}}$ and $q_{\mathrm{bl}}$ occur during time steps $4$ and $6$, respectively. For weight-two $X$-type stabilizers on the top boundary, the CNOT gates acting on $q_{\mathrm{bl}}$ and $q_{\mathrm{br}}$ occur during time steps $6$ and $5$, respectively. For weight-two $X$-type stabilizers on the bottom boundary, the CNOT gates acting on $q_{\mathrm{tl}}$ and $q_{\mathrm{tr}}$ occur during time steps $4$ and $2$, respectively.

These schedule assignments are propagated unchanged into routing-space regions until a cut line is encountered. Note that logical patches undergoing a $Y$ extension have their own gate scheduling rules which we describe below. The gate schedules used at $Y$ extensions are treated as static like any other logical qubit patches and should not be modified while traversing such regions. 

\textbf{Cut lines: }A cut line is a connected diagonal chain of stabilizers separating two regions that require different fault-propagation directions. Across a cut line, the schedule assigned to at least one stabilizer type changes from \textbf{e} to \textbf{f} or from \textbf{f} to \textbf{e}.

We distinguish between two types of cut lines:

\begin{itemize}
\item \textbf{$X$-cut lines:} only the schedules assigned to standard weight-four $X$-type stabilizers change across the cut line.

\item \textbf{Mixed cut lines:} the schedules assigned to both standard weight-four $X$-type and standard weight-four $Z$-type stabilizers change across the cut line.
\end{itemize}

The purpose of a cut line is to maintain fault propagation perpendicular to the local minimum-weight representatives of the relevant logical observables. Whenever the orientation of those representatives changes, a cut line is introduced to generate the corresponding schedule transition which maintains perpendicular error propagation.

An example is shown in \cref{fig:CuttingCorners}, where the parity of $X_L\otimes Z_L$ is measured via lattice surgery. The merged patch contains multiple corners where routing corridors change direction. Along the lower horizontal corridor, minimum-weight representatives of $X_L$ are vertical strings. After traversing the corner, the minimum-weight representatives become horizontal strings. Consequently, the preferred fault-propagation direction changes, requiring a transition between \textbf{e} and \textbf{f} schedules for $X$-type stabilizers. The red line labeled $c_1$ in \cref{fig:CuttingCorners} indicates the corresponding cut line.

Although some stabilizers immediately adjacent to a cut line may propagate faults parallel to a local minimum-weight representative, we verified numerically that the merged patch remains distance preserving. In particular, for a distance-$d$ surface code, at least $(d+1)/2$ faults are still required to produce a logical failure. 

Apart from areas used to perform a $Y$ extension, maintaining an \textbf{f} schedule for $Z$ stabilizers in the routing space region does not lower the effective distance of the code. This is due to our choice of initialization and readout bases for routing space data qubits, and the resulting construction of logical observables following the methods of \cref{sec:Logic_Auto_Gen}. Weight-two $Z$ errors arising from a single fault on a $Z$ stabilizer would be either parallel to a logical observable of weight $\ge 2d$, or perpendicular to a weight-$d$ logical $Z$ observable. 

\textbf{Schedule-transition rule for $X$-cut lines: }Suppose an $X$-cut line separates two regions $R_1$ and $R_2$.

If standard weight-four $X$-type stabilizers in $R_1$ are assigned an \textbf{e} schedule, then standard weight-four $X$-type stabilizers in $R_2$ are assigned an \textbf{f} schedule. Conversely, if standard weight-four $X$-type stabilizers in $R_1$ are assigned an \textbf{f} schedule, then standard weight-four $X$-type stabilizers in $R_2$ are assigned an \textbf{e} schedule. The schedules assigned to standard weight-four $Z$-type stabilizers remain unchanged when crossing an $X$-cut line.

\textbf{Schedule-transition rule for mixed cut lines: }Suppose a mixed cut line separates two regions $R_1$ and $R_2$.

If standard weight-four $X$-type stabilizers in $R_1$ are assigned an \textbf{e} schedule, then standard weight-four $X$-type stabilizers in $R_2$ are assigned an \textbf{f} schedule. Conversely, if standard weight-four $X$-type stabilizers in $R_1$ are assigned an \textbf{f} schedule, then standard weight-four $X$-type stabilizers in $R_2$ are assigned an \textbf{e} schedule. Similarly, if standard weight-four $Z$-type stabilizers in $R_1$ are assigned an \textbf{e} schedule, then standard weight-four $Z$-type stabilizers in $R_2$ are assigned an \textbf{f} schedule. Conversely, if standard weight-four $Z$-type stabilizers in $R_1$ are assigned an \textbf{f} schedule, then standard weight-four $Z$-type stabilizers in $R_2$ are assigned an \textbf{e} schedule.

\textbf{Placement of cut lines: }For merged patches that do not contain twist defects, cut lines are placed according to the following rules.

\textbf{Rule 1 ($X$-cut lines bounded by weight-three stabilizers).}
Suppose a routing corridor transitions between a horizontal segment of height $d$ and a vertical segment of width $d$. If both corner locations contain either a weight-three $X$-type stabilizer or a weight-three domain wall, construct an $X$-cut line along the diagonal chain of standard weight-four $X$-type stabilizers connecting the two corner locations. Standard weight-four $X$-type stabilizers along the cut line are assigned an \textbf{f} schedule. The schedule-transition rule for $X$-cut lines is then applied across the cut line. Examples are shown by cut lines $c_1$ and $c_3$ in \cref{fig:CuttingCorners}.

\begin{figure*}
     \centering
 \subfloat[\label{fig:Junction_example_v1EX} ]{\includegraphics[width=.5\textwidth]{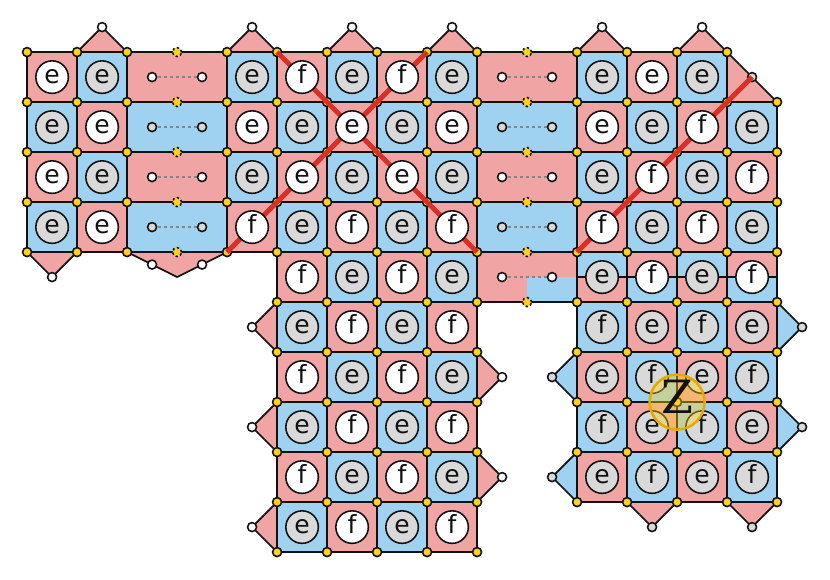}}
     \caption{ Example of two intersecting $X$-cut lines placed at a junction, along with the relevant gate schedules placed at standard weight-four stabilizers following the construction in \cref{subsec:standard_weight_4_gate_schedule}. } 
     \label{fig:Junction_example_v1}
\end{figure*}

\textbf{Rule 2 ($X$-cut lines bounded by weight-four $Z$ stabilizers).} Suppose a routing corridor transitions between a horizontal segment of height $d$ and a vertical segment of width $d$, and both corner locations contain standard weight-four $Z$-type stabilizers. Let $S_c(Z)$ denote the corner $Z$-type stabilizer adjacent to two weight-two $X$-type boundary stabilizers. Let $S_c(X)$ denote the standard weight-four $X$-type stabilizer adjacent to $S_c(Z)$ on the same horizontal row. Construct an $X$-cut line along the diagonal chain of standard weight-four $X$-type stabilizers passing through $S_c(X)$. Standard weight-four $X$-type stabilizers along the cut line are assigned an \textbf{e} schedule. The schedule-transition rule for $X$-cut lines is then applied across the cut line. An example is shown by cut line $c_2$ in \cref{fig:CuttingCorners}.

\textbf{Rule 3 ($X$-cut lines at junctions).}

\textbf{Junctions.}
A junction is a routing-space region where a horizontal corridor and a vertical corridor intersect. More precisely, a junction is a connected rectangular subregion containing a $d\times d$ array of data qubits such that a horizontal routing strip of height $d$ and a vertical routing strip of width $d$ overlap on this subregion. We refer to this $d\times d$ overlap as the junction cell.

Equivalently, in a grid representation of the merged patch, a junction cell is the intersection of a maximal horizontal routing strip and a maximal vertical routing strip, both of width $d$ in data-qubit units. The boundary of the junction cell has four distinct sides, corresponding to the four corridor arms incident on the junction: north, south, east, and west.

At each junction, two intersecting $X$-cut lines are introduced. These cut lines form an $X$-shaped pair of diagonals through the junction cell.

Let $S_{\mathrm{tl}}$ denote the top-left-most standard weight-four stabilizer contained in the junction cell. If $S_{\mathrm{tl}}$ is $X$-type, set $S_1(X) = S_{\mathrm{tl}}$. If $S_{\mathrm{tl}}$ is $Z$-type, let $S_1(X)$ be the standard weight-four $X$-type stabilizer immediately to the right of $S_{\mathrm{tl}}$.

The first $X$-cut line is the diagonal chain of $X$-type stabilizers beginning at $S_1(X)$ and extending in the south-east direction until $d-1$ $X$-type stabilizers have been included. The final stabilizer in this chain may be either a standard weight-four $X$-type stabilizer or an elongated $X$-type stabilizer. All \textit{standard weight-four} $X$-type stabilizers on this cut line are assigned an \textbf{e} schedule. Gate schedules for elongated stabilizers presented in \cref{subsec:elongated_rec_gate_schedule} are unaffected.

Similarly, let $S_{\mathrm{tr}}$ denote the top-right-most standard weight-four stabilizer contained in the junction cell. If $S_{\mathrm{tr}}$ is $X$-type, set $S_2(X) = S_{\mathrm{tr}}$. If $S_{\mathrm{tr}}$ is $Z$-type, let $S_2(X)$ be the standard weight-four $X$-type stabilizer immediately to the left of $S_{\mathrm{tr}}$.

The second $X$-cut line is the diagonal chain of $X$-type stabilizers beginning at $S_2(X)$ and extending in the south-west direction until $d-1$ $X$-type stabilizers have been included. Again, the final stabilizer in this chain may be either a standard weight-four $X$-type stabilizer or an elongated $X$-type stabilizer. All \textit{standard weight-four} $X$-type stabilizers on this cut line are assigned an \textbf{e} schedule. Gate schedules for elongated stabilizers presented in \cref{subsec:elongated_rec_gate_schedule} are unaffected.

Now, let
\[
C_{\searrow}
=
\{c^{\searrow}_0,c^{\searrow}_1,\ldots,c^{\searrow}_{d-2}\}
\] denote the $X$-cut line running from north-west to south-east, ordered from its north-west endpoint to its south-east endpoint. Similarly, let
\[
C_{\swarrow}
=
\{c^{\swarrow}_0,c^{\swarrow}_1,\ldots,c^{\swarrow}_{d-2}\}
\] denote the $X$-cut line running from north-east to south-west, ordered from its north-east endpoint to its south-west endpoint.

Suppose the two cut lines intersect at a stabilizer \[
p=c^{\searrow}_k=c^{\swarrow}_\ell.
\]

We define four boundary sets:

\begin{align}
V_{\mathrm{top}}
&=
\{c^{\searrow}_i : 0 \le i \le k\}
\cup
\{c^{\swarrow}_j : 0 \le j \le \ell\},
\\[1ex]
V_{\mathrm{bottom}}
&=
\{c^{\searrow}_i : k \le i \le d-2\}
\cup
\{c^{\swarrow}_j : \ell \le j \le d-2\},
\\[1ex]
V_{\mathrm{left}}
&=
\{c^{\searrow}_i : 0 \le i \le k\}
\cup
\{c^{\swarrow}_j : \ell \le j \le d-2\},
\\[1ex]
V_{\mathrm{right}}
&=
\{c^{\searrow}_i : k \le i \le d-2\}
\cup
\{c^{\swarrow}_j : 0 \le j \le \ell\}.
\end{align}

The sets $V_{\mathrm{top}}$ and $V_{\mathrm{bottom}}$ form the upper and lower $V$-shaped boundaries of the $X$-cut line at the junction, respectively. The sets $V_{\mathrm{left}}$ and $V_{\mathrm{right}}$ form the left and right sideways $V$-shaped boundaries of the $X$-cut line at the junction, respectively.

Standard weight-four $X$-type stabilizers in the connected component above $V_{\mathrm{top}}$ are assigned an \textbf{f} schedule. This assignment is propagated away from the junction throughout that component until another cut line is encountered. Similarly, standard weight-four $X$-type stabilizers in the connected component below $V_{\mathrm{bottom}}$ are assigned an \textbf{f} schedule and retain this assignment until another cut line is encountered.

Standard weight-four $X$-type stabilizers in the connected component to the left of $V_{\mathrm{left}}$ are assigned an \textbf{e} schedule. This assignment is propagated away from the junction until another cut line is encountered. Likewise, standard weight-four $X$-type stabilizers in the connected component to the right of $V_{\mathrm{right}}$ are assigned an \textbf{e} schedule and retain this assignment until another cut line is encountered.

Lastly, consider the standard weight-four $X$ stabilizers that the cut lines terminate on at a junction. We assign an \textbf{f} schedule to those stabilizers. An example is shown in \cref{fig:Junction_example_v1}.

\begin{figure*}
     \centering
\subfloat[\label{fig:horizontal_elongated_circuitEX} ]{\includegraphics[width=.5\textwidth]{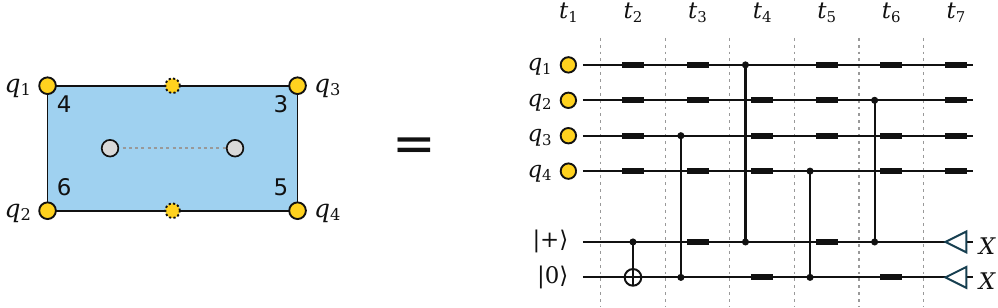}}
\subfloat[\label{fig:elongated_flagEX} ]{\includegraphics[width=.5\textwidth]{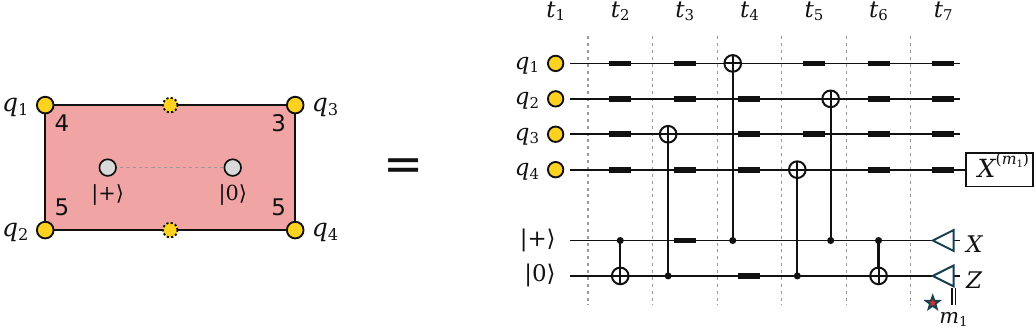}}
\vfill
\subfloat[\label{fig:vertical_elongated_circuitEX} ]{\includegraphics[width=.8\textwidth]{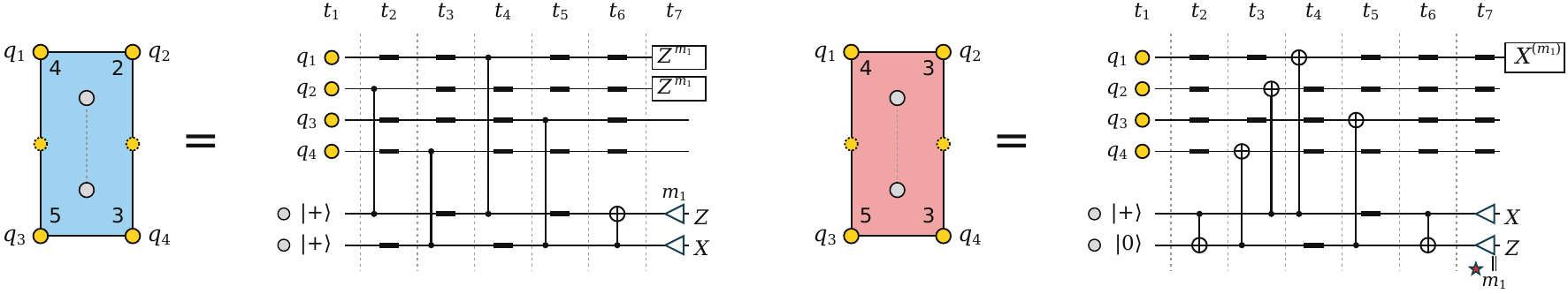}}
\vfill
\subfloat[\label{fig:twist_circuitsEX} ]{\includegraphics[width=.7\textwidth]{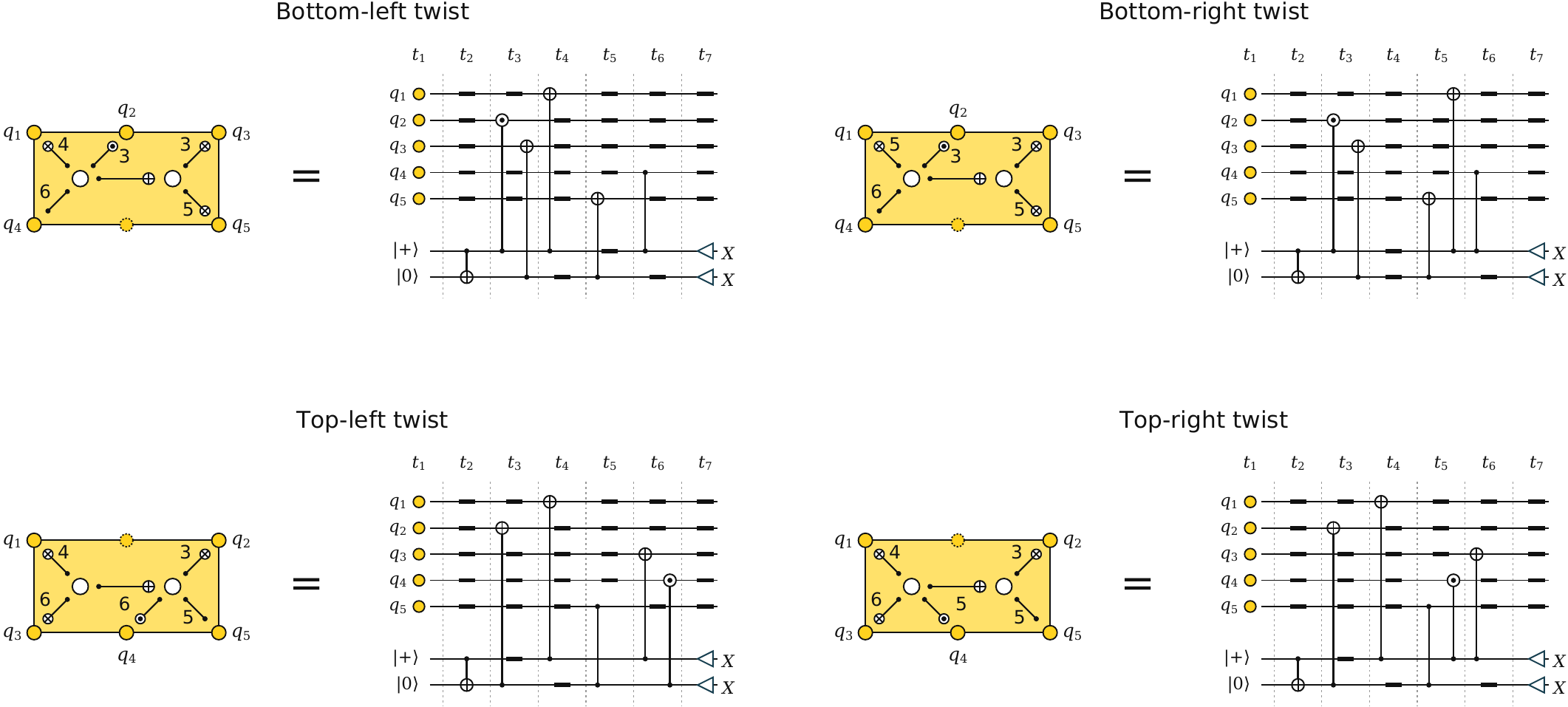}}
\vfill
\subfloat[\label{fig:Elongated_Domain_XXXZEX} ]{\includegraphics[width=.4\textwidth]{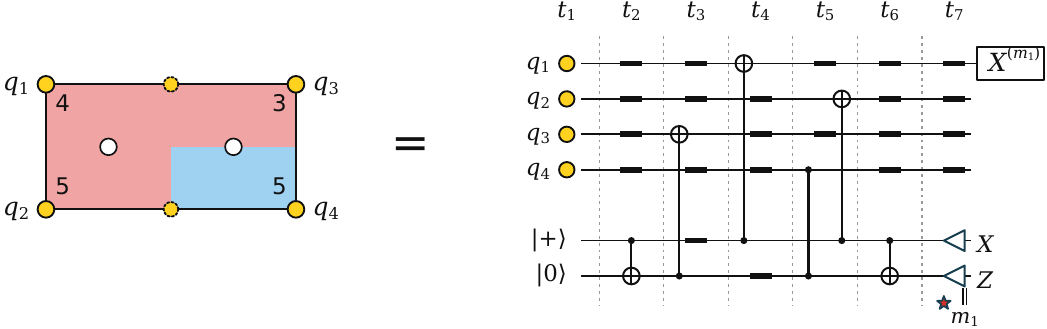}}
\subfloat[\label{fig:trapezoid_checkXXXXEX} ]{\includegraphics[width=.4\textwidth]{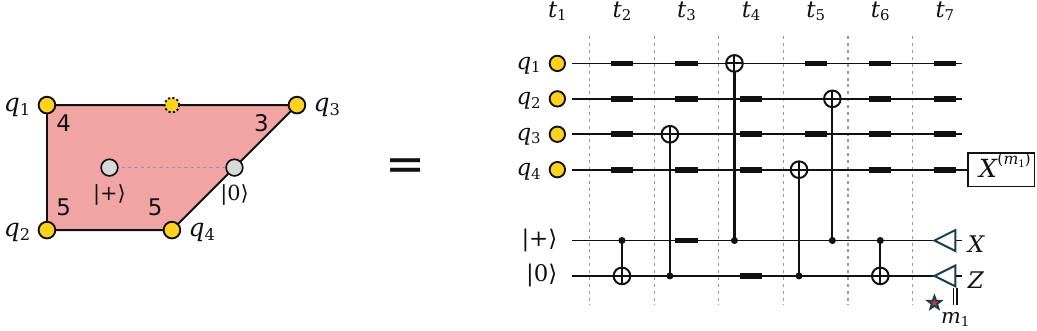}}
     \caption{ (a) Circuit used to measure horizontal elongated $Z$ stabilizers. (b) Flag circuit used for elongated $X$-type stabilizers. The flag qubit is prepared in $|0\rangle$. If it is measured non-trivially, an $X$ correction is performed on the data qubit labeled $q_4$. Black rectangles illustrate idling qubits. (c) Circuits used to measure vertical elongated stabilizers. Vertical $X$ stabilizers normally use a flag circuit as in (b). When a vertical elongated $X$ stabilizer shares a data qubit with a horizontal elongated $X$ stabilizer, we use the unflagged analogue of the vertical $Z$ circuit described in \cref{subsec:elongated_rec_gate_schedule}. (d) Circuits used to measure twist defects. We include circuits used in $Y$-extension patches for all four logical qubits in a unit cell. (e) Flag-based circuit used to measure a weight-four elongated domain wall of the form $X \otimes X \otimes X \otimes Z$. (f) Flag-based circuit for a trapezoidal elongated weight-four $X$ stabilizer. } 
     \label{fig:elongated_flag}
\end{figure*}

\begin{figure*}
     \centering
 \subfloat[\label{fig:Continuation_weight_2EX} ]{\includegraphics[width=.95\textwidth]{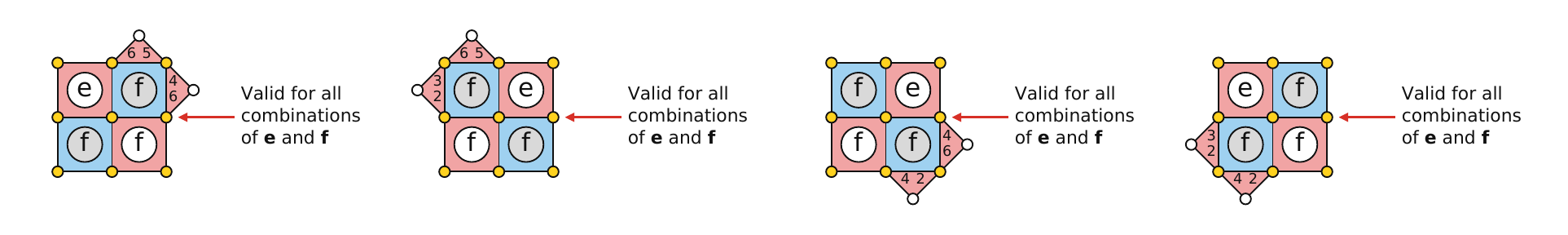}}
     \caption{  Illustration of gate schedules used for weight-two stabilizers. We use the same gate schedule that was used for standard surface code patches.  } 
     \label{fig:Weight_3_schedule}
\end{figure*}

\subsection{Gate schedules for elongated stabilizers and twist defects}
\label{subsec:elongated_rec_gate_schedule}

In this section we provide the gate schedules for elongated rectangles and twist defects. In what follows, we label the top-left, top-right, bottom-left and bottom-right data qubits in the support of the elongated checks (whether they are horizontal or vertical) as $q_{\text{tl}}$, $q_{\text{tr}}$, $q_{\text{bl}}$ and $q_{\text{br}}$. The time steps for the two qubit gates acting on qubits $[q_{\text{tl}}, q_{\text{tr}}, q_{\text{bl}}, q_{\text{br}}]$ is denoted $[t_1, t_2, t_3, t_4]$.

We start with horizontal elongated stabilizers that are $X$-type and $Z$-type. For $Z$-type elongated stabilizers, the two qubit gates acting on $[q_{\text{tl}}, q_{\text{tr}}, q_{\text{bl}}, q_{\text{br}}]$ are always done in time steps $[4,3,6,5]$. The full circuit used is shown in \cref{fig:horizontal_elongated_circuitEX}. For horizontal elongated stabilizers that is $X$-type, we instead uses the schedule $[4,3,5,5]$. Additionally, the second ancilla qubit is used as a flag qubit \cite{ChaoFlag1,Chamberland2018flagfaulttolerant,ChaoFlag3}. When the flag qubit prepared in $|0\rangle$ is measured as $-1$ instead of $+1$, an $X$ correction is performed on $q_4$. It can be verified that a weight-two $X$ error arising from a single fault in such a circuit will always be horizontal. This choice is made to ensure errors do not propagate parallel to logical observables since the placement of horizontal elongated $X$ stabilizers will always be at locations containing logical observables with weight-$d$ minimum weight representatives formed by vertical $X$ strings. The circuit is shown in \cref{fig:elongated_flagEX}. Note that if a standard weight-four $Z$-type stabilizer is to the immediate left of an elongated weight-four $X$ stabilizer along a horizontal row, we always set its gate schedule to $\textbf{f}$.

Next we consider vertical elongated stabilizers that are $X$-type and $Z$-type. For $Z$-type elongated stabilizers, the two-qubit gates acting on $[q_{\text{tl}}, q_{\text{tr}}, q_{\text{bl}}, q_{\text{br}}]$ are always done in time steps $[4,2,5,3]$. An illustration of the circuit is shown in \cref{fig:vertical_elongated_circuitEX}. For $X$ stabilizers, we normally use the flag circuit shown in \cref{fig:vertical_elongated_circuitEX} with the gate schedule $[4,3,5,3]$. As in the previous flag circuit, if the $Z$ measurement is non-trivial, an $X$ correction is applied to qubit $q_1$. This correction ensures that a weight-two $X$ error arising from a single fault will always form a vertical string. For this flagged circuit, if the gate schedule of the standard weight-four $Z$ stabilizer above the vertical elongated $X$ stabilizer is \textbf{f}, change it to \textbf{e}.

If a vertical elongated $X$ stabilizer shares a data qubit with a horizontal elongated $X$ stabilizer, we instead use the unflagged analogue of the vertical $Z$ circuit with schedule $[4,2,5,3]$. Both ancillas are prepared in $|+\rangle$, and the controlled-$Z$ data interactions are replaced by CNOTs with ancilla controls and data targets. The final ancilla CNOT is directed from the lower to the upper ancilla: the lower ancilla's $X$ measurement gives the stabilizer outcome, while a non-trivial $Z$ measurement of the upper ancilla triggers $X$ corrections on $q_{\text{tl}}$ and $q_{\text{tr}}$ in place of the conditional $Z$ corrections. The preceding \textbf{f}-to-\textbf{e} replacement is not applied in this case.

We conclude this section with the gate scheduling used for twist defects. The gate schedule depends on the logical qubit placement in a unit cell. For a bottom-left logical qubit, we use $[4,3,3,6,5]$. For a bottom-right logical qubit we use $[5,3,3,6,5]$. For a top-left logical qubit we use $[4,3,6,6,5]$. Finally for a top-right logical qubit we use $[4,3,6,5,5]$. The four circuits are shown in \cref{fig:twist_circuitsEX}. Lastly, we also provide the circuit an gate schedules for the elongated $X \otimes X \otimes X \otimes Z$ domain wall and a weight-four $X$ stabilizer forming a trapezoidal shape (which is required for certain multi-qubit Pauli measurements) in \cref{fig:Elongated_Domain_XXXZEX,fig:trapezoid_checkXXXXEX}. 

\subsection{Gate schedules for boundary stabilizers}
\label{subsec:boundary_stab_gate_schedule}

In this section we provide the gate schedule for all boundary stabilizers added at a merged lattice used to perform lattice surgery. We've already seen gate schedules for weight-two stabilizers of standard surface code patches in \cref{subsec:standard_weight_4_gate_schedule}. We now extend to all boundary stabilizers used in routing space region. 

We start with weight-three stabilizers. Let $S_c$ be the weight-three stabilizer being assigned the gate schedule. Consider the standard weight-four stabilizer $S$ along the diagonal of $S_c$ sharing a data qubit with $S_c$. $S$ will either have an \textbf{e} or \textbf{f} schedule. The schedule assigned to $S_c$ is chosen to match the \textbf{e} or \textbf{f} schedule assigned to $S$ with a time step for the missing data qubit in $S_c$ removed. For instance, suppose the data qubits in $S_c$ are $[q_{\text{tl}}, q_{\text{tr}}, q_{\text{br}}]$ and $S$ has en \textbf{e} schedule which is $[4,2,6,5]$. Then the schedule assigned to $S_c$ is $[4,2,5]$.

We conclude this section with the gate schedule assigned to weight-two stabilizers in the routing space region. We essentially extend the weight-two gate schedule used for a logical qubit patch to all weight-two stabilizers. More precisely, if a weight-two stabilizer has a top-right and bottom-right data qubit in its support, we use the $[3,2]$ schedule. If it has bottom-left and bottom-right data qubits in its support, we use the $[6,5]$ schedule. If it has the top-left and bottom-left data qubits in its support, we use the $[4,6]$ schedule. Lastly, if it has the top-left and top-right data qubits in its support, we use the $[4,2]$ schedule. We note that the schedule is unaltered even if the weight-two stabilizers intersect at a corner stabilizer as shown in \cref{fig:Weight_3_schedule}. Also, the above schedule is applied regardless of whether the weight-two stabilizer is $X$-type, $Z$-type or a domain wall. 

\subsection{Gate schedules for $Y$-extension regions.}
\label{subsec:gate_schedule_y_ext}

In order to ensure fault-tolerance up to the full code distance of the surface code during $Y$ measurements, extraordinary care must be taken in setting the gate schedule for the standard weight-four $X$ and $Z$ type stabilizers in a $Y$-extension patch. In this section, we provide a detailed description on how the gate schedule is chosen.

In what follows, all code distances are considered to be odd and at least five. We define
\begin{align}
    \nu = \frac{d-3}{2}.
    \label{eq:y-extension-nu}
\end{align}
All row and column counts in this section refer to rows and columns of \emph{syndrome stabilizers} (equivalently, their ancilla locations), rather than rows and columns of data qubits.  We use Cartesian directions, so that the $x$- and $y$-coordinates increase toward the east and north, respectively.

The legacy $Y$-extension patch is treated as a fixed gate-schedule island. In particular, a routing-space walk that assigns schedules to an external corridor using the rules from previous sections skips the interior of the $Y$-extension patch. The schedules specified below are therefore intrinsic to the $Y$-extension patch and do not depend on which
external routing path, if any, subsequently touches its boundary.

\begin{figure*}
     \centering
\subfloat[\label{fig:q4y_d9_BLEX} ]{\includegraphics[width=.4\textwidth]{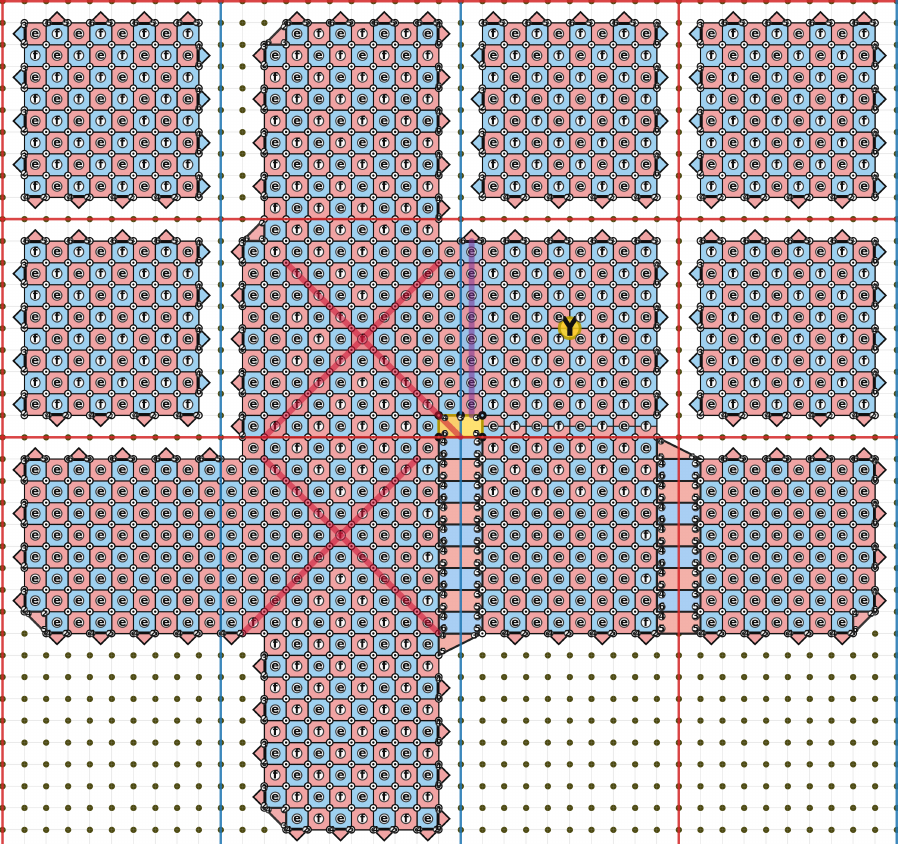}}
\subfloat[\label{fig:q5y_d9_BREX} ]{\includegraphics[width=.4\textwidth]{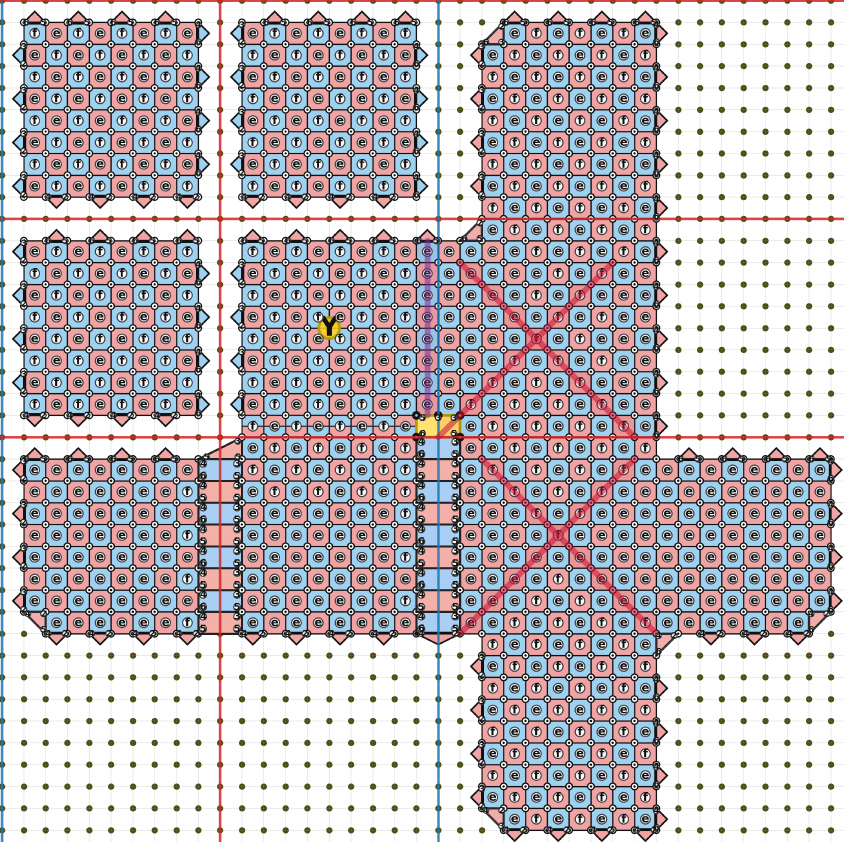}}
\vfill
\subfloat[\label{fig:q6y_d9_TLEX} ]{\includegraphics[width=.4\textwidth]{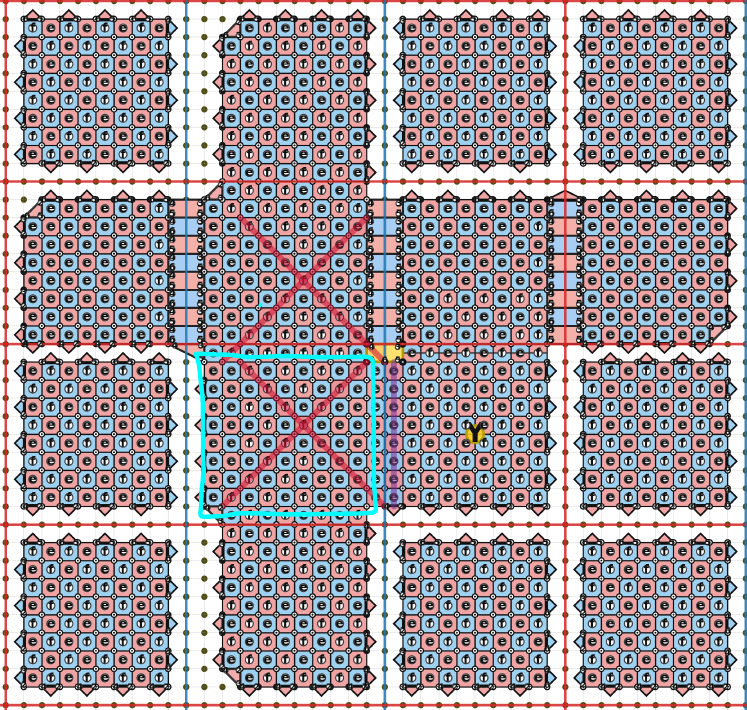}}
\subfloat[\label{fig:q7y_d9_TREX} ]{\includegraphics[width=.4\textwidth]{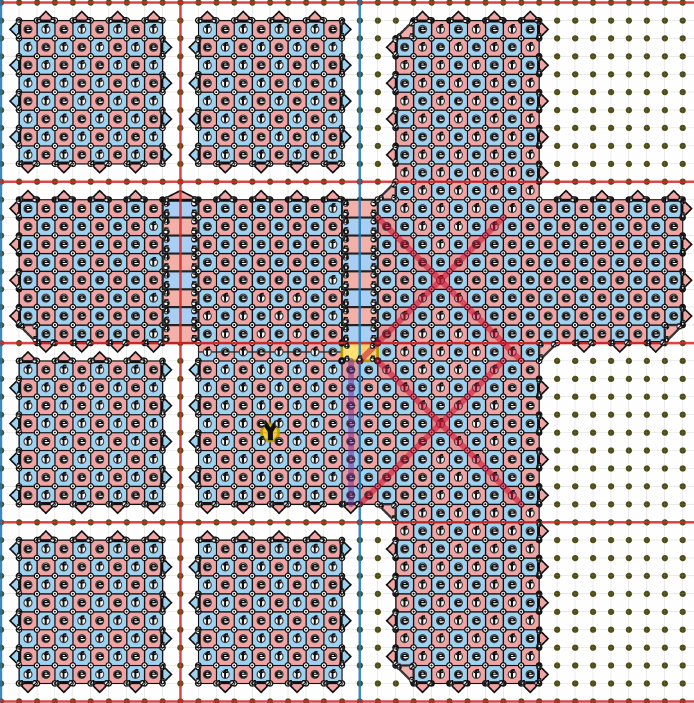}}
     \caption{ $Y$-extension patches including protruding lobes for $d=9$ surface codes. Red crosses indicate junction cut lines used to denote transitions in gate schedules. (a) Bottom-left logical qubit patch in a unit cell. (b) Bottom-right logical qubit patch in a unit cell. (c) Top-left logical qubit patch in a unit cell. (d) Top-right logical qubit patch in a unit cell.  }
     \label{fig:Example_VICL_Y}
\end{figure*}

For later reference, the horizontal elongated weight-four schedules shown in \cref{fig:horizontal_elongated_circuitEX} are labeled as
\begin{equation}
\begin{aligned}
    \mathbf h_X&=(4,3,5,5), &
    \mathbf h_Z&=(4,3,6,5),
\end{aligned}
\label{eq:elongated-y-schedules}
\end{equation}
where $\mathbf h_P$ denote horizontal elongated checks of Pauli type $P\in\{X,Z\}$, respectively.  

\paragraph{Orientation convention.}
A logical patch is labeled by its position within a unit cell.  A left patch exposes its western $X$ boundary, a right patch exposes its eastern $X$ boundary, a top patch exposes its northern $Z$ boundary, and a bottom patch exposes its southern $Z$ boundary.  The horizontal row of domain-wall checks lies immediately outside the exposed $Z$ boundary.  The twist defect and alternating elongated-check stack lie at the corner at which this row meets the exposed $X$ boundary.

The two extension-side quadrants are denoted $Q_{\mathrm p}$ and $Q_{\mathrm s}$.  The \emph{primary} quadrant $Q_{\mathrm p}$ occupies the same data-qubit rows as the original logical patch, whereas the \emph{secondary} quadrant $Q_{\mathrm s}$ lies on the opposite side of the domain-wall row.  The region $Q_{\mathrm{patch}}$ is the extension beyond the exposed $Z$ boundary but between the columns spanned by the original logical patch.  These regions, together with the relevant halves of the paired junction cuts introduced below, are listed in \cref{tab:y-extension-orientations}.

\begin{table*}
    \centering
    \small
    \setlength{\tabcolsep}{4pt}
    \begin{tabular}{c c c c c c c}
        \toprule
        patch & $X$-extension & $Q_{\mathrm p}$ & $Q_{\mathrm s}$
        & $Q_{\mathrm{patch}}$
        & primary wall arm & secondary wall arm\\
        \midrule
        bottom-left  & west & northwest & southwest & southeast
                     & $V_{\mathrm{bottom}}$ & $V_{\mathrm{top}}$ \\
        bottom-right & east & northeast & southeast & southwest
                     & $V_{\mathrm{bottom}}$ & $V_{\mathrm{top}}$ \\
        top-left     & west & southwest & northwest & northeast
                     & $V_{\mathrm{top}}$ & $V_{\mathrm{bottom}}$ \\
        top-right    & east & southeast & northeast & northwest
                     & $V_{\mathrm{top}}$ & $V_{\mathrm{bottom}}$\\
        \bottomrule
    \end{tabular}
    \caption{Orientation data for the four legacy $Y$-extension patches. The labels $V_{\mathrm{top}}$ and $V_{\mathrm{bottom}}$ denote the upper and lower halves of the paired diagonal junction cuts.}
    \label{tab:y-extension-orientations}
\end{table*}

Thus, the bottom-right construction is the east--west reflection of the bottom-left construction, and the top-right construction is the east--west reflection of the top-left construction.  

In what follows, a junction contains two diagonal $X$-cut lines: one running northwest--southeast and one running northeast--southwest. In what follows, these virtual cuts are confined to the fixed $Y$-extension patch.  Let $I$ denote their common intersection. The portions of the two cuts above and below $I$ form $V_{\mathrm{top}}$ and $V_{\mathrm{bottom}}$, respectively. The portions of the two cuts to the left and to the right of $I$ form $V_{\mathrm{left}}$ and $V_{\mathrm{right}}$, respectively.

\subsubsection{Fixed schedule before the junction refinements}
\label{sec:y-fixed-schedule}

The portion of the \(Y\)-extension patch belonging to the original logical
patch retains the ordinary assignment
\begin{equation}
    X_4\longmapsto \mathbf e,
    \qquad
    Z_4\longmapsto \mathbf f.
    \label{eq:y-original-patch-schedule}
\end{equation}

Let \(D\) denote the horizontal domain-wall row.  A weight-four mixed
domain-wall stabilizer is assigned the same schedule family as the standard
weight-four stabilizer adjacent to it along the relevant diagonal and lying
in the original logical patch.  Equivalently,
\begin{align}
    X\text{ on the original-patch side of }D
        \Longrightarrow \mathbf e, \\
    Z\text{ on the original-patch side of }D
        \Longrightarrow \mathbf f.
    \label{eq:y-domain-wall-schedule}
\end{align}

Let $C_{\mathrm{mix}}^{(1)}$ be the column of stabilizers immediately
outside the exposed $X$ boundary and spanning the rows of the original
logical patch (it is illustrated with a purple line in \cref{fig:Example_VICL_Y}). Further, let the column of stabilizers next to $C_{\mathrm{mix}}^{(1)}$ that span the same rows and which is away from the original patch be $C_{\mathrm{mix}}^{(2)}$. Every standard weight-four stabilizer on
$C_{\mathrm{mix}}^{(1)}$ and $C_{\mathrm{mix}}^{(2)}$ uses an $\mathbf e$ schedule.  

Let $S(Z)$ be a standard weight-four $Z$ stabilizer. For all four $Y$-extension patch orientations, if $S(Z)$ is in $Q_{\mathrm{patch}}$, $Q_{\mathrm p}$ and $Q_{\mathrm s}$, an $\mathbf e$ schedule is assigned to it except if there is an elongated $X$ stabilizer $\mathbf h_X$ that has support on the top-right and bottom-right data qubits in the support of $S(Z)$. In this case we assign an $\mathbf f$ schedule to $S(Z)$.

In what follows, we provide the gate scheduling rule for the remaining $X$ stabilizers in each of the four $Y$-extension patch orientations.

\textbf{Bottom-left patch: }Let $S(X)$ be a standard weight-four $X$ stabilizer. If $S(X)$ is in $Q_{\mathrm{patch}}$, then an $\mathbf e$ schedule is assigned to it unless $S(X)$ is in the first $\nu$ rows (from top to bottom), in which case we assign an $\mathbf f$ schedule to $S(X)$. 

\begin{figure*}
     \centering
\subfloat[\label{fig:XYMeasEX} ]{\includegraphics[width=.5\textwidth]{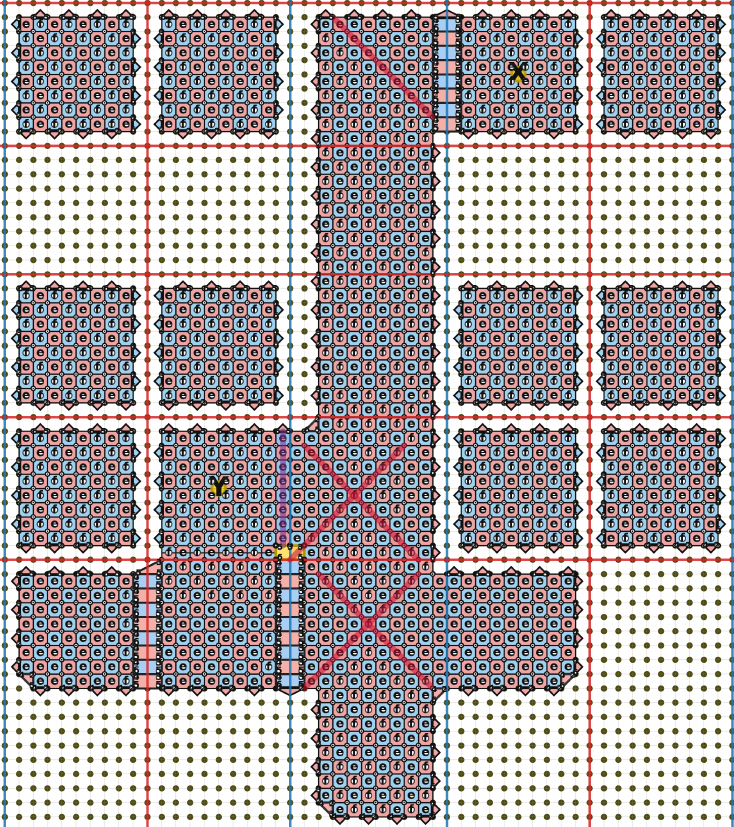}}
     \caption{ Example of a $Y \otimes X$ measurement where a protruding lobe along the vertical corridor is extended to perform the logical $X$ measurement on the patch located at the top-right of the grid.  }
     \label{fig:XYMeas}
\end{figure*}

Now suppose $S(X)$ is in $Q_{\mathrm p}$. For the first $\nu$ columns in $Q_{\mathrm p}$ (moving westward), $S(X)$ is assigned an $\mathbf e$ schedule unless it is below $V_{\mathrm{bottom}}$ or on the cut lines forming $V_{\mathrm{bottom}}$, in which case it's assigned an $\mathbf f$ schedule. For the remaining columns, if $S(X)$ is above $V_{\mathrm{bottom}}$ or below $V_{\mathrm{top}}$, it's assigned an $\mathbf f$ schedule. Further, if $S(X)$ is along a cut line, its assigned an $\mathbf f$ schedule. Let $C_{\text{int}}$ be the column in which the cut lines intersect. Suppose $S(X)$ is to the left of $C_{\text{int}}$, and between $V_{\mathrm{top}}$ and $V_{\mathrm{bottom}}$. For the first $\nu$ columns to the left of $C_{\text{int}}$, $S(X)$ is assigned an $\mathbf f$ schedule. For the remaining columns in $Q_{\mathrm p}$, $S(X)$ is assigned an $\mathbf e$ schedule.

Now suppose $S(X)$ is in $Q_{\mathrm s}$. If $S(X)$ is at the intersection of the cut lines, or along the cut lines that forms $V_{\mathrm{top}}$, $S(X)$ is assigned an $\mathbf f$ schedule. If its along the cut lines that forms $V_{\mathrm{bottom}}$, its assigned an $\mathbf e$ schedule, except for the endpoint of the left-most cut line of $V_{\mathrm{bottom}}$, in which case we assign an \textbf{f} schedule. If $S(X)$ is above $V_{\mathrm{top}}$ or below $V_{\mathrm{bottom}}$, it's assigned an $\mathbf f$ schedule. Let $C_{\text{int}}$ be the column in which the cut lines intersect. If $S(X)$ is to the right or left of $C_{\text{int}}$ and between $V_{\mathrm{top}}$ and $V_{\mathrm{bottom}}$, it's assigned an $\mathbf e$ schedule. An example for a $d=9$ surface code is shown in \cref{fig:q4y_d9_BLEX}. Note that additional protruding lobes are included in the figure which we explain in \cref{sec:protruding_lobes}.

\textbf{Bottom-right patch: }The rules for assigning gate schedules to standard weight-four $X$ stabilizers are the mirror image of the rules used above for the bottom-left patch. An example for a $d=9$ surface code is shown in \cref{fig:q5y_d9_BREX}. Note that additional protruding lobes are included in the figure which we explain in \cref{sec:protruding_lobes}.

\textbf{Top-left patch: }Let $S(X)$ be a standard weight-four $X$ stabilizer. Suppose further that $S(X)$ is in $Q_{\mathrm{patch}}$. If $S(X)$ is in the first $\nu_2 = \max{((d-5)/2,0)}$ columns (moving eastward), an $\mathbf e$ schedule is assigned to $S(X)$. Now if $S(X)$ is in the first $\nu$ rows (moving northward) but outside the first $\nu_2$ columns, an $\mathbf f$ schedule is assigned to $S(X)$. Otherwise, we assign an $\mathbf e$ schedule to $S(X)$.

\begin{figure*}
     \centering
\subfloat[\label{fig:YYMeasEX} ]{\includegraphics[width=.5\textwidth]{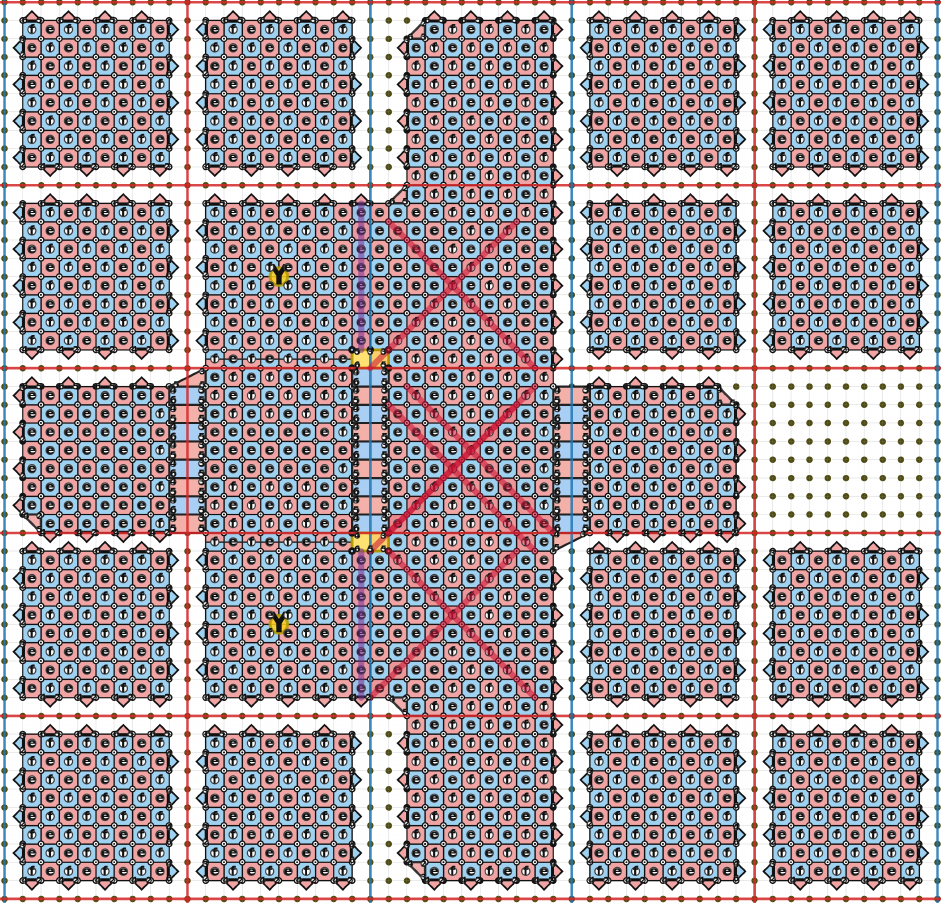}}
     \caption{ Example of a $Y \otimes Y$ measurement with overlapping $Q_{\mathrm s}$ and $Q_{\mathrm{patch}}$ regions. In the overlapping region, half of the rows of stabilizers closest to a given patch retain the rules assigned to that patch.  }
     \label{fig:YYMeas}
\end{figure*}

Now suppose $S(X)$ is in $Q_{\mathrm s}$. If $S(X)$ is in the first column (moving westward), an $\mathbf e$ schedule is assigned to $S(X)$. Otherwise, if $S(X)$ is in the last $\nu_2$ rows (moving northward) or $S(X)$ is above $V_{\mathrm{top}}$ or below $V_{\mathrm{bottom}}$, an $\mathbf f$ schedule is assigned to $S(X)$. If $S(X)$ is at the intersection of the cut lines or along $V_{\mathrm{bottom}}$, an $\mathbf f$ schedule is assigned to $S(X)$. If $S(X)$ is along $V_{\mathrm{top}}$ (but not in the last $\nu_2$ rows or first column) or to the left of the left-most cut lines forming $V_{\mathrm{bottom}}$ and $V_{\mathrm{top}}$, an $\mathbf e$ schedule is assigned to $S(X)$.

If $S(X)$ is in the row to the left of the twist defect, an $\mathbf f$ schedule is assigned to $S(X)$.

Now suppose $S(X)$ is in $Q_{\mathrm p}$. If $S(X)$ is at the intersection of the cut lines, or on or above $V_{\mathrm{top}}$, an $\mathbf f$ schedule is assigned to $S(X)$. Now suppose $S(X)$ is in the first $\nu$ columns (moving westward) and below $V_{\mathrm{top}}$, or to the right of the right-most cut lines forming  $V_{\mathrm{top}}$ and  $V_{\mathrm{bottom}}$, an $\mathbf e$ schedule is assigned to $S(X)$. If $S(X)$ is below $V_{\mathrm{bottom}}$ and not in the first $\nu$ columns, an $\mathbf f$ schedule is assigned to $S(X)$. If $S(X)$ is on the right-most cut line forming $V_{\mathrm{bottom}}$, an $\mathbf e$ schedule is assigned to $S(X)$. If $S(X)$ is on the left-most cut line forming $V_{\mathrm{bottom}}$, an $\mathbf f$ schedule is assigned to $S(X)$. Now suppose $S(X)$ is to the left of the left-most cut lines forming $V_{\mathrm{top}}$ and $V_{\mathrm{bottom}}$. If $S(X)$ is $\nu$ columns from the intersection of the junction cut lines, an $\mathbf f$ schedule is assigned to $S(X)$. Otherwise, an $\mathbf e$ schedule is assigned to $S(X)$.

An example is shown in \cref{fig:q6y_d9_TLEX}. Note that additional protruding lobes are included in the figure which we explain in \cref{sec:protruding_lobes}.

\textbf{Top-right patch: }The rules for assigning gate schedules to standard weight-four $X$ stabilizers are the mirror image of the rules used above for the top-left patch. An example for a $d=9$ surface code is shown in \cref{fig:q7y_d9_TREX}. Note that additional protruding lobes are included in the figure which we explain in \cref{sec:protruding_lobes}.

\subsection{Adding protruding lobes for improved effective code distance}
\label{sec:protruding_lobes}

As shown in \cref{fig:Example_VICL_Y}, the $Y$-extension patches have protruding lobes along the boundaries of $Q_{\mathrm p}$, $Q_{\mathrm s}$, and $Q_{\mathrm{patch}}$ that face the routing space corridors. For the untrimmed constructions tested at $d \le 11$, our exhaustive numerical search found that these lobes, together with the gate schedules described in \cref{subsec:gate_schedule_y_ext}, preserve fault-tolerant $Y$ measurements at the full code distance: no pattern of at most $(d-1)/2$ circuit faults caused a logical failure. Without these lobes and gate schedules, the distance reduction arises primarily from the twist defect during the first syndrome-measurement round of the merged patch. Because the initial twist measurement has a random outcome, errors that anticommute with the twist need not produce a detectable syndrome in that round. When combined with faults elsewhere in the circuit, these errors can produce the same syndrome as errors arising from fewer faults, potentially causing the decoder to select a correction that results in a logical failure. This ambiguity is responsible for the ``distance-reducing'' behavior of twist defects. A similar phenomenon was observed in Ref.~\cite{Chamberland22b}, although the proposed solution there was limited to $d_x=3$.

Next we discuss the structure of the added lobes. Let $L_x\times L_y$ denote the number of data-qubit columns and rows occupied by a protruding lobe.  Before trimming, the transverse span of every lobe is $d$, while its longitudinal span is $d+1$, with one exception.  Thus, every eastward- or westward-growing lobe has dimensions $(d+1)\times d$, and every northward- or southward-growing lobe has dimensions $d\times(d+1)$, except for the southward lobe that continues the legacy vertical arm of a bottom-left or bottom-right $Y$-extension patch, which has dimensions $d\times d$.  Equivalently, all vertical lobes of top-left and top-right patches have size $d\times(d+1)$, whereas bottom-left and bottom-right patches have a northward lobe of size $d\times(d+1)$ and a southward legacy-arm lobe of size $d\times d$.

In such lobes, all standard weight-four $Z$ stabilizers are assigned an $\mathbf e$ schedule unless it is adjacent to a weight-four elongated $X$ stabilizer (to the left along the same row) in which case we assign an $\mathbf f$ schedule. If the lobe is in a horizontal routing space corridor, all standard weight-four $X$ stabilizers are assigned an $\mathbf e$ schedule. However if the lobe is in a vertical routing space corridor, all standard weight-four $X$ stabilizers are assigned an $\mathbf f$ schedule.

\begin{figure*}
    \centering
    \includegraphics[width=.95\textwidth]{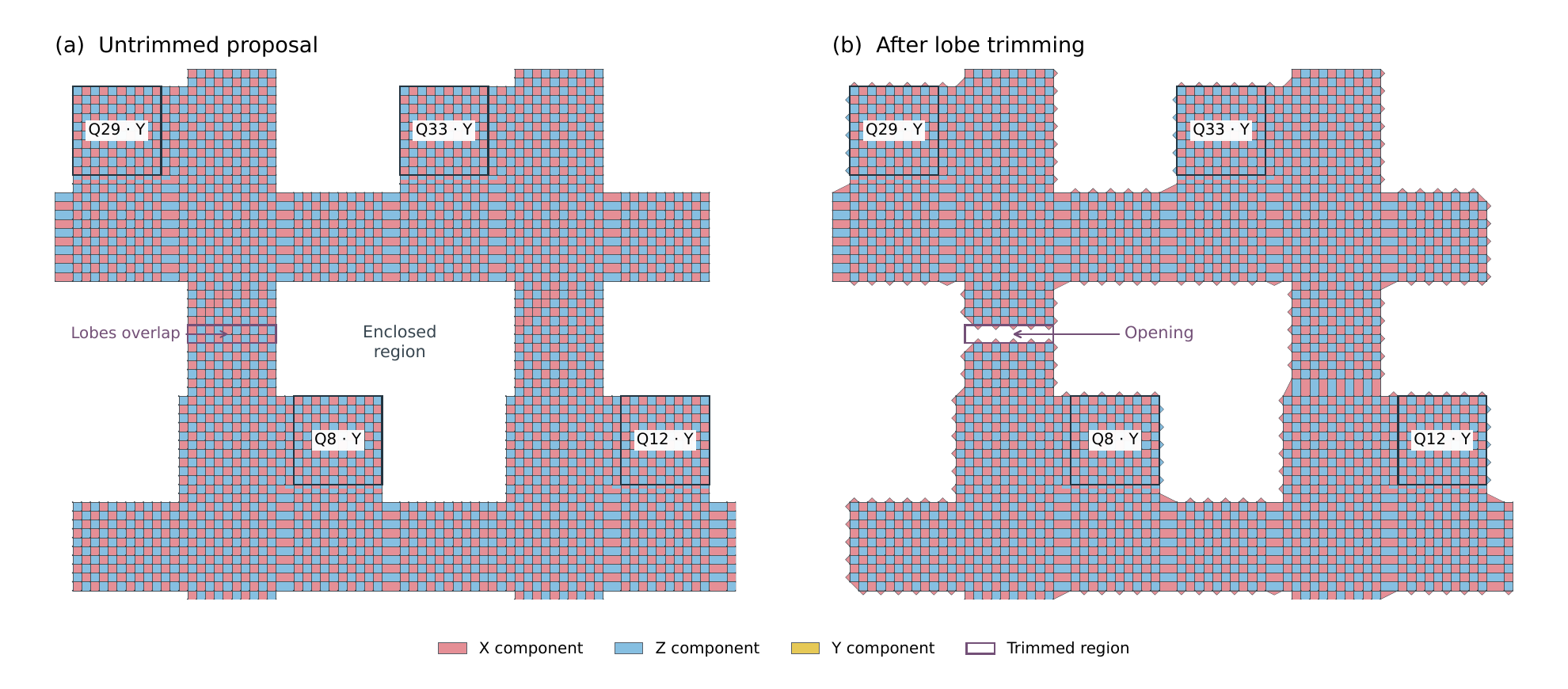}
    \caption{Trimming protruding lobes for the measurement $Y_{Q_8}\otimes Y_{Q_{12}}\otimes Y_{Q_{29}}\otimes Y_{Q_{33}}$ at nominal distance $d=11$. (a) The untrimmed lobes form a closed loop around an unmerged region; the proposal is shown before boundary stabilizers are added. (b) Shortening two facing lobes opens the loop while preserving the legacy $Y$-extension checks and the connections required by the measurement. The completed boundary stabilizers are included. Purple outlines identify the trimmed region. Idle logical patches, logical observables, and cut lines are omitted.}
    \label{fig:lobe_trimming}
\end{figure*}

One advantage of our protocol is that the added lobes fit in the available routing space corridors and thus do not require a change in the physical qubit layout shown in \cref{fig:Layout_Example}. Furthermore, stabilizers can be added to the lobes to perform multi-qubit Pauli measurements of additional Pauli operators. An example of a $Y \otimes X$ measurement is shown in \cref{fig:XYMeas}.   

When protruding lobes from neighboring $Y$ extensions meet, they can form a closed loop around an unmerged region. Selected lobes are then shortened to create an opening, as illustrated in \cref{fig:lobe_trimming}. The legacy $Y$-extension checks remain unchanged, and the opening is chosen to retain the logical-boundary connections required by every factor of the Pauli measurement. If a contact is needed for an additional $X$ or $Z$ factor, an alternative opening must be found. The effective distance of a trimmed construction must be checked separately, and any reduction below the nominal distance $d$ recorded.

In Patchwork, the protruding-lobe construction can be requested explicitly by passing \texttt{--Y\_extra\_pad "True"} to \texttt{ls\_driver.py}. The three layout options are resolved in the following order of precedence: \texttt{--ancilla\_twist\_free}, \texttt{--grid\_double}, and \texttt{--Y\_extra\_pad}. If \texttt{--ancilla\_twist\_free "True"} is specified, Patchwork uses the original routing corridors of height $d+1$ and width $d+3$ and disables both doubled corridors and protruding lobes. Otherwise, if \texttt{--grid\_double "True"} is specified, Patchwork uses routing corridors of height and width $2d$ and disables \texttt{--Y\_extra\_pad}, even if the latter was explicitly set to \texttt{"True"}. Finally, when both higher-priority options are \texttt{"False"}, as they are by default, Patchwork automatically enables \texttt{--Y\_extra\_pad} regardless of whether it was omitted or explicitly set to \texttt{"False"}; thus, protruding lobes are included automatically for the standard twist-defect-based $Y$-measurement protocol.

\subsection{Gate scheduling protocol for overlapping $Y$ measurements}
\label{sec:neighbor_Y}

In \cref{subsec:gate_schedule_y_ext} we provided a gate-scheduling procedure for all components of a $Y$-extension patch. However in the case where neighboring $Y$ measurements are performed, some of the gate scheduling rules applied to $Q_{\mathrm p}$, $Q_{\mathrm s}$ and $Q_{\mathrm{patch}}$ regions can overlap. An example of such overlaps during a multi-qubit Pauli $Y$ measurement is shown in \cref{fig:YYMeas}.

Suppose a $Y \otimes Y$ measurement is performed between two patches $P_1$ and $P_2$. We first consider the case where overlapping stabilizers from the $Y$-extension patches occur along horizontal rows of stabilizers (as in \cref{fig:YYMeas}). We apply the gate scheduling rules of \cref{subsec:gate_schedule_y_ext} for $P_1$ to half the rows of stabilizers closest to $P_1$. The remaining half rows of stabilizers closest to $P_2$ receive the gate schedules originally assigned to $P_2$ and derived using the rules in \cref{subsec:gate_schedule_y_ext}.

We also note that there may be cases where a $Y$-extension patch overlaps with a protruding lobe from another $Y$-extension patch. In such a case, the gate schedule used for the legacy $Y$-extension patch takes precedence. That is, it cannot be overridden by the gate schedule used for a protruding lobe.  We further note that if a multi-qubit Pauli measurement includes $Y$ terms in addition to $X$ and $Z$ terms, the $X$ and $Z$ terms will not impact the gate schedule used in the legacy $Y$-extension patch which is considered fixed and immutable. If stabilizers from an $X$ or $Z$ measurement reach the protruding lobes of a legacy $Y$-extension patch, they are simpler merged with the lobe using the standard checkerboard pattern, and using the look-ahead rules to prevent stabilizer type clashes. 

Suppose that the  $Y \otimes Y$ measurement between $P_1$ and $P_2$ has instead overlapping columns of stabilizers. In this case, we perform the same steps described in the paragraph above but with rows replaced by columns.

\section{Space-time partitioning of merged surface code patches for parallel decoding}
\label{sec:spacetime_partition}

\begin{figure*}
     \centering
\subfloat[\label{fig:neighbor_X_pairEX} ]{\includegraphics[width=.3\textwidth]{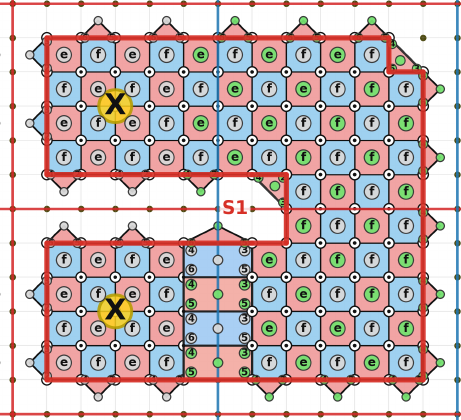}}
\subfloat[\label{fig:q33y_q4x_d5EX} ]{\includegraphics[width=.4\textwidth]{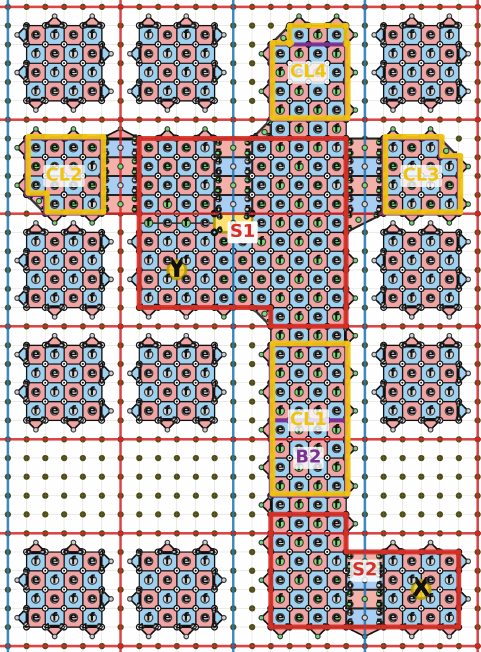}}
\vfill
\subfloat[\label{fig:seed_wavefront_expansionEX} ]{\includegraphics[width=.8\textwidth]{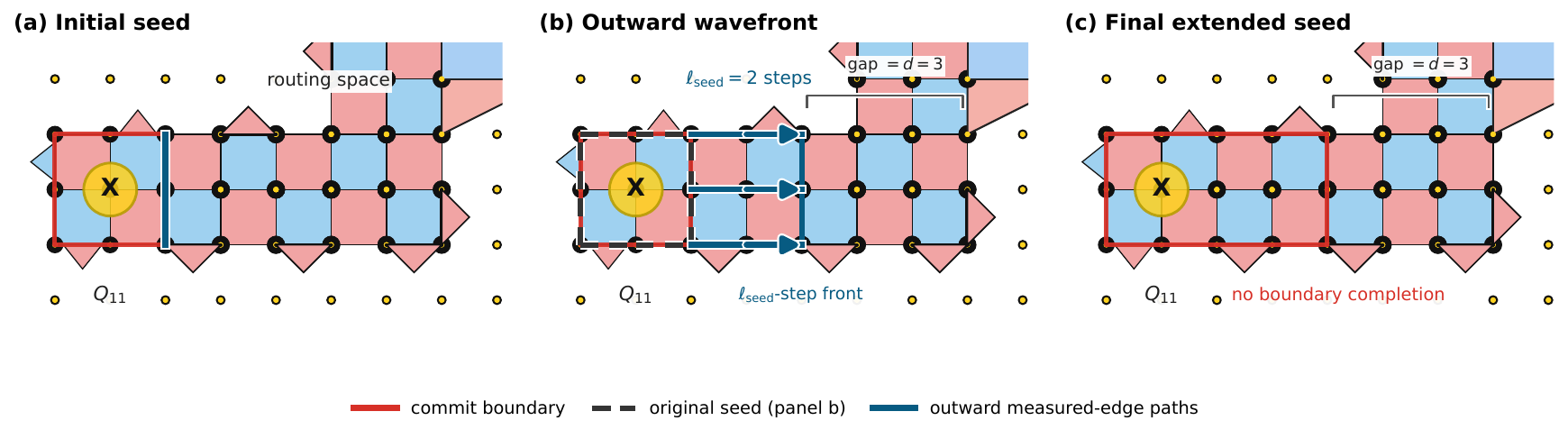}}
     \caption{ (a) Two neighboring logical qubits undergoing an $X \otimes X$ measurement. The seed commit regions for both logical qubits are merged. (b) Example of a $Y \otimes X$ measurement. The southern protruding lobe of the $Y$-extension patch is used in a routing space corridor to connect to the $X$ boundary of the second logical qubit. The commit region reserved for the $Y$-extension patch extends by one row in the southern corridor. (c) Example of a seed commit using the wavefront algorithm to generate an extended seed commit region. $(d+1)/2$ active steps are performed emanating from the face of the initial seed facing the routing space corridor to form the extended seed commit boundary. In this $d=3$ example, the remaining gap to the merged-patch boundary equals $d$, so no further boundary extension is made. }
     \label{fig:q33y_q4x_d5}
\end{figure*}

\begin{figure*}
     \centering
\subfloat[]{\includegraphics[width=.8\textwidth]{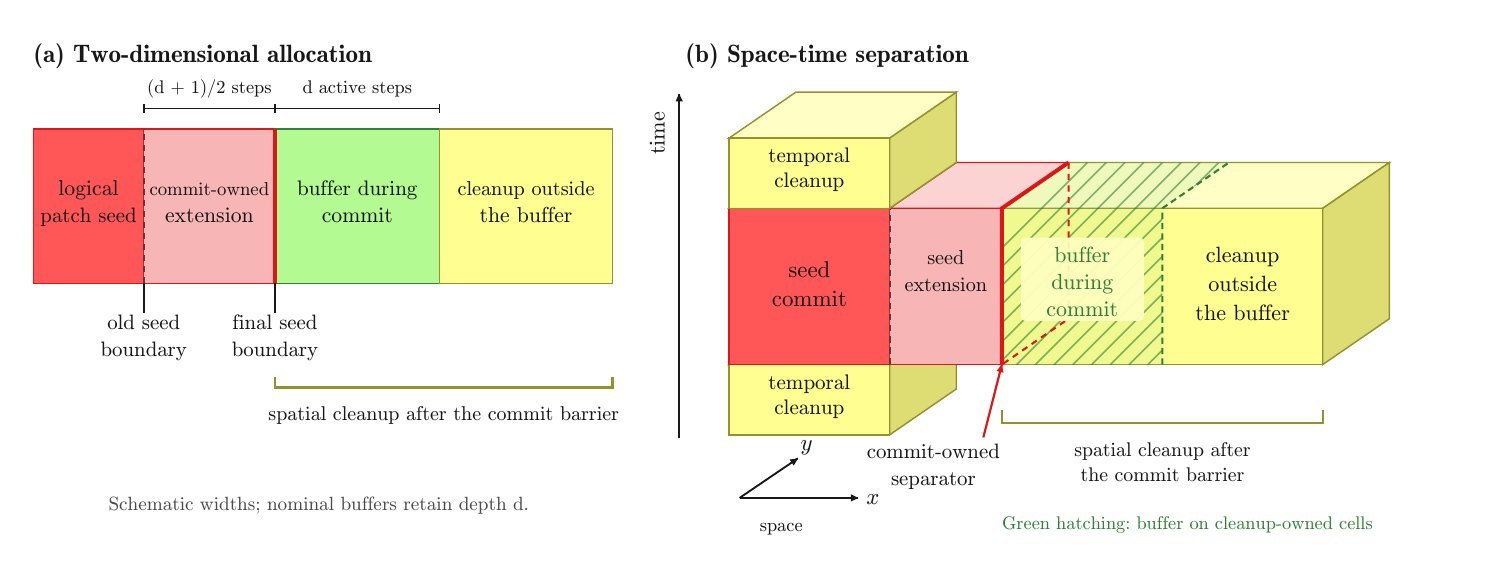}}
     \caption{ Construction of the wavefront-extended seed commit, buffer, and spatial cleanup regions using a seed extension of length $(d+1)/2$ and a buffer of length $d$. (a) The wavefront is propagated from the exposed seed boundary into the routing space, and the active data on and inside its closed measured enclosure are included in the seed commit region shown in red. The buffer is then constructed from the new seed boundary. It is shown in green overlapping the yellow spatial cleanup region: during commit decoding, these cells supply additional syndrome information, while corrections remain restricted to the commit region. After the commit corrections have been applied, the cleanup decoder acts on the residual syndrome in the same cells. (b) The corresponding space-time construction. The commit extension separates the temporal cleanup adjoining the logical patch from the spatial cleanup in the routing corridor. Independent correction execution is checked using the effects of the allowed corrections on the DEM detectors, as described in the text. Widths are schematic. }
     \label{fig:seed-extension}
\end{figure*}

Decoding in real time is necessary to avoid the exponential backlog problem~\cite{TerhalBacklog,ChambsLocalNN22}. For pure quantum memories, this problem was addressed in Refs.~\cite{CampbellParallelV1,AlibabaParallel} by partitioning the syndrome-measurement history into commit regions, their associated buffer regions, and cleanup regions. The buffers are chosen to be sufficiently large that errors in the commit and cleanup regions can be corrected independently up to the full code distance. Distributing these tasks across classical resources allows the syndrome history to grow while keeping the size of the individual decoding tasks controlled. With sufficient parallel resources enough throughput can be achieved, preventing the decoding time from growing exponentially with the number of consecutive non-Clifford gates in a quantum algorithm.

During lattice surgery, the merged patch grows with the number of participating logical qubits and the routing space connecting them. Decoding this patch monolithically increases the workload assigned to a single decoder and can increase decoding latency. At a fixed code distance, our objective is to distribute the additional workload across classical resources while controlling the size of individual decoding tasks, so that larger multi-qubit Pauli measurements can be decoded without a corresponding increase in latency. To this end, we introduce a spatial block-wise decoding algorithm that can be combined with the temporal block-wise decoding protocols of Refs.~\cite{CampbellParallelV1,AlibabaParallel}.

Previous spatial decoding schemes employ fixed window layouts~\cite{Fuhui_Lin_2025} or construct decoding tasks from
a specified network of logical blocks and their interfaces, with buffers grown on the decoding graph~\cite{Bombin2023Modular}. Here, \texttt{patchwork} provides an explicit wavefront algorithm that constructs commit, buffer, and cleanup regions directly from the merged-patch geometry, adapting the partition to irregular boundaries, routing corridors, branches, and $Y$-measurement extensions.

We first describe how the commit, buffer, and cleanup regions are constructed within a two-dimensional slice of a connected merged surface-code patch. We then explain how these spatial regions are extended into space-time for full detector-error-model decoding. Appendix~\ref{app:graph} gives the complete mathematical formulation of the spatial construction, and the discussion below points to the equations defining each major stage of the algorithm.

\subsection{Seed commit regions}
\label{subsec:seed_commit_regions}

Let $M$ denote one connected merged surface-code patch with odd code distance $d\ge 3$. We first place a seed commit region on each participating logical-qubit patch. For an $X$- or $Z$-type seed, this initial footprint contains every active data qubit on or inside the perimeter of the logical-qubit patch. A seed placed on a logical-qubit patch undergoing a $Y$ measurement is treated separately: the seed commit region is chosen to contain the full $Y$-extension patch described in \cref{subsec:gate_schedule_y_ext,sec:protruding_lobes,sec:neighbor_Y}, excluding the protruding lobes. These fixed logical-patch and $Y$-extension footprints are defined before any neighboring seeds are paired. This initial construction is defined in \cref{eq:A13} of \cref{app:seed-regions} in Appendix~\ref{app:graph}.

When the $Y$-extension patches of neighboring logical qubits overlap, their data qubits are initially assigned to provisional seed commit regions. The cleanup separators used to divide horizontal and vertical chains of $Y$ measurements are described in \cref{subsec:y_cleanup_separators}.

We next determine whether pairs of neighboring seed commit regions can be combined before applying the generic routing-space extension to the $X$- and $Z$-type seeds described in \cref{subsec:path_extend_seed}. For this pairing step, two seeds are considered neighboring when they belong to the same multi-qubit Pauli measurement and can be joined through the measured merged patch by a connection containing at most $d+3$ active steps. The connection must also satisfy the geometric conditions given in \cref{app:neighboring-seed-pairs} of Appendix~\ref{app:graph}. Two neighboring seeds can be combined only if either both are $Y$-type or neither is $Y$-type, the resulting region does not contain data belonging to a third seed, and its largest horizontal or vertical span remains within the larger span bound assigned to the two seeds. The initial span bounds are $L_{X/Z}=3d+3$ for $X$- or $Z$-type seeds and $L_Y=4d+4$ for $Y$-type seeds. When shared $Y$-extension data have already been divided among provisional seeds, the corresponding bounds are adjusted as specified in \cref{eq:A13-pair-headroom}. An example of merging commit regions between neighbors is shown in \cref{fig:neighbor_X_pairEX} for two neighboring logical qubits undergoing an $X\otimes X$ measurement.

Each seed can participate in at most one such combination at this stage. Consequently, a chain of $n$ seeds for which every neighboring connection satisfies these conditions is divided into adjacent pairs, together with one unpaired seed when $n$ is odd. For example, a chain of five mutually compatible seeds produces two combined pairs and one unpaired seed. If one or more connections do not satisfy the pairing conditions, additional seeds can remain unpaired. These combinations are provisional: if the enlargement described in \cref{subsec:path_extend_seed} causes a pair to exceed its assigned span bound, we check again whether a cleanup separator can divide it into smaller commit regions, using the procedure described in \cref{subsec:seed_cleanup_separators}. The complete pairing procedure is defined in \cref{app:neighboring-seed-pairs} of Appendix~\ref{app:graph}, with the resulting seed family given in \cref{eq:A13a}.

Now consider protruding lobes of $Y$-extension patches. These lobes are normally excluded from the seed commit region and remain part of the routing space. If the selected lattice-surgery routing passes through a particular lobe, the first complete straight row or column of measured data immediately outside the fixed $Y$-extension footprint can be included in the seed commit region. This face must have active width $d$, and the measured routing corridor must continue outward from it. All subsequent layers of that lobe, together with every lobe not used by the selected routing, remain part of the routing space. Including this first routed face enlarges the footprint of the corresponding $Y$ seed; it does not combine that seed with another seed. An example is shown in \cref{fig:q33y_q4x_d5EX}, with the conditions given in \cref{app:seed-footprint-normalization} of Appendix~\ref{app:graph}.

Since commit regions are already assigned to the full $Y$-extension patch, rather than just the original logical qubit prior to extending the patch, the generic depth-$(d+1)/2$ wavefront extension described next in \cref{subsec:path_extend_seed} is applied only to $X$- and $Z$-type seed commit regions.

\subsection{Wavefront-extended seed commits and buffer regions}
\label{subsec:path_extend_seed}

For an $X$- or $Z$-type seed, the logical-patch footprint is not yet the final seed commit region. Before constructing a buffer or child commit, we take the part of the seed boundary exposed to a measured routing corridor and propagate a wavefront outward through $\ell_{\rm seed}=(d+1)/2$ active steps, as defined in \cref{eq:A-seed-depth}. For seeds already combined at the preceding pairing stage, this propagation starts from the exposed boundary of the combined footprint. We retain the measured stabilizers and data qubits between that boundary and the outer front that determine the enclosure. A short side branch that cannot contribute to the enclosure is removed. The retained boundary paths and terminal fronts are joined by horizontal and vertical boundaries (forming a rectilinear enclosure), and all active data qubits on or inside the resulting measured enclosure are included in the proposed extension. An advantage of the wavefront approach is that we do not need to enumerate all possible paths of length $\ell_{\rm seed}$ through the measured routing corridor. A simple illustration of the wavefront algorithm to generate an extended seed commit region is shown in \cref{fig:seed_wavefront_expansionEX}.

This construction is performed separately for each initial seed commit region. Extensions from different seeds can overlap or become connected through measured stabilizers. We resolve these overlaps and connections by merging the corresponding seed commits or placing a cleanup separator between them, subject to the conditions described in \cref{subsec:seed_cleanup_separators}. Such cleanup separators prevent merged seed commit regions from becoming too large. When seeds are merged, all active data qubits in the combined extended region, together with the measured stabilizers supported entirely on those qubits, belong to a single seed commit region. The wavefront and rectilinear completion are defined in \cref{app:wavefront-enclosure} of Appendix~\ref{app:graph}.

If an exposed face of the extended seed lies fewer than $d$ active data-qubit steps from the boundary of the merged patch, we further extend that face until it reaches the patch boundary, while maintaining a rectilinear enclosure. The proximity test uses the nominal distance $d$, and there is no minimum transverse width for the face. Only the face facing that boundary is extended, following measured connections within the merged patch. This avoids leaving a narrow cleanup region between the seed commit and the patch boundary. Closing this gap also blocks a connection around the seed that could otherwise join separate cleanup regions into one larger region. The precise conditions for this additional extension are given in \cref{app:seed-promotion} of Appendix~\ref{app:graph}.

\begin{figure*}
     \centering
\subfloat[]{\includegraphics[width=.8\textwidth]{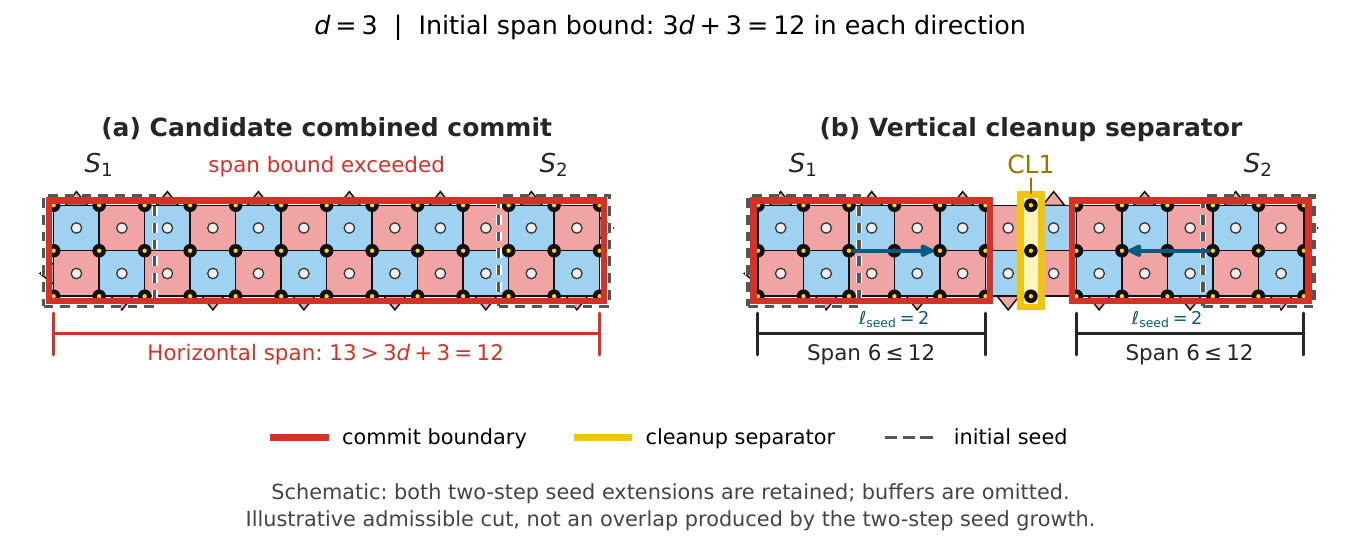}}
     \caption{ Schematic of a possible cleanup separator for an oversized candidate at $d=3$. A full column of cleanup data separates the candidate into two allowed spans while preserving each protected two-step seed extension. Buffers are omitted. }
     \label{fig:seed_cleanup_separator_d3}
\end{figure*}

Once all extended seed commit regions have been formed, we construct their buffer regions from the portions of their perimeters exposed to the routing space corridors. Starting from each such portion, we propagate a wavefront through $d$ active data-qubit steps and form a rectilinear enclosure using the same construction as for the seed extension. Thus only the seed-extension depth is shortened; buffers retain depth $d$. Unlike the seed extension, the buffer is not merged with the seed commit region. It provides additional syndrome information for decoding the seed commit, but corrections are applied only to data qubits within the commit region. Buffer data outside all commit regions remain part of the cleanup regions whose construction we describe below. After the commit corrections have been applied, these cleanup regions are decoded using the residual syndrome. The buffer construction and its distance requirement are given in \cref{eq:A27,eq:A28} and \cref{app:buffer-sufficiency} of Appendix~\ref{app:graph}.

So far we've discussed seed commit regions and their buffer constructions in the space domain. In the temporal domain, we use the parallel window-decoding methods of Refs.~\cite{CampbellParallelV1,AlibabaParallel}, with the simplification that each multi-qubit Pauli measurement is performed over $d$ rounds and consecutive measurements are separated by $d$ idle rounds. These idle rounds provide temporal buffer information for the neighboring seed commits. After the corresponding commit corrections have been applied, the same idle interval is treated as temporal cleanup. An idle interval shared by two consecutive measurements is created only once, as specified in \cref{eq:A50a,eq:A50b}.

\Cref{fig:seed-extension} shows why the seed extension is assigned to the commit region. In (a), the first length-$(d+1)/2$ extension belongs to the seed commit, while the next length-$d$ region is its buffer. In (b), the commit extension separates the temporal cleanup adjoining the logical patch from the spatial cleanup in the routing corridor. If this first extension were only buffer data, it would remain cleanup-owned and could connect the spatial and temporal cleanup regions. They would then need to be decoded together. 

\subsection{Cleanup separators between $X$- and $Z$-type seed commit regions}
\label{subsec:seed_cleanup_separators}

The proposed extensions of all seeds are calculated from the same original seed geometry, so one proposed extension does not block the construction of another. For an $X$- or $Z$-type seed, the initial bound on its horizontal and vertical spans is $3d+3$. If its own required extension exceeds this bound, the bound is increased just enough to contain that extension. Seeds whose proposals overlap or become connected through a measured stabilizer can be combined when the combined region fits within the largest of their individual bounds. Otherwise, we search for a cleanup separator. Buffer allowance does not increase this ownership bound. The precise conditions are given in \cref{eq:A19a,eq:A19b,eq:A19c,eq:A20}.

A separator consists of a horizontal or vertical line of active routing data, one coordinate wide, spanning the full local corridor. When necessary, one line of each orientation can form an L-, T-, or cross-shaped separator. Each seed retains its complete length-$\ell_{\rm seed}=(d+1)/2$ extension, together with any additional extension to the merged-patch boundary. The separator may lie immediately outside this protected region, but cannot remove any of its data qubits. Thus, in a straight corridor without an additional boundary extension, the first available separator layer is one active step beyond the length-$\ell_{\rm seed}$ extension. Its position must still allow a complete cut satisfying the connectivity and stabilizer-ownership conditions in \cref{eq:A21,eq:A22,eq:A23,eq:A24,eq:A25}. Buffers and child commits retain their length-$d$ construction. A schematic example is shown in \cref{fig:seed_cleanup_separator_d3}.

We first search for a separator across the full group of conflicting seeds and then, if necessary, between its directly conflicting oversized pairs. After each accepted cut or combination, the remaining conflicts are checked again. If no valid separator can divide a conflicting pair, that pair is combined and its span bound is increased only as much as required. Other seeds retain their own bounds and may still be separated from the combined pair. This search and the local exception to the span bound are specified in \cref{eq:A25a,eq:A25b,eq:A25c}.

Seed pairs formed before the wavefront extension are also reconsidered if their extended region exceeds the original pairing bound. Their original constituent seeds are then extended separately by the same $\ell_{\rm seed}$, and the routing data that originally joined the pair become available for a cleanup separator. If the pair has already joined a larger group, all affected constituent seeds are checked together. An existing cleanup gap may provide the separation without a new cut. A pair that remains within its bound is not divided merely because a cut could fit. The complete rule is given in \cref{app:seed-pair-replay,eq:A25d}.

A single row or column of routing data between facing seed commits is initially left unassigned so that it remains available for cleanup. If it is not reserved as a separator, it may be added to the smaller seed, subject to the size and geometry conditions in \cref{app:deferred-seed-strips,eq:A25e}. After seed proposals are combined, we also check for a stabilizer whose committed support belongs to one seed but whose remaining support lies in separate cleanup components. Those remaining data are included in that seed only when they are not reserved for cleanup or assigned to another region, and the resulting seed satisfies its span bound. These ownership checks are completed before buffers are constructed.

Accepted separators remain cleanup-owned, except where a child commit must occupy a routing junction crossed by a separator, as described in \cref{subsec:child_commit_regions}. The construction is applied separately to each connected merged patch associated with a multi-qubit Pauli measurement.

\subsection{Cleanup separators in horizontal and vertical chains of $Y$ measurements}
\label{subsec:y_cleanup_separators}

\begin{figure*}
     \centering
\subfloat[\label{fig:q2y_q3y_q6y_q7y_q10y_q11y_d5EX} ]{\includegraphics[width=.7\textwidth]{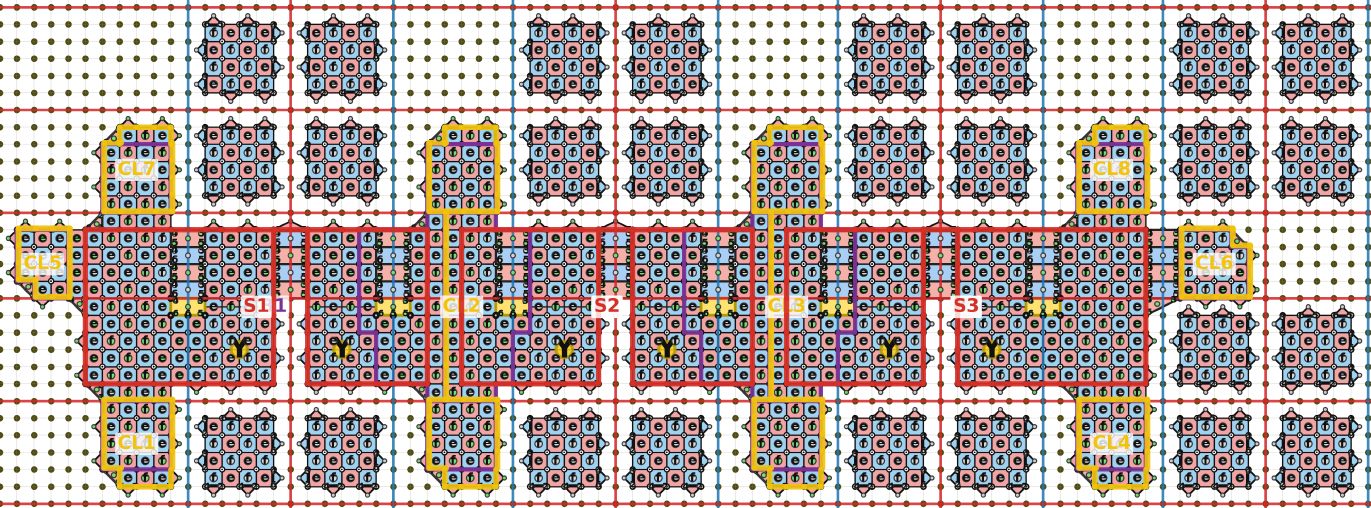}}
\vfill
\subfloat[\label{fig:q53y_q56y_q39y_q42y_q37y_q40y_q23y_q26y_q21y_q24y_q7y_q10y_d5EX} ]{\includegraphics[width=.35\textwidth]{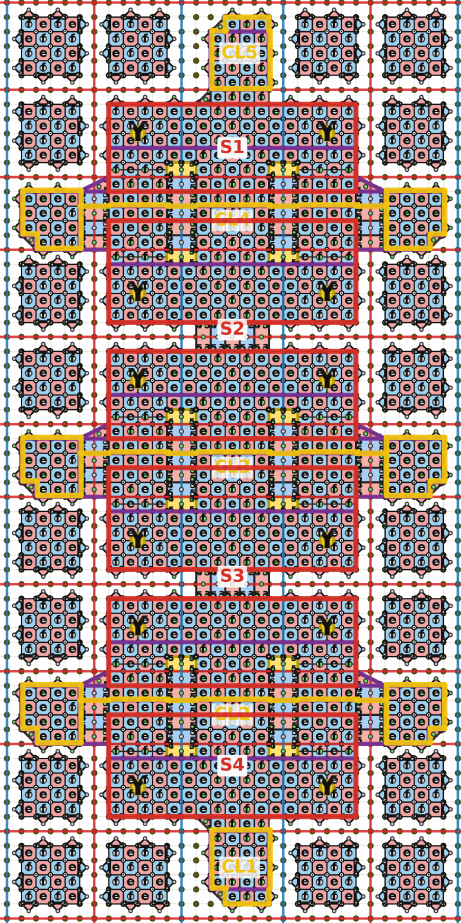}}
     \caption{ (a) Example of the commit and cleanup regions for a horizontal chain of $Y$ measurements. Neighboring $Y$-extension patches connected by horizontal elongated stabilizers are assigned to the same commit region. At interfaces where neighboring $Y$-extension patches overlap without being connected by a horizontal elongated stabilizer, a vertical cleanup region is placed midway between the two logical-qubit patches. The resulting commit regions $S1$, $S2$, and $S3$ are separated by the vertical cleanup regions $CL2$ and $CL3$, preventing the complete chain from forming one commit region. (b) Same as (a) but for a vertical chain of $Y$ measurements.}
     \label{fig:q2y_q3y_q6y_q7y_q10y_q11y_d5}
\end{figure*}

For a horizontal chain of $Y$ measurements, neighboring patches joined by horizontal elongated stabilizers belong to the same commit region. Between these connected groups, the $Y$ extensions can overlap without a horizontal elongated stabilizer joining them. We place vertical cleanup separators in these overlap regions, separating the neighboring groups without cutting through the horizontal elongated stabilizers. An example is shown in \cref{fig:q2y_q3y_q6y_q7y_q10y_q11y_d5EX}.

We proceed from left to right. At each eligible interface, the cut is placed midway between the facing boundaries of the two logical-qubit patches. If there are two middle columns, we choose the column closest to the patch on the left. The cut covers the full height of the overlapping $Y$-extension region. It must lie entirely in measured routing data, avoid the logical-patch cores, and leave each retained side connected to its logical patch. An elongated stabilizer directed across the proposed separation must neither touch nor bridge the cut. An elongated stabilizer parallel to the cut is permitted, provided the connectivity and stabilizer-ownership conditions are satisfied. Any measured stabilizer acting on both retained seed commits must also act on at least one data qubit in the cleanup separator. This prevents a stabilizer supported entirely on the two commits from forcing them to merge again. These conditions are given in \cref{eq:A26a,eq:A26b,eq:A26c}.

The accepted column is removed from the seed commit regions and reserved for cleanup. It can supply buffer syndrome information, but it is not absorbed by the later thin-strip adjustments. The resulting cuts prevent the final combination of neighboring commits from continuing across the complete horizontal chain and growing indefinitely. This is a separate construction from the separator search for $X/Z$ seed extensions: only the specified midpoint column is tested. If it is invalid, that interface remains uncut. The update is defined in \cref{eq:A26d}.

The same construction applies to vertical chains of $Y$ measurements, using horizontal cleanup separators. We proceed from top to bottom and place each cut midway between the facing boundaries of the two logical-qubit patches. If there are two middle rows, we choose the row closest to the upper patch. The cut spans the full active width of the overlapping $Y$-extension region and must satisfy the same connectivity and stabilizer-ownership conditions described above. For both orientations, a cleanup separator may pass through several neighboring overlapping $Y$-extension regions. We check the resulting segments together to ensure that they form a complete separator across the corridor. If any required segment is invalid, the entire separator is rejected; we do not retain an incomplete cut. This joint check is defined in \cref{app:y-seed-separators}. The accepted row is removed from the seed commit regions and reserved for cleanup. As for horizontal chains, only the specified midpoint is tested; if it is invalid, that interface remains uncut. An example is shown in \cref{fig:q53y_q56y_q39y_q42y_q37y_q40y_q23y_q26y_q21y_q24y_q7y_q10y_d5EX}.

\subsection{Child commit regions}
\label{subsec:child_commit_regions}

\begin{figure*}
     \centering
     \includegraphics[width=.98\textwidth]{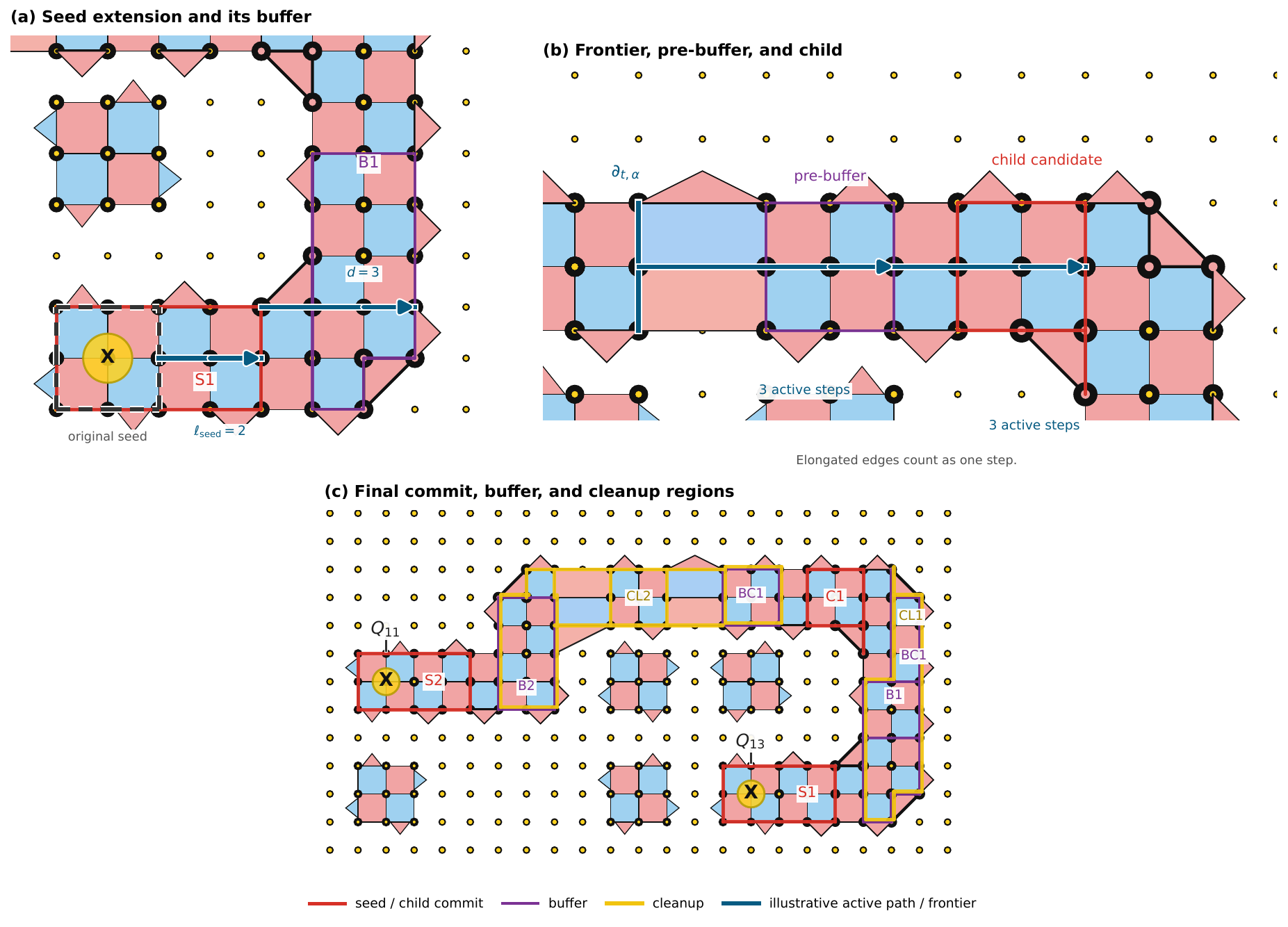}
     \caption{Wavefront construction for an $X_{Q_{11}}\otimes X_{Q_{13}}$ measurement at $d=3$ on a $(2,2)$ cluster grid. (a) The original $Q_{13}$ seed is shown with a dashed perimeter. Its two-step extension is commit-owned, and its depth-three buffer is constructed from the new exposed boundary. Illustrative paths distinguish the two active steps used for seed growth from the three active steps used for its buffer. (b) Starting from the $Q_{11}$ seed's buffer, eastward propagation reaches $\partial_{t,\alpha}$. The next three active steps give the pre-buffer, followed by the child candidate. An elongated measured edge counts as one step; its inactive middle coordinate does not contribute a step. (c) Final commit, buffer, and cleanup regions. The child remains separate from both the $Q_{11}$ and $Q_{13}$ seeds. Red, purple, and yellow boundaries denote commits, buffers, and cleanup, respectively. The regions labeled $BC1$ form the child's buffer. Buffer data outside the commits remain cleanup-owned.}
     \label{fig:wavefront-construction}
\end{figure*}

Consider a multi-qubit Pauli measurement between two logical qubits that are far apart on the two-dimensional grid. Although the extended seed commit regions cover part of the routing space corridor, a large gap may remain between them. If this gap were left entirely to cleanup, it could produce a large region that must be decoded together. We therefore introduce a child commit algorithm that places additional commit regions along the routing space corridor to limit the size of the remaining cleanup regions.

We first distinguish how a seed is connected to the routing space. An endpoint seed lies at the end of a corridor. A branch seed is attached to the side of a corridor that continues past the attachment. Starting from an endpoint seed, we begin at the outer boundary of its buffer and propagate a wavefront through $d$ active steps away from the seed. The extremal terminal faces define the candidate frontier $\partial_{t,\alpha}$, where $\alpha$ is the direction of the corridor leaving that face. Starting one active step beyond this frontier, the next $d$ active sites along the forward rails define an intermediate region called the \textit{pre-buffer}. A further $d$-step construction from the outer boundary of the pre-buffer gives the next child commit candidate. This procedure can follow a straight corridor, turn a corner, or follow the active boundary of an elongated or trapezoidal stabilizer. At each stage, we retain the paths that can complete the prescribed number of active steps and use their terminal fronts to construct the region's boundary. A path that terminates too early does not produce a child commit. The three stages, including the distinction between proposing a frontier and completing the pre-buffer or child, are defined in \cref{eq:A29,eq:A30,eq:A31}.

Once a child commit has been placed, we recompute its buffer from the final child boundary. In a straight corridor, the next construction starts from the outer boundary of this buffer on the side opposite the pre-buffer used to construct the child. When there are several outgoing routing faces, we consider their unexplored buffer boundaries separately. If the construction reaches an endpoint seed before another child can be placed, it terminates along that path without placing a child. We then consider the other unexplored routing faces of the reached seed, starting from their corresponding buffer boundaries. Any unexplored routing faces of previously placed children also remain available. Buffer recomputation and continuation upon reaching an endpoint seed are described in \cref{eq:A34,eq:A35} and \cref{app:wavefront-certificates}. 

A short attachment to a branch seed does not start a second construction along the same corridor. We continue along the main corridor without treating that short side attachment as an independent route. Endpoint seeds are used as starting points one at a time. Another endpoint can be tried if the first cannot make progress, or if it is needed to reach a junction left unexplored after the active paths have ended. The detailed ordering is given in \cref{eq:A14,eq:A28a,eq:A32}.

When the construction reaches a junction that is not already occupied by a commit, a child commit is placed over the complete junction. If an ordinary child would be placed within active distance $d$ of the junction, the child commit is shifted to cover the complete junction instead. After its buffer is constructed, child placement continues independently along every unexplored corridor leaving the junction. Failure to continue through one corridor does not stop the others.

A cleanup separator passing through a junction does not prevent this child from being placed. The child is assigned the full junction region, and the portion of the separator inside that region becomes commit-owned. The rest of the separator remains cleanup-owned on the sides of the child. In this way, placing the child breaks the separator into separate pieces without leaving part of the junction unassigned. The junction is identified from the full measured routing geometry, before removing the separator data. This rule applies only to a true junction with no existing commit there; a short attachment to a branch seed does not create an additional child. The complete construction is given in \cref{app:junction-children}, \cref{eq:A31a,eq:A31b,eq:A31c}.

\begin{figure*}
     \centering
\subfloat[]{%
    \IfFileExists{images/q38y_q26x_q27z__q42z_q20z_q22x_d5.pdf}{%
        \includegraphics[width=.8\textwidth]{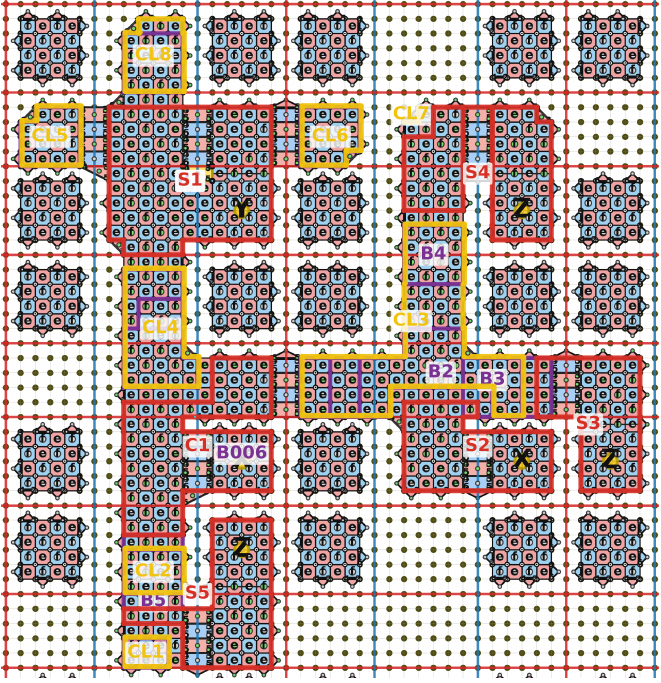}%
    }{%
        \PackageWarningNoLine{patchwork}{Missing figure asset; draft placeholder used}%
        \fbox{\parbox[c][3cm][c]{.78\textwidth}{\centering
            Figure asset missing from this checkout\\[1ex]
            \footnotesize\texttt{\detokenize{images/q38y_q26x_q27z__q42z_q20z_q22x_d5.pdf}}}}%
    }%
}
     \caption{ Example two-dimensional slice for a $P=Y_{Q_{38}}\otimes X_{Q_{26}}\otimes Z_{Q_{27}}\otimes Z_{Q_{42}}\otimes Z_{Q_{20}}\otimes X_{Q_{22}}$ measurement. Red and yellow borders denote commit and cleanup regions, respectively. Purple borders denote the buffer regions associated with the commits.}
     \label{fig:q38y_q26x_q27z__q42z_q20z_q22x_d5}
\end{figure*}

As each child is placed, we check whether a small local adjustment is needed. If a seed and child are separated by only one or two complete rows or columns of routing data, they can be combined with that strip when the resulting region remains within the applicable span bound. A strip deliberately reserved as a cleanup separator is not absorbed in this way. A junction child can also be enlarged by the minimum adjacent row or column needed to prevent a measured stabilizer from connecting two intended cleanup regions, provided that the child remains within its span bound. Whenever a commit is moved, enlarged, or combined with another commit, its buffer is reconstructed from the new boundary before placing any further children. These adjustments are defined in \cref{eq:A33,eq:A34,eq:A35}.

\cref{fig:wavefront-construction} illustrates this construction in a measured routing corridor. Panel (a) shows the extended seed and its buffer. Panel (b) shows the frontier $\partial_{t,\alpha}$, pre-buffer, and child commit candidate obtained by propagating eastward from the $Q_{11}$ seed's buffer. Panel (c) shows the resulting three commit regions, with the child remaining separate from both seeds.

\subsection{Final commit and cleanup regions}
\label{subsec:final_cleanup_regions}

After all seed and child commit regions have been constructed and their local adjustments completed, we make a final check of the support of every measured stabilizer. If a stabilizer acts only on data belonging to two or more commit regions, those commits must be combined so that one decoder can correct it. This combination is repeated as necessary until no fully commit-owned stabilizer is shared by distinct commits. If the required combination exceeds a previous span bound, its ownership bound is increased only to the size of the combined region. Combining commits assigns their existing data to one decoder; it does not fill the rectangle between them or add unrelated routing data. The final combination rule is given in \cref{eq:A36,eq:A37,eq:A38,eq:A39,eq:A39a}.

Suppose a remaining single row, column, or data qubit is completely enclosed by a single commit region and does not touch another commit. If every stabilizer touching these data qubits has all its other support in that same commit, we include them in the commit region provided that the enlargement remains within the applicable span bound. Reserved cleanup separator cells are excluded from this adjustment. All affected buffers are then reconstructed.

Only now are the final cleanup regions formed. Earlier references to cleanup cuts or cleanup-owned data specify data reserved for cleanup, without yet assigning them to separate cleanup decoders. Let $C_{\mathrm{all}}$ be the union of all final commit regions. The remaining active data qubits, $V_M\setminus C_{\mathrm{all}}$, are divided into connected regions. These give the cleanup regions.

Separate pieces of an accepted L-, T-, or cross-shaped cleanup separator can be connected by a measured stabilizer even when they are not connected by neighboring data qubits. We assign these pieces to one cleanup region when that stabilizer connection was accepted during the separator construction and all the data qubits on which it acts outside the commits remain reserved for cleanup separators. The precise conditions are given in \cref{eq:A42a,eq:A42b}. A connection removed when a child commit occupies the junction cannot be used to reconnect the remaining cleanup pieces.

A measured stabilizer belongs to a commit only when its entire active support lies in that commit. Otherwise, it belongs to the cleanup region containing its support outside the commits. Every stabilizer is checked to have exactly one such owner. A stabilizer crossing a commit boundary may still provide syndrome information through a buffer, but it remains cleanup-owned. In DEM decoding, the selected corrections are represented by their detector and logical-observable effects. The full detector effects of the commit responses are included in the residual syndrome passed to cleanup, as described in \cref{sec:spacetime_corrections}. The complete partition and ownership conditions are given in \cref{app:cleanup-ownership} of Appendix~\ref{app:graph}, \cref{eq:A40,eq:A41,eq:A42,eq:A43,eq:A44,eq:A45,eq:A46,eq:A47}.

\begin{figure*}
     \centering
\subfloat[]{\includegraphics[width=.8\textwidth]{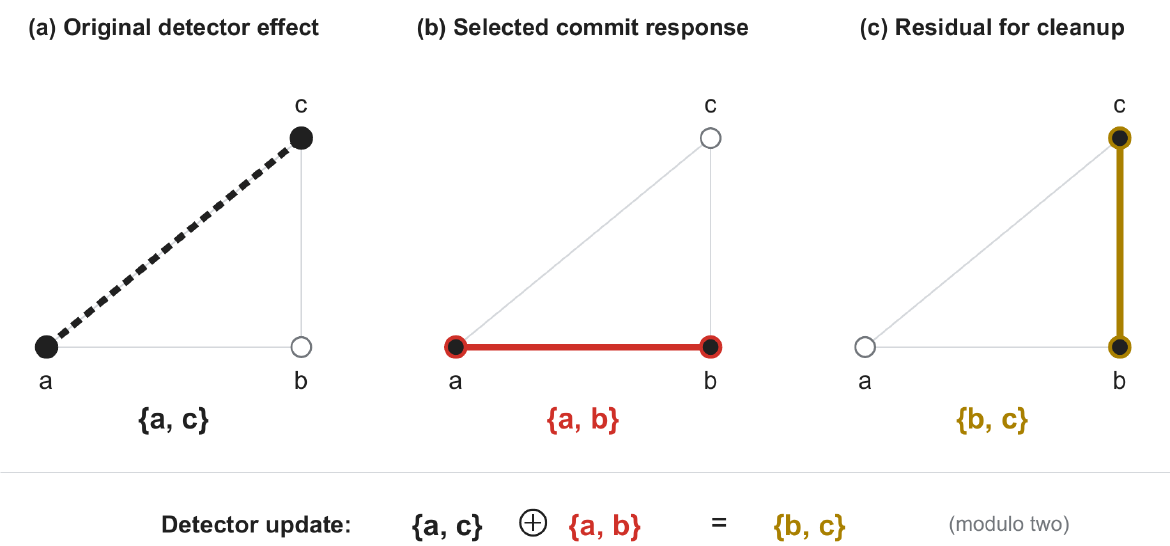}}
     \caption{ Illustration of the detector-syndrome update after commit decoding. (a) The original detector effect is $\{a,c\}$. (b) For the illustrated example, the selected commit response has detector effect $\{a,b\}$. (c) Combining these effects modulo two cancels $a$ and leaves $\{b,c\}$ for cleanup. Filled circles denote nonzero detector effects in each panel. The panels show full detector effects; the associated logical-observable effects are tracked alongside them.}
     \label{fig:spacetime_dem_partition}
\end{figure*}

For a fixed logical-patch layout and measurement pattern, our numerical studies are consistent with approximately $O(d^3)$ scaling of the wavefront commit/cleanup construction over the distances tested. This is numerical evidence, not a proven worst-case bound for arbitrary routing geometries. The improvement comes from finding and filling the measured enclosure and combining equivalent propagation states, rather than enumerating every path inside it. \Cref{app:wavefront-scaling} of Appendix~\ref{app:graph} separates this observation from the cost of individual wavefront steps and from circuit and DEM construction.

An example of a full commit and cleanup construction for a multi-qubit Pauli measurement using the techniques described above is shown in \cref{fig:q38y_q26x_q27z__q42z_q20z_q22x_d5}.

\subsection{Space-time decoding}
\label{subsec:spacetime_decoding}

Each multi-qubit Pauli measurement is performed over $d$ syndrome-measurement rounds. The spatial commit, buffer, and cleanup regions described above are therefore defined throughout those $d$ rounds, forming three-dimensional decoding volumes. We consider an architecture where each participating logical qubit undergoes $d$ idle rounds before and $d$ idle rounds after a multi-qubit Pauli measurement. For consecutive measurements involving the same logical qubit, the intervening $d$ idle rounds serve both purposes and are included only once. These idle rounds provide the temporal buffers of seed commit regions and later the temporal cleanup regions already described. A single commit can contain several logical-qubit patches, but the temporal intervals are still associated with the corresponding logical qubits. These volumes are defined in \cref{eq:A48,eq:A49,eq:A50,eq:A50a,eq:A50b}.

All commit regions are decoded in parallel using the syndrome information in their commit and buffer volumes. Cleanup decoding follows after the commit corrections have been incorporated into the syndrome. The construction of the corresponding detector-error models (DEMs), the assignment of corrections, and the conditions under which cleanup regions are decoded together are described in \cref{sec:spacetime_corrections}.

\section{Space-time DEM partitioning}
\label{sec:spacetime_corrections}

In \cref{sec:spacetime_partition}, we assigned data qubits and measured stabilizers to commit and cleanup regions. We construct their DEMs from the syndrome-extraction circuit and noise model using Stim~\cite{Gidney2021stimfaststabilizer}. Each modeled error specifies a probability, the detectors it flips, and its logical-observable effects. For graph-based decoders such as PyMatching~\cite{HiggottPyMatch} or Union Find~\cite{DelfosseUnionFind}, one or two detector endpoints define a boundary edge or an ordinary edge, respectively; larger effects require a supported graphlike decomposition.

Each commit decoder observes its commit and buffer volumes. The buffer provides additional syndrome information that helps the decoder distinguish between possible errors. The local DEM includes this context, while only error components assigned to the commit contribute to its correction response. These responses record the full detector and logical-observable effects of the selected corrections.

All commit decoders act on the original observed syndrome. Their selected detector effects are combined modulo two and applied to the global syndrome, including any changes to cleanup detectors. Cleanup then acts on the remaining syndrome. \Cref{fig:spacetime_dem_partition} illustrates this update: if the original detector effect is $\{a,c\}$ and the selected commit response has detector effect $\{a,b\}$, the residual effect is $\{b,c\}$. The selected response depends on the local syndrome and DEM. In the equations below, $c_i$ denotes such a correction response, with its detector and logical-observable effects tracked together.

Cleanup observations assigned to the same correction region are decoded together. Distinct cleanup regions are also decoded together if their allowed corrections can change a common detector outcome. This grouping is transitive: if one region is connected in this way to two others, all three belong to the same decoding task. Sharing DEM information alone does not require regions to be grouped. The precise criterion is given in \cref{eq:A51a}.

Both commit and cleanup DEMs are constructed before runtime decoding. During compilation, we determine the commit decoders' responses to the syndromes associated with individual graphlike components of the original DEM. For a component $e$ with detector and logical effects $(\boldsymbol{d}_e,\boldsymbol{\ell}_e)$, let $c_i[e]$ be the correction effect selected by commit decoder $i$ in response to that component's syndrome. Its remaining effect is
\begin{align}
\boldsymbol{r}_e
&=
(\boldsymbol{d}_e,\boldsymbol{\ell}_e)
\oplus
\bigoplus_i
\bigl(\Delta_D(c_i[e]),\Delta_L(c_i[e])\bigr),
\label{eq:cleanup_replayed_effect}
\end{align}
where $\Delta_D$ and $\Delta_L$ give the full detector and logical-observable changes represented by a selected correction. These remaining effects define the cleanup likelihood model.

For example, suppose a fault of probability $p$ flips detectors $D_a$ and $D_c$, while the selected commit correction flips $D_a$ and $D_b$. The remaining detector effect is
\begin{align}
\{D_a,D_c\}\mathbin{\triangle}\{D_a,D_b\}
&=
\{D_b,D_c\},
\label{eq:cleanup_boundary_example}
\end{align}
where $\triangle$ denotes cancellation modulo two. The fault therefore contributes an edge connecting $D_b$ and $D_c$ to the cleanup DEM with its original probability $p$: the correction changes the represented effect, not the underlying fault probability.

For a local decoder $R$, let $\mathcal{F}_R(a)$ contain independent error mechanisms of probabilities $p_j$ that contribute the same local edge and correction response $a$, including its logical-frame effect. For example, a Pauli error during an idle step and a particular Pauli fault associated with a CNOT gate at another circuit location can produce the same local edge and correction response, including the same logical-frame effect, while occurring with different probabilities. Their combined probability and matching weight are
\begin{align}
q_R(a)
&=
\frac{1}{2}
\left[
1-\prod_{j\in\mathcal{F}_R(a)}(1-2p_j)
\right],
\nonumber\\
w_R(a)
&=
\log\!\left(\frac{1-q_R(a)}{q_R(a)}\right),
\label{eq:local_dem_probabilities}
\end{align}
before any correlation-aware reweighting. The product accounts for cancellation when an even number of these mechanisms occur. Commit models group the original locally visible effects and assign correction responses to components owned by the commit decoder; cleanup models group the remaining effects in \cref{eq:cleanup_replayed_effect}. Different mechanisms can therefore contribute to their boundary edges, giving different effective edge probabilities.

A single fault can produce effects in both commit and cleanup regions. Assigning those effects to different regions does not make them separate faults: each contribution retains the probability of the original fault. Correlation-aware decoding accounts for their common origin within a local decoder. When correlated effects are handled by separate workers, their joint occurrence is not fully represented by the separate local models.

The cleanup probabilities are precomputed using the commit responses to individual DEM effects. Several effects occurring together can lead to different commit corrections, so these probabilities approximate the noise remaining after commit decoding. At runtime, however, the residual syndrome is computed from the actual corrections selected for the full observed syndrome. In particular, all commit decoders act on the original observed syndrome $\boldsymbol{s}$. If $c_i$ is the correction effect selected by commit decoder $i$, cleanup receives
\begin{align}
\boldsymbol{s}_{\mathrm{res}}
&=
\boldsymbol{s}
\oplus
\bigoplus_i \Delta_D(c_i).
\label{eq:cleanup_residual_syndrome}
\end{align}
This update includes changes to cleanup detectors caused by the selected commit corrections. Cleanup decodes the residual using its preconstructed DEM, and the logical-observable effects of all selected commit and cleanup corrections are included in the final logical-frame update.

\begin{figure*}
     \centering
\subfloat[\label{fig:Exp4_M1EX} ]{\includegraphics[width=1.0\textwidth]{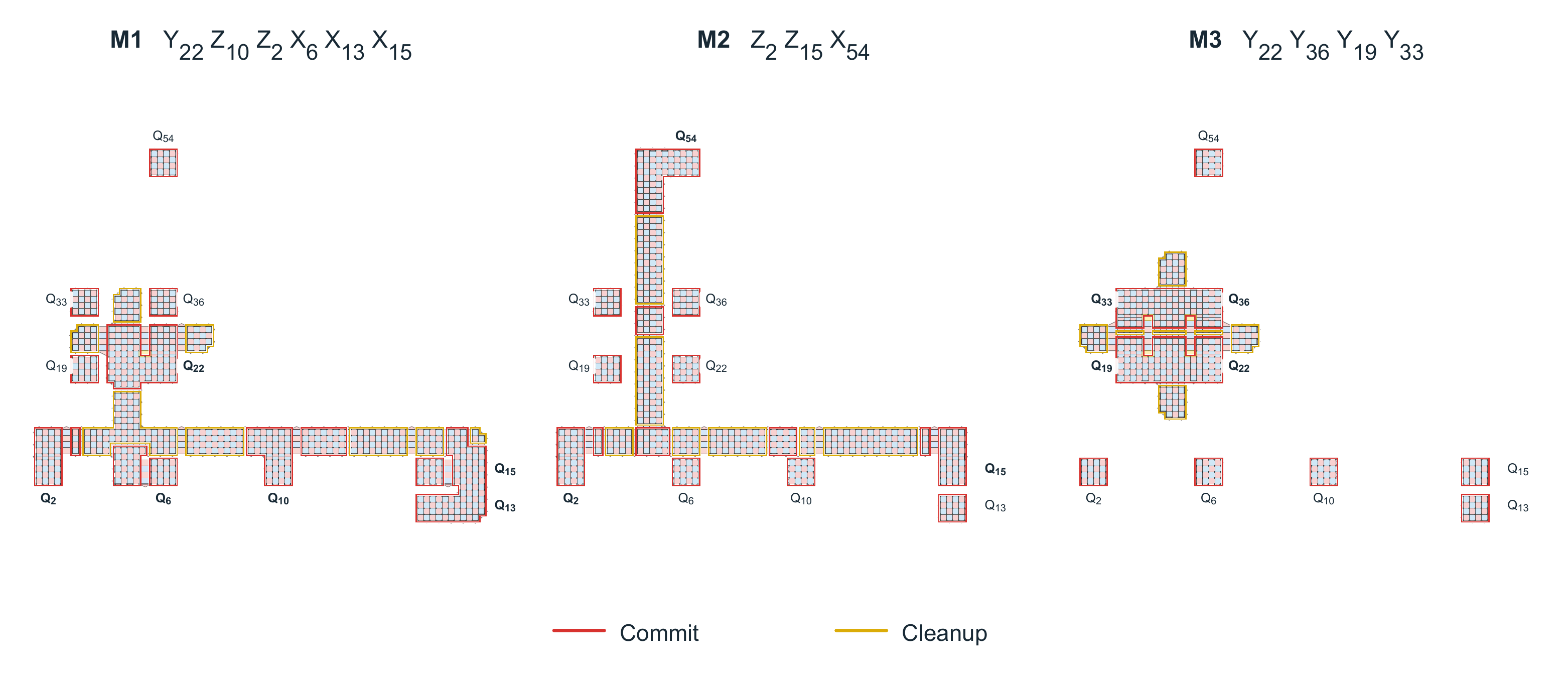}}
\vfill
\subfloat[\label{fig:Exp4_M2EX} ]{\includegraphics[width=.5\textwidth]{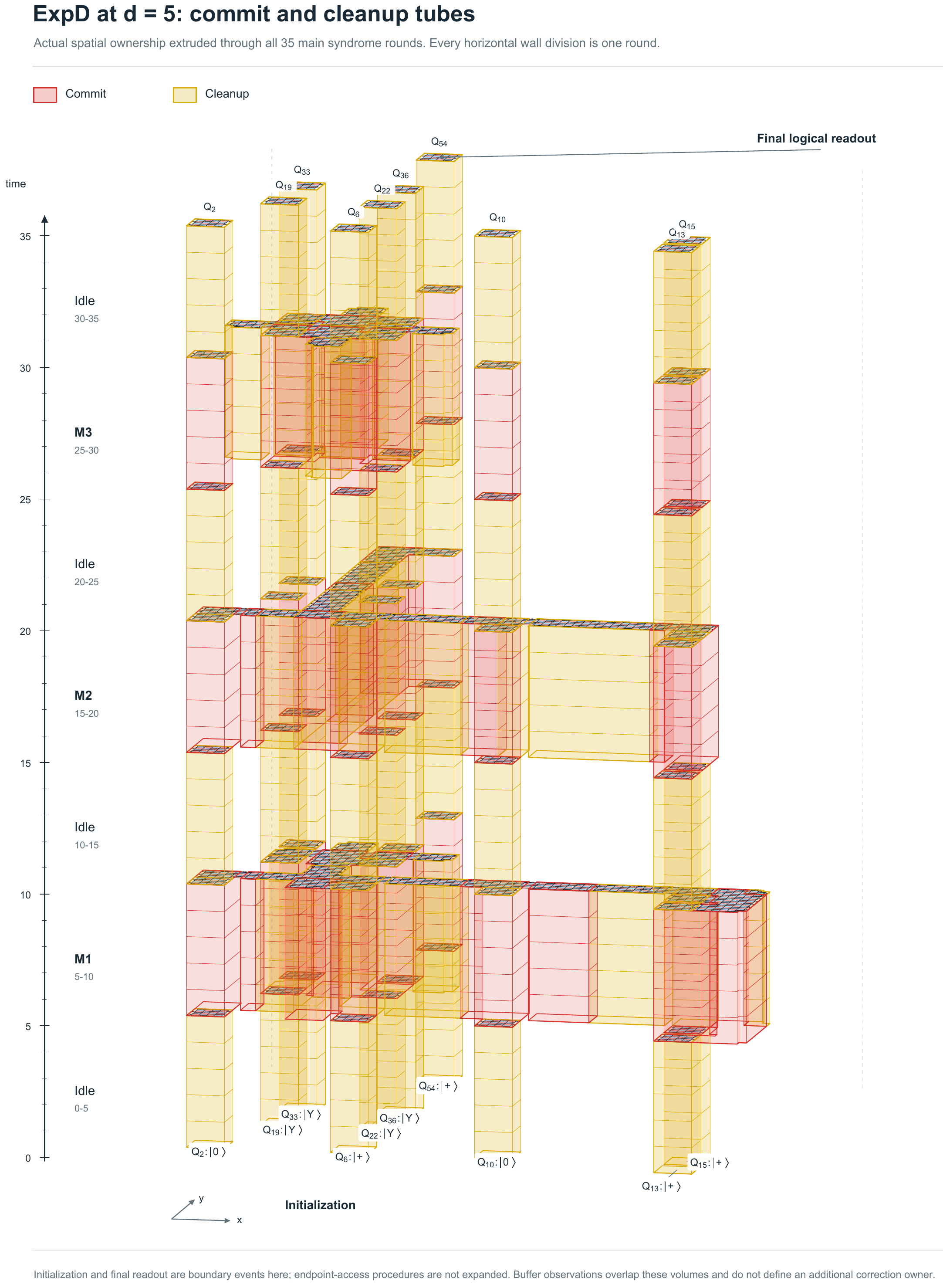}}
     \caption{ (a) Illustration for the three measurements shown in \cref{eq:exp4PsM1,eq:exp4PsM2,eq:exp4PsM3} for \texttt{ExpD}. For each multi-qubit Pauli measurement, we show both the merged surface code patch as well as the other logical qubits participating in other measurements for \texttt{ExpD}. Each measurement is performed for $d$ rounds (with $d=5$ in these illustrations) with $d$ idling rounds in between, before and after the final measurement. (b) Full three-dimensional visualization of \texttt{ExpD}, including the initialization and readout bases. }
     \label{fig:Exp4_Measurements}
\end{figure*}

\begin{figure*}
     \centering
\subfloat[\label{fig:exp4_ler_d5_d7_d9_loglogEX} ]{\includegraphics[width=.8\textwidth]{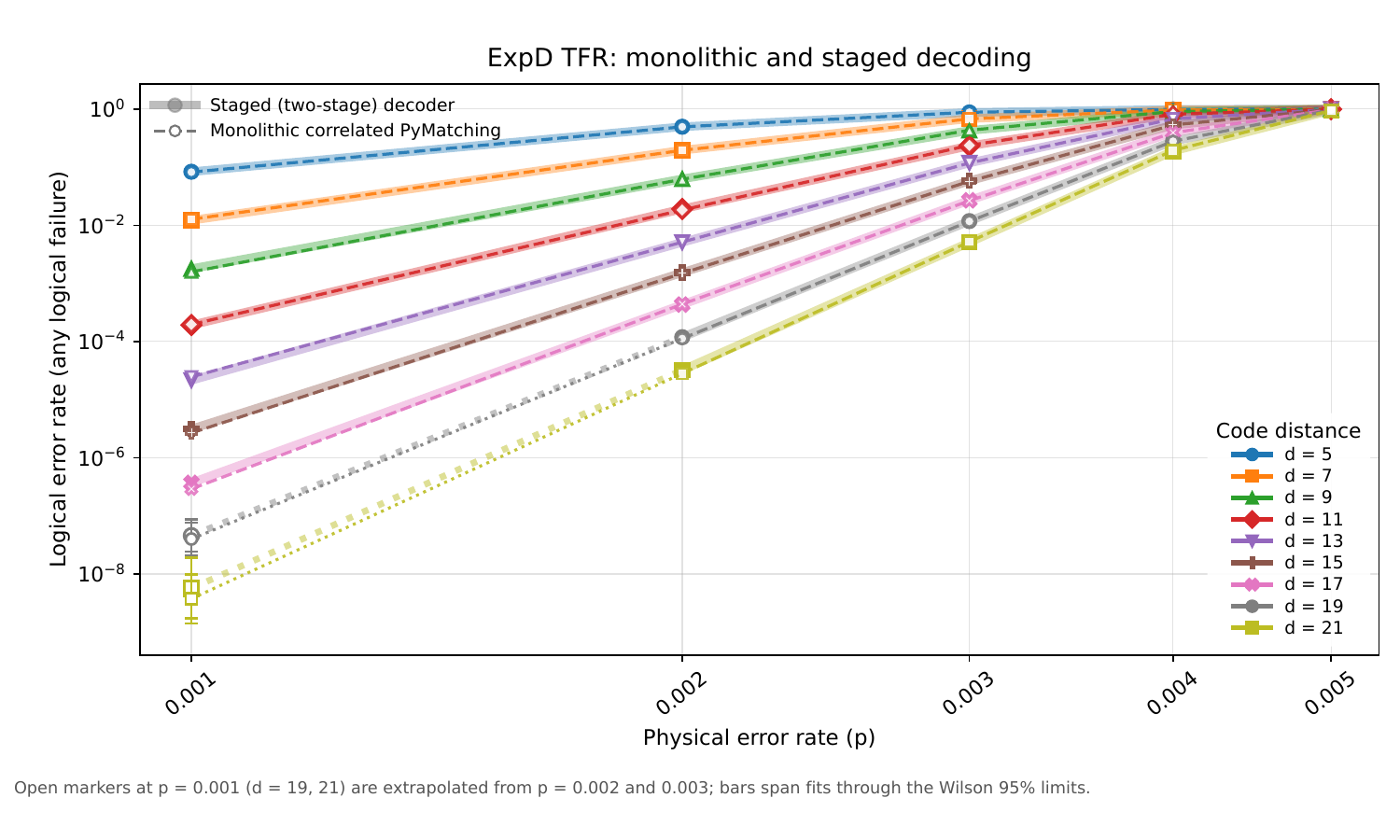}}
     \caption{ TFR of the three consecutive multi-qubit Pauli measurements given in \cref{eq:exp4PsM1,eq:exp4PsM2,eq:exp4PsM3}. The state-preparation bases and final readout bases for each logical qubit are given in \cref{tab:exp4_logical_qubits}. TFR corresponds to the probability that \textit{any} of the logical qubits to receive a spacelike logical failure or for any of the three multi-qubit Pauli measurements to have a timelike logical failure. }
     \label{fig:LERresult}
\end{figure*}

\section{Numerical results}
\label{sec:numerics}

\subsection{LER's for sequential multi-qubit Pauli measurements}
\label{subsec:Sequence_PPM}

In this section we demonstrate the methods discussed in \cref{sec:spacetime_partition,sec:spacetime_corrections} with a simple experiment consisting of three consecutive multi-qubit Pauli operators. We label this experiment \texttt{ExpD}. In \texttt{ExpD}, we calculate the \textit{total} failure rate (TFR) of the experiment. That is, the probability that any of the multi-qubit Pauli measurements have a logical timelike failure or any of the logical qubits contain a logical spacelike failure after readout.

\begin{table}[htbp]
 \centering
 \begin{tabular}{|c|c|c|}
 \hline
 Logical qubit & Preparation state &  Final readout basis \\
 \hline
 $Q_2$  & $| 0 \rangle$ & $Z$ \\
 $Q_6$  & $| + \rangle$ & $X$ \\
 $Q_{10}$ & $| 0 \rangle$ & $Z$ \\
 $Q_{13}$ & $| + \rangle$ & $X$ \\
 $Q_{15}$ & $| + \rangle$ & $Z$ \\
 $Q_{19}$ & $| Y \rangle$ & $Y$ \\
 $Q_{22}$ & $| Y \rangle$ & $Y$ \\
 $Q_{33}$ & $| Y \rangle$ & $Y$ \\
 $Q_{36}$ & $| Y \rangle$ & $Y$ \\
 $Q_{54}$ & $| + \rangle$ & $X$ \\
 \hline
 \end{tabular}
 \caption{Basis used to prepare each logical qubit, as well as the basis used for final readout of each logical qubit. Final readout is performed after the $d$ idling rounds performed following the final multi-qubit Pauli measurement. }
 \label{tab:exp4_logical_qubits}
\end{table}

\begin{table}[htbp]
    \centering
    \begin{tabular}{|c|c|}
        \hline
        Type & Parities, each ideally $0$ \\
        \hline
        Two PPM outcomes & $m_1,m_3$ \\
        \hline
        \shortstack{One combined\\PPM/readout parity}
        & $m_2 \oplus r_{15}^{Z}$ \\
        \hline
        Nine final readouts &
        \begin{tabular}{@{}c@{}}
            $r_2^Z, r_6^X, r_{10}^Z,$ \\
            $r_{13}^X, r_{19}^Y, r_{22}^Y,$ \\
            $r_{33}^Y, r_{36}^Y, r_{54}^X$
        \end{tabular} \\
        \hline
    \end{tabular}
    \caption{The twelve independent deterministic logical parity observables for \texttt{ExpD}. Here, $m_i$ denotes the outcome bit of the $i$th multi-qubit Pauli measurement, and $r_j^B$ denotes the final readout bit of logical qubit $Q_j$ in basis $B$, with bit $0$ corresponding to eigenvalue $+1$. Every listed parity is zero in the absence of faults.}
    \label{tab:exp4_deterministic_observables}
\end{table}

\texttt{ExpD} consists of measuring the following three multi-qubit Pauli operators
\begin{align}
    P_1 &= Y_{Q_{22}} \otimes Z_{Q_{10}} \otimes Z_{Q_{2}} \otimes X_{Q_{6}} \otimes X_{Q_{13}} \otimes X_{Q_{15}}, \label{eq:exp4PsM1}  \\
    P_2 &= Z_{Q_{2}} \otimes Z_{Q_{15}}, \otimes X_{Q_{54}} \label{eq:exp4PsM2} \\
    P_3 &= Y_{Q_{22}} \otimes Y_{Q_{36}} \otimes Y_{Q_{19}} \otimes Y_{Q_{33}}. 
    \label{eq:exp4PsM3}
\end{align}
A total of ten logical qubits participate in the experiment. Each multi-qubit Pauli measurement is measured over $d$ rounds, with $d$ idle rounds in between each measurement, before the first measurement and after the final measurement. The basis used for initializing the logical qubits and final readout is shown in \cref{tab:exp4_logical_qubits}. We note that without the $d$ idling rounds between two consecutive multi-qubit Pauli measurements, we would not be able to assign spatial commit regions to all the logical qubits involved in the multi-qubit Pauli measurement. To see this, suppose a logical qubit $Q_l$ is assigned a seed commit region for the measurement $P_j$ (performed over $d$ rounds) and then $Q_l$ also participates in the measurement $P_{j+1}$. For the $d$ rounds of $P_{j+1}$, $Q_l$ would require a spatial cleanup region assignment instead of a seed commit assignment. Doing so would completely change the wavefront algorithm we've used which assumes all logical qubits are assigned a seed commit region during all rounds of a multi-qubit Pauli measurement. This does not mean however that we cannot use an alternative wavefront algorithm where a portion of the logical qubits are assigned a seed cleanup region rather than a seed commit region. Such investigations are left to future work. 

\subsubsection{Noise model}
\label{subsec:noise_model}

All results in this section use a uniform circuit-level depolarizing noise model with a single parameter $p$ (the \texttt{single\_p} model of Appendix~\ref{app:expd-spec}), applied to the syndrome-extraction circuit that Patchwork compiles for \texttt{ExpD}. Every operation of the circuit is followed, or in the case of measurements preceded, by an independent stochastic Pauli fault:
\begin{itemize}[leftmargin=*]
\item \emph{Two-qubit gates.} Each \textsc{cnot} is followed by a two-qubit Pauli channel in which each of the fifteen non-identity Paulis $\{I,X,Y,Z\}^{\otimes 2}\setminus\{II\}$ occurs with probability $p/15$ (total probability $p$).
\item \emph{Single-qubit gates and idling.} Each Hadamard, and each qubit that is idle during a layer in which other qubits are acted on, is followed by a single-qubit Pauli channel with $X$, $Y$, and $Z$ each occurring with probability $p/3$ (total probability $p$).
\item \emph{Reset.} A qubit prepared in $|0\rangle$ suffers an $X$ error, and a qubit prepared in $|+\rangle$ a $Z$ error, each with probability $2p/3$.
\item \emph{Measurement.} A $Z$-basis measurement is preceded by an $X$ error, and an $X$-basis measurement by a $Z$ error, each with probability $2p/3$, so that the outcome is flipped with that probability.
\item \emph{Data qubits during ancilla reset and measurement.} While the ancillas are reset or measured, each data qubit undergoes a single-qubit Pauli channel with $X$, $Y$, and $Z$ each occurring with probability $2p/3-4p^{2}/9$, the composition of two idle steps at $p/3$.
\end{itemize}
The same rates apply to every patch, routing corridor, and syndrome round; there is no leakage, crosstalk, or spatial variation. The DEM is obtained from this noisy circuit by propagating every fault to the detectors and logical observables it flips~\cite{Gidney2021stimfaststabilizer}, and all decoding experiments sample detector outcomes and observable values from that DEM, so the monolithic and staged decoders of \cref{fig:LERresult} see the same error model at every $(d,p)$. We sweep $p\in\{0.001,0.002,0.003,0.004,0.005\}$ and $d\in\{5,7,\ldots,21\}$.

Let $m_i$ denote the outcome bit of $P_i$, and $r_j^{B}$ the final readout bit of $Q_j$ in basis $B$. The deterministic observables are given in \cref{tab:exp4_deterministic_observables}.

An illustration for the full experiment is shown in \cref{fig:Exp4_Measurements}. In \cref{fig:Exp4_M1EX}, the merged surface patches for each of the three measurements in \cref{eq:exp4PsM1,eq:exp4PsM2,eq:exp4PsM3} are shown along with the spatial commit and cleanup regions. Each logial qubit not participating in the multi-qubit Pauli being measured, but which participates in a later multi-qubit Pauli is shown. In \cref{fig:Exp4_M2EX}, the full three-dimensional illustration of \texttt{ExpD} including temporal commit and cleanup regions are shown, with all 35 syndrome rounds (the figure uses $d=5$). The initialization and readout bases following \cref{tab:exp4_logical_qubits} are also shown. 

A plot showing the TFRs for various code distances is shown in \cref{fig:LERresult}. In particular, the TFR is computed as
\begin{align}
p_{\mathrm{L}}^{\mathrm{total}}
&=
\Pr\!\left[
\exists\, a \in \{1,\ldots,12\} :
\widehat{o}_a \neq o_a
\right],
\label{eq:Tot_LER_ExpD}
\end{align}
where $o_a$ is the parity outcome of one of the deterministic logical observables in \cref{tab:exp4_deterministic_observables}.

The computed TFRs in \cref{eq:Tot_LER_ExpD} are obtained from decoding the commit regions followed by decoding the cleanup regions. TFRs obtained from full monolithic decoding nearly match the TFRs of parallel block-wise decoding. Exact comparisons are given in \cref{tab:expd_mono_vs_staged}. We also found by sampling faults of a given weight that only failures of size greater than $(d-1)/2$ resulted in errors that were correct by the monolithic decoder and missed by parallel block-wise decoding. These discrepancies occurring only from high order failures is expected in a fault-tolerant implementation of spatio-temporal parallel block-wise decoding. The complete specification of \texttt{ExpD} as a Patchwork program, and the decoding workflow used to obtain \cref{fig:LERresult}, are given in Appendix~\ref{app:expd-spec}.

\begin{table*}[htbp]
 \centering
 \footnotesize
 \setlength{\tabcolsep}{4pt}
 \begin{tabular}{|c|c|c|c|c|c|c|}
 \hline
 $d$ & $p$ & Shots & Mono & Staged & Ratio & Staged-only/mono-only \\
 \hline
 5  & 0.001 & $10^5$          & 8{,}319  & 8{,}442  & 1.015 & 221 / 98  \\
 5  & 0.002 & $10^5$          & 49{,}427 & 49{,}753 & 1.007 & 789 / 463 \\
 7  & 0.001 & $2\times10^5$   & 2{,}396  & 2{,}422  & 1.011 & 55 / 29   \\
 7  & 0.002 & $2\times10^5$   & 38{,}852 & 39{,}165 & 1.008 & 712 / 399 \\
 9  & 0.001 & $2\times10^6$   & 3{,}090  & 3{,}129  & 1.013 & 56 / 17   \\
 9  & 0.002 & $2\times10^5$   & 12{,}501 & 12{,}611 & 1.009 & 232 / 122 \\
 11 & 0.001 & $1.5\times10^7$ & 2{,}973  & 3{,}007  & 1.011 & 51 / 17   \\
 11 & 0.002 & $6\times10^5$   & 10{,}973 & 11{,}051 & 1.007 & 166 / 88  \\
 13 & 0.001 & $1.3\times10^8$ & 2{,}994  & 3{,}018  & 1.008 & 43 / 19   \\
 13 & 0.002 & $6\times10^5$   & 3{,}129  & 3{,}138  & 1.003 & 35 / 26   \\
 \hline
 \end{tabular}
\caption{Comparison of monolithic and staged decoding of \texttt{ExpD}
using identical shots. Both decoders use correlated minimum-weight perfect
matching on the same compiled program and sampled detector outcomes.
Monolithic decoding uses the full DEM, while staged decoding processes
commit regions followed by cleanup regions, as described in
\cref{sec:spacetime_partition,sec:spacetime_corrections}.
The Mono and Staged columns count shots with at least one incorrectly
decoded logical parity from \cref{tab:exp4_deterministic_observables}.
The ratio is the staged failure count divided by the monolithic failure
count. The last column reports failures unique to the staged decoder
and failures unique to the monolithic decoder, separated by a slash.}
 \label{tab:expd_mono_vs_staged}
\end{table*}

\subsection{DEM scaling with multi-qubit Pauli measurement sizes}
\label{subsec:DEM_PPM}

\begin{figure*}
     \centering
\subfloat[\label{fig:spatial_areaEX} ]{\includegraphics[width=.8\textwidth]{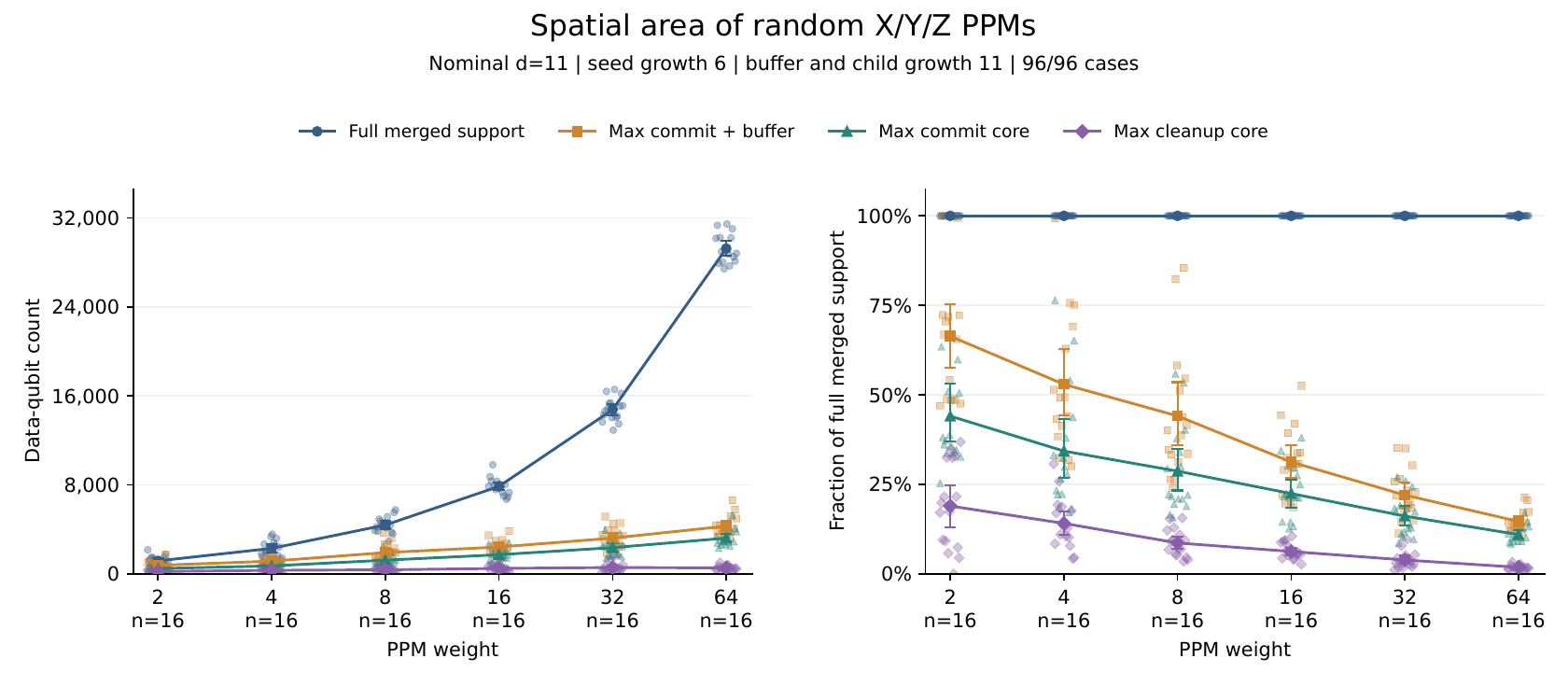}}
     \caption{ Spatial areas for random multi-qubit Pauli measurements (PPMs) of weights 2--64 at nominal distance $d=11$. Left: area of the full merged patch, the largest commit core, the largest commit region including its buffer, and the largest spatial cleanup core, measured by their numbers of data qubits. Right: the corresponding areas normalized by the full merged-patch area for each instance. Faint markers show the 16 instances at each weight; lines connect their means, and error bars indicate 95\% bootstrap confidence intervals. }
     \label{fig:spatial_area}
\end{figure*}

To study how the spatial size of individual decoding regions grows with the weight of a lattice-surgery measurement, we generate 16 random PPMs at each weight $w\in\{2,4,8,16,32,64\}$, with each Pauli factor chosen uniformly from $X$, $Y$, and $Z$. We use nominal distance $d=11$, seed-commit extensions of $(d+1)/2$, and length-$d$ buffers. \Cref{fig:spatial_area} compares the full merged-patch area with the largest commit core, commit region including its buffer, and spatial cleanup core for each instance. The area is measured by
the number of occupied data-qubit sites. We also normalize each local area by the full merged-patch area for the same instance.

The local regions grow substantially more slowly than the full merged patch over the tested range. From weight 2 to 64, the mean full-patch area increases by approximately $24.8$ times, whereas the mean largest commit region including its buffer increases by only $5.5$ times. The mean fraction occupied by the largest commit core decreases from $40.8\%$ to $10.9\%$; including its buffer, this fraction decreases from $66.4\%$ to $14.7\%$. The corresponding fraction for the largest spatial cleanup core decreases from $24.3\%$ to $2.2\%$. Thus, although local regions are not strictly constant in size, spatial partitioning increasingly reduces the largest region relative to the monolithic merged patch as the PPM weight grows.

\section{Conclusion}
\label{sec:conclusion}

In this work, we introduced explicit rules for performing lattice surgery in narrow-corridor layouts and described their implementation in the upcoming \texttt{patchwork} software framework. We developed a gate-scheduling framework based on cut lines to control fault propagation as logical observables change orientation within merged surface code patches. We also introduced dedicated gate schedules and protruding lobes for fault-tolerant $Y$ measurements.

For the $Y$-measurement constructions tested at $d \le 11$, exhaustive numerical searches supported preservation of the full effective code distance. This conclusion applies to the validated constructions; adaptations that shorten protruding lobes can reduce the effective distance. The total failure probabilities (TFRs) obtained for \texttt{ExpD}, described in \cref{subsec:Sequence_PPM} and containing $Y$ measurements, provide additional numerical evidence of reliable performance at larger code distances.

The second part of this work developed a wavefront algorithm that constructs spatial commit, buffer, and cleanup regions and combines them with temporal regions to support parallel decoding of lattice surgery. We described how to construct local DEMs within this framework and evaluated a sequence of three multi-qubit Pauli measurements at multiple code distances. Staged decoding using space-time commit and
cleanup regions achieved TFRs close to those of monolithic decoding over the parameters tested.

Future work will integrate \texttt{patchwork} with the CUDA-Q Logical framework \cite{mccaskey2026cuda} and investigate real-time decoding for experiments involving approximately 100 logical qubits. This will connect the automated lattice-surgery constructions and decoding partitions developed here with an implementation whose
decoding latency can be evaluated directly.

\newpage
\appendix

\addtocontents{toc}{\let\protect\tocsection\protect\apptocsection}

\section{Wavefront construction of commit, buffer, and cleanup regions}
\label{app:graph}

In this appendix, we give a mathematical description of the spatial commit and cleanup construction summarized in the main text. The construction is applied independently to each connected merged-patch component associated with one multi-qubit Pauli measurement, including when several such measurements are performed in parallel. All regions and paths below are therefore restricted to one measured merged patch. The construction follows the measured boundary connections and closes their rectilinear enclosure before assigning ownership. It avoids enumerating every path inside the enclosure. The observed approximately cubic distance scaling is stated as numerical evidence, not as a general complexity theorem.

\subsection{Measured stabilizer complex}

Let $M$ denote a connected merged surface-code patch with nominal code distance $d\ge 3$. We write $V_M$ for its active data qubits and $\mathcal S_M$ for its measured stabilizers. Every stabilizer $s\in\mathcal S_M$ is represented by its Pauli-labeled active support $\supp(s)\subseteq V_M$. The stabilizer-support hypergraph is
\begin{align}
\mathcal H_M&=\bigl(V_M,\{\supp(s):s\in\mathcal S_M\}\bigr). \label{eq:A1}
\end{align}
The hypergraph determines ownership and detects stabilizers shared by proposed regions.

Let $E_{\rm patch}$ contain the unit horizontal and vertical pairs lying inside one participating logical patch. For each measured stabilizer support, every axis-aligned pair is joined. All three pairs of a weight-three corner or domain-wall support are joined, even when one pair is diagonal in the drawing grid. These edges define the structural graph $G_M^{\rm str}=(V_M,E_M^{\rm str})$. Wavefront propagation and buffer-distance calculations additionally use the slanted physical hull edge of a weight-four trapezoid, giving $G_M^{\rm walk}=(V_M,E_M^{\rm walk})$:
\begin{align}
E_M^{\rm str}&=E_{\rm patch}\cup\bigcup_{s\in\mathcal S_M}
\left(E_{\rm axial}(s)\cup E_{\rm tri}(s)\right),\notag\\
E_M^{\rm walk}&=E_M^{\rm str}\cup\bigcup_{s\in\mathcal S_M}E_{\rm trap}(s). \label{eq:A2}
\end{align}
No inactive drawing-grid coordinate is introduced as a vertex. In particular, an inactive middle coordinate of an elongated check does not contribute a step. An active walk path is a sequence of distinct active data qubits
\begin{align}
\gamma&=(v_0,v_1,\ldots,v_\ell),\qquad (v_{r-1},v_r)\in E_M^{\rm walk}, \label{eq:A3}\\
|\gamma|_M&=\ell. \label{eq:A4}
\end{align}
For $A,B\subseteq V_M$, the active graph distance is
\begin{align}
d_M(A,B)&=\min\{|\gamma|_M:v_0\in A,\ v_\ell\in B,\ \gamma\subseteq G_M^{\rm walk}\}. \label{eq:A5}
\end{align}

\subsection{Contours, exposed faces, and wavefront enclosures}
\label{app:wavefront-enclosure}

Let $k$ denote the requested propagation depth in this subsection: $k=\ell_{\rm seed}$ for $X/Z$ seed promotion, with $\ell_{\rm seed}$ defined in \cref{eq:A-seed-depth}, and $k=d$ for buffers. Here, a full-distance path means a path of the requested length $k$. The object needed for assigning a commit or buffer is its closed measured enclosure, not a list of all paths that can be drawn inside it. We therefore distinguish the construction of this enclosure from the measured paths retained to establish a terminal direction or a protected extension depth. A filled interior data qubit does not, merely by being filled, become a path endpoint.

For a data region $R\subseteq V_M$, let $\Gamma(R)$ denote its rectilinear contour inside the measured patch. The data-qubit perimeter and strict interior are
\begin{align}
\partial_\Gamma R&=R\cap\operatorname{Vert}(\Gamma(R)), \label{eq:A6}\\
\Int_\Gamma(R)&=R\setminus\partial_\Gamma R. \label{eq:A7}
\end{align}
An exposed face is a connected perimeter segment together with its initial outward direction:
\begin{align}
f&=(F_f,\nu_f),\qquad F_f\subseteq\partial_\Gamma R. \label{eq:A8}
\end{align}
Different exposed faces are kept distinct, including when they face the same compass direction. A diagonal measured connection can contribute to each of its nonzero cardinal directions, but does not join otherwise unrelated source faces.

For reference, a family $\mathcal P$ of measured simple paths has the following sweep and terminal sets:
\begin{align}
\Sweep(\mathcal P)&=\bigcup_{\gamma\in\mathcal P}V(\gamma), \label{eq:A9}\\
\Term(\mathcal P)&=\{v_{|\gamma|_M}:\gamma\in\mathcal P\}, \label{eq:A10}\\
\Sweep^\circ(\mathcal P;A)&=\Sweep(\mathcal P)\setminus A. \label{eq:A11}
\end{align}
These definitions do not require that all simple paths be enumerated. Below, $\mathcal W_{f,h}^{(k)}$ denotes the full-distance measured paths certified for source half-edge $h$ on face $f$. They establish the directions and depths used by the enclosure construction. This family is not asserted to contain every admissible simple path.

\subsubsection{Outward layers and the part of the graph determining the enclosure}

Fix a commit $C$, one exposed face $f$, and the absorbing foreign-commit data $Z$. First split $G_M^{\rm walk}[V_M\setminus(C\cup Z)]$ into open routing components. For each such component, attach the connected components of $G_M^{\rm walk}[Z]$ adjacent to it, giving one propagation piece $U$. First hops entering an absorbing component directly are grouped into their own absorbing-only piece. This prevents distinct open corridors from being joined through a foreign commit during the enclosure calculation. Let $H_f(U)$ contain the measured half-edges $h=(u,v)$ belonging to this piece, with $u\in F_f$ and initial outward component $\nu_f$. Put $A_f=\{v:(u,v)\in H_f(U)\}$; this first active step has depth one. On $U$, use the directed relation
$$
\begin{aligned}
x\longrightarrow y
\quad\Longleftrightarrow\quad&
\{x,y\}\in E_M^{\rm walk},\\
&x,y\in U,\\
&\bigl(x\in Z\Longrightarrow y\in Z\bigr).
\end{aligned}
$$
Thus a path may continue inside an absorbing region but cannot leave it. Such data do not become owned by the source commit. Propagation after the initial edge can turn through the measured corridor; it is not constrained to move monotonically along $\nu_f$.

Writing $N^+(X)$ for the outgoing neighbors under this relation, define the minimum-distance layers and the distance cap by
\begin{align}
L_1&=A_f,\qquad
L_{r+1}=N^+(L_r)\setminus\bigcup_{j=1}^{r}L_j,\notag\\
Q_k&=\bigcup_{r=1}^{k}L_r,\qquad T_k=L_k. \label{eq:A12a}
\end{align}
Only active measured edges count towards $r$. In particular, crossing an authenticated elongated edge costs one step even when its drawing spans an inactive coordinate.

Suppose first that $T_k\ne\varnothing$. In the undirected measured graph induced by $Q_k$, construct the block-cut forest: its block vertices are the maximal biconnected subgraphs, including bridges, and its articulation vertices are the shared attachment points. For each anchor in $A_f\cup T_k$, mark its articulation node if it is an articulation vertex, and otherwise its unique block node. Repeatedly remove unmarked nodes of degree zero or one. Let $K_f$ be the union of the retained blocks and anchor vertices. This operation keeps the alternative boundary connections between the source and the front, while removing a spur or cyclic side pocket attached through a single point that cannot contribute to that connection. It does not enumerate the possible paths inside a retained block.

Draw the retained vertices and authenticated rectilinear edges on the doubled coordinate grid. A long axial edge is drawn across an inactive coordinate only when the corresponding measured elongated stabilizer supplies that edge. A diagonal graph edge can be used for propagation but cannot on its own create a horizontal or vertical barrier. Every horizontal/vertical raster intersection must be an endpoint of both measured segments. If this test fails, the retained graph core can remain available, but no interior fill or terminal metadata is published from that barrier. For an accepted barrier, let $\mathscr R_f$ be its rasterization and let $\operatorname{Ext}(\mathscr R_f)$ be the unbounded four-neighbor component of its complement. The initial enclosed data are
\begin{align}
E_f^{\,0}
&=K_f\cup
\{q\in Q_k:2q\notin\operatorname{Ext}(\mathscr R_f)\}.
\label{eq:A12b}
\end{align}
The exterior flood is evaluated in a finite rectangle with one empty layer around the complete barrier. Only active data in $Q_k$ can be admitted by this interior-filling step. Inactive drawing coordinates are never data qubits.

\subsubsection{Terminal directions and local rectilinear completion}

Geometric face-front records and certified full-distance path records are distinct. An ordinary depth-$k$ front has measured full-distance witnesses. A clipped front, defined below, may instead inherit the initial outward normal as a geometric face label; that label is not proof of a length-$k$ terminal. Detour and returning-path terminal additions, and every assertion of the requested extension depth, require separately certified full-distance paths. Source labels are retained even when geometric wavefront states are combined. The following restricted path families supply these labels without enumerating all paths in the distance cap.

For one first half-edge $h=(u,v)$, let $\delta_h(q)$ be the shortest depth from $u$ through $h$ in the retained directed core, with $\delta_h(v)=1$. Shortest-path witnesses follow edges that increase $\delta_h$ by one. Additional witnesses can replace one such edge $a\to e$ by the other three edges $a\to b\to c\to e$ of a measured unit square, provided $\delta_h(e)=\delta_h(a)+1$, $\delta_h(b),\delta_h(c)\ge\delta_h(a)$, and the subsequent depth-increasing suffix avoids $b,c$. The replacement adds two active steps, not two arbitrary units of geometric distance.

A simple contour witness consists of a shortest prefix to the unique minimum-depth entry of an isolated degree-two contour, one proper arc of that contour, and a depth-increasing suffix outside the contour. The combined length must be $k$, and the combined path must have distinct vertices. Traversing the full contour back to its entry is excluded. These constructions retain actual hop positions; a detour is not relabeled by its shortest distance.

An elongated measured connection can leave one missing lateral line between terminal records of the same source. When the two records share that elongated support, are separated by two lateral coordinates, and by one or two forward coordinates, the missing line is checked from its outwardmost candidate inward. A candidate can be certified by a returning path with one fixed coordinate of nondecreasing signed progress. After its first transition, lateral motion cannot reverse until progress advances. The state records depth, current vertex, and lateral sign. The recovered full-distance path is explicitly checked for distinct vertices, legal measured edges, absorption, and its claimed final direction. Only the first certified candidate on that line is retained. This supplies terminal evidence; it does not enlarge $E_f^{\,0}$ by itself.

The resulting certified family $\mathcal W_{f,h}^{(k)}$ records, for each origin, the endpoint, the signed cardinal components of the final edge, and the vertices occurring at each actual hop. Combine the supported full-distance terminal records with any admitted clipped geometric front records, keeping their different meanings. For a terminal direction $\nu$, choose its perpendicular direction $\tau$ and retain the outermost endpoint at each lateral coordinate $\tau\cdot q$. Sort these endpoints by lateral coordinate. A terminal fan is a maximal chain for which consecutive lateral increments are one or two, the nonzero forward increments have magnitude at most two, and their signs agree along the chain. A coordinate jump of two requires a common measured elongated support and an inactive skipped coordinate line in the local rectangle. Complete only the active data in the rectangle of that fan, intersected with its outside component.

Local corner completion is also permitted when three corners of a unit rectilinear square are certified by compatible source records and the fourth is active measured data in the same component. The general three-terminal corner test uses a connected compatibility group with horizontal and vertical directions represented. Two records are compatible when they come from the same physical face or their original source-start sets overlap; pairwise overlap of all records is not required. Records are not combined merely because they reach the same routing component.

Neither completion can extend one coordinate beyond a complete contiguous terminal rail of at least $k$ active sites through the candidate row or column. The special corner joining perpendicular fronts of the same half-edge can pass this test only at the intersection of their extremal coordinates, with the other three corners already present in that source's original support or provisional fan completion. All tests use the original certified records and the single provisional fan pass. Newly completed corners are not used to start another round of filling.

Write $J_f$ for the union of the admitted fan and corner additions. The face enclosure is
\begin{align}
\mathcal E_k(C;f,U)&=E_f^{\,0}\cup J_f. \label{eq:A12c}
\end{align}
The terminal additions lie in $U$ but need not lie in $Q_k$. This is why the construction is neither an unrestricted graph-distance flood nor a bounding rectangle of all reached data. We use $\operatorname{cl}^{\square}_M$ for the corresponding local measured rectilinear completion, with source, component, and terminal-direction records understood:
\begin{align}
X&\subseteq\operatorname{cl}^{\square}_M(X)\subseteq V_M. \label{eq:A12}
\end{align}
In the directed child construction, this notation closes the certified directed sweep and its terminal rails within their outgoing component. It does not authorize joining disconnected corridors or filling an arbitrary global bounding box.

\subsubsection{Finite corridors and enclosed pockets}

A finite corridor need not contain a minimum-distance layer $L_k$ even when it admits a simple path of length $k$. If at least two first-hop anchors form one connected rail and the cap has at least $k$ vertices, its deepest nonempty layer can supply a clipped geometric front. Disconnected first-hop rails cannot pool their reaches to create this front. A clipped front alone is not a full-distance path certificate.

If the finite component is a rectilinear cactus, its block-cut tree consists of bridges and simple rings. Within a bridge block the admissible transfer is its edge. Within a ring block, the admissible transfer between distinct entry and exit vertices is either of the two simple arcs. For an entry and exit at the same vertex, only the zero-length transfer is permitted. Propagate achievable lengths, truncated at $k$, through the block-cut tree, combining lengths by addition and alternatives by union. A path can end at any vertex within its final block. Retaining predecessors and propagating success backward gives its full-distance endpoints and hop-indexed support without a full-ring return or repeated articulation vertex.

A finite connected component not covered by this rule can instead be certified by a bounded local subset calculation with state $(r,v,H)$, where $H$ is the visited subset of that component. The transition to $w$ is admitted only along an admitted local rectilinear measured edge and when $w\notin H$, after which $H$ becomes $H\cup\{w\}$. An endpoint or prefix is used only if a complete length-$k$ certificate survives. The dense calculation applies to components of at most 12 vertices. For a larger component lacking a geometric front, a sparse calculation stores only the visited vertices of each state, at most $k$, and can recertify the same source after the enclosure is restricted. Both calculations share a limit of 250,000 certificate transitions across the enclosure request. This bound is retained when several sources are checked; it is not reset for each source. These bounded subset calculations do not imply a polynomial worst-case bound.

If the complete reachable outside component has fewer than $k$ vertices, it cannot supply the $k$ distinct vertices required after the source half-edge. This proves that the corresponding full-distance path family is empty. Otherwise, exhausting a certificate limit does not prove emptiness. A missing front that requires such a certificate is reported as unsupported; an optional certificate can be withheld when an independent geometric front already exists. A nonempty seed enclosure must still have a certified full-distance origin before it is promoted. There is no automatic return to an unrestricted path enumeration.

For a nonempty region $R$, the inclusive coordinate spans remain
$$
\begin{aligned}
w_x(R)&=\max_{q\in R}x(q)-\min_{q\in R}x(q)+1,\\
w_y(R)&=\max_{q\in R}y(q)-\min_{q\in R}y(q)+1.
\end{aligned}
$$
and $\max\Span(R)=\max\{w_x(R),w_y(R)\}$. These spans count the drawing-coordinate interval; active depth counts measured edges.

\subsection{Seed regions}
\label{app:seed-regions}

Let $S_\lambda^{\rm patch}$ contain the active data qubits on or inside the perimeter of participating logical patch $\lambda$. When $\lambda$ undergoes a $Y$ measurement, this footprint already contains the full $Y$-extension patch described in \cref{subsec:gate_schedule_y_ext,sec:protruding_lobes,sec:neighbor_Y}, excluding its protruding lobes. Thus the fixed $Y$-extension is present before neighboring seeds are paired. The initial patch-seed family is
\begin{align}
\mathcal C_{\rm patch}&=\{S_\lambda^{\rm patch}:\lambda\text{ participates in the measurement}\}. \label{eq:A13}
\end{align}
Overlapping $Y$-extension footprints are assigned disjoint provisional ownership before the remaining neighboring seeds are considered. Within each multi-qubit Pauli measurement, we first divide the overlapping $Y$-seed footprints into connected components. For each component, the logical-patch centers are ordered along the direction in which their centers have the greatest spread, choosing the horizontal direction when the two spreads are equal and using the orthogonal coordinate to break ties. Consecutive seeds in this order are grouped into provisional pairs, leaving one provisional singleton when the component contains an odd number of logical patches. Midpoint planes between neighboring pair blocks divide the shared $Y$-extension footprint, and each pair or remaining singleton retains its logical-patch data together with the portion assigned to it. Because this preliminary division can change the span of the data assigned to a provisional seed, its span bound is recalculated so that the division does not change how much further that seed is permitted to grow. These pair and singleton blocks provide unique geometric ownership during the routing construction; they do not necessarily define distinct final commit decoders. Before buffers are constructed, an admissible interface through a horizontal or vertical $Y$ chain can instead be reserved for cleanup using \cref{app:y-seed-separators}. Interfaces crossed by a required elongated stabilizer are not cut. At every remaining interface with no residual cleanup support, the final stabilizer-ownership coalescence in \cref{eq:A38,eq:A39} combines the corresponding provisional blocks transitively.

\subsubsection{Pairing neighboring seed commit regions}
\label{app:neighboring-seed-pairs}

We now define the neighboring-seed relation used only during the initial pairing step. Let $g(\lambda)$ identify the multi-qubit Pauli measurement containing logical patch $\lambda$, and let $\tau(\lambda)=Y$ when $\lambda$ undergoes a $Y$ measurement and $\tau(\lambda)=\overline Y$ otherwise. The initial span bound of a singleton seed is
\begin{align}
L_\lambda^{(0)}&=
\begin{cases}
3d+3,&\tau(\lambda)=\overline Y,\\
4d+4,&\tau(\lambda)=Y.
\end{cases} \label{eq:A13-pair-span}
\end{align}
When overlapping $Y$-extension footprints have already been divided into a provisional block $R$, let $\Lambda_R$ be the set of its constituent logical patches. The bound is recalculated so that the preliminary division does not change the additional span that the seed can subsequently acquire:
\begin{align}
L_R&=\max\Span(R)\notag\\
&\quad+\max_{\lambda\in\Lambda_R}\left[
L_\lambda^{(0)}-\max\Span\!\left(S_\lambda^{\rm patch}\right)
\right]_+ . \label{eq:A13-pair-headroom}
\end{align}
Here $[x]_+=\max\{x,0\}$. The quantity in square brackets is the difference between the original bound of one constituent seed and its original span. Adding the largest such difference to the new span of $R$ means that assigning shared $Y$-extension data changes the present size of the provisional seed but not the additional span it can subsequently acquire.

Let $\mathcal C_{\rm pre}$ be the provisional seed family after the overlapping $Y$-extension data have been assigned. Consider two members $A$ and $B$ that are still singletons at this stage, and denote their logical patches by $\lambda$ and $\mu$, respectively. Let $\gamma_{AB}^{\rm str}$ be the canonical shortest path joining them through active measured data belonging to their multi-qubit Pauli measurement, using the structural graph $G_M^{\rm str}$. We call this connection ``strip-like'' when
\begin{align}
|\gamma_{AB}^{\rm str}|_M&\le d+3,\notag\\
\min\!\left\{
w_x\!\left(V(\gamma_{AB}^{\rm str})\right),
w_y\!\left(V(\gamma_{AB}^{\rm str})\right)
\right\}&\le d+3. \label{eq:A13-pair-connection}
\end{align}
The first inequality limits the number of active steps between the seeds, while the second excludes a connection that spreads too far in both coordinate directions.

Let $P_{AB}$ be the complete candidate pair footprint. For two $Y$-type seeds, it contains $A$, $B$, and the active data on $\gamma_{AB}^{\rm str}$. For two non-$Y$-type seeds, it additionally contains the complete measured strip aligned with the facing seed boundaries and spanning the interval between them. If this completed strip spans the bounding box of one connected measured component, that complete component is included. The pair is admissible precisely when
\begin{align}
g(\lambda)&=g(\mu),\qquad
\tau(\lambda)=\tau(\mu),\notag\\
P_{AB}\cap\!\bigcup_{D\in
\mathcal C_{\rm pre}\setminus\{A,B\}}D&=\varnothing,\notag\\
\max\Span(P_{AB})&\le\max\{L_A,L_B\}.
\label{eq:A13-pair-admissible}
\end{align}
Thus the completed pair footprint cannot contain data belonging to a third seed, and its largest horizontal or vertical span cannot exceed the larger bound assigned to the two constituents. In this appendix, ``neighboring seed commit regions'' refers to singleton seeds satisfying \cref{eq:A13-pair-connection,eq:A13-pair-admissible}. This relation is distinct from the one-coordinate alignment condition used for the local footprint adjustment below.

The pairing step repeatedly considers every admissible pair. Eligible horizontal pairs are considered before eligible vertical pairs. Within an orientation, candidates are ordered first by the position of the corridor perpendicular to the pairing direction and then along the corridor, from left to right for a horizontal pair and from bottom to top for a vertical pair. Remaining ties are resolved by the original seed order. The first pair is accepted and its two seeds are removed from further pairing, after which the admissible candidates are recomputed. The procedure terminates when no admissible pair remains. Thus, a chain of $n$ seeds for which every required neighboring pairing is admissible is divided into $\lfloor n/2\rfloor$ pairs and one unpaired seed when $n$ is odd. An inadmissible connection can leave additional seeds unpaired.

Let $\widehat S_\lambda^{\rm patch}$ denote the fixed footprint of logical patch $\lambda$ after the midpoint assignment of any overlapping $Y$-extension data. Let $\Pi$ be the resulting deterministic set of disjoint pairs, let $\operatorname{Vert}(\Pi)$ be the logical patches occurring in those pairs, and let $P_{\lambda\mu}$ be the complete footprint assigned to one pair, including its admitted measured connecting strip in the generic pairing step. The paired family is
\begin{align}
\overline{\mathcal C}_{\rm seed}
&=\left\{\widehat S_\lambda^{\rm patch}:\lambda\notin\operatorname{Vert}(\Pi)\right\}
\cup\left\{P_{\lambda\mu}:(\lambda,\mu)\in\Pi\right\}. \label{eq:A13a}
\end{align}
An example of two seeds combined into one region is shown in \cref{fig:neighbor_X_pairEX}. These pairs remain provisional. The original constituent footprints and their bounds are retained so that an oversized extended pair, including a larger aggregate that subsequently contains it, can be reconsidered by \cref{app:seed-pair-replay}. 

\subsubsection{Local enlargement of seed commit regions}
\label{app:seed-footprint-normalization}

Index the members of $\overline{\mathcal C}_{\rm seed}$ as $\overline S_1,\ldots,\overline S_{N_{\rm seed}}$. A protruding lobe of a $Y$-extension patch is normally excluded from the seed commit region. If the selected routing strip uses a particular lobe, exactly the first complete, straight, measured row or column immediately outside the fixed $Y$-extension footprint can be assigned to the seed containing that logical patch. An example is shown in \cref{fig:q33y_q4x_d5EX}. The source face and this first outward face must each contain exactly $d$ contiguous active data qubits, and every corresponding lane must continue into deeper measured routing data. Every later layer of that lobe, and every lobe not used by the selected routing, remains routing data. Let $Y_i^{\rm route}$ denote the union of the admitted first routed-lobe faces for the logical patches contained in $\overline S_i$, with $Y_i^{\rm route}=\varnothing$ for a non-$Y$ seed or when no lobe face is admitted.

A complete row or column of routing data between two aligned seeds is not assigned at this stage. It remains available during the ordered cleanup-separator searches below. Its later treatment is specified in \cref{app:deferred-seed-strips}. The immutable seed family therefore contains the admitted routed $Y$-lobe faces but no strip assigned by that later adjustment:
\begin{align}
S_i^0&=\overline S_i\cup Y_i^{\rm route},\qquad
\mathcal C_{\rm seed}^0=\{S_1^0,\ldots,S_{N_{\rm seed}}^0\}. \label{eq:A13b}
\end{align}
The generic $X/Z$ seed promotion described below skips every $Y$ seed. During the $X/Z$ conflict-resolution pass, set $S_i^\star=S_i^0$ and take the protected requirement of a $Y$ seed to be $S_i^0$ throughout this pass. The separate $Y$-separator pass can remove only the complete midpoint row or column satisfying \cref{eq:A26a,eq:A26c}.

\subsection{Wavefront promotion of X/Z seeds}
\label{app:seed-promotion}

We now give the seed-allocation rule that precedes every buffer and child construction. For odd nominal distance $d$, the seed extension depth is
\begin{align}
\ell_{\rm seed}&=\frac{d+1}{2}. \label{eq:A-seed-depth}
\end{align}
Buffers, pre-buffers, child propagation and junction placement continue to use nominal $d$.

Let $S_i^0$ be an $X$- or $Z$-type immutable seed. If neighboring logical seeds have already been paired, $S_i^0$ is their combined footprint, including its admitted connecting strip, and growth starts from that combined perimeter. For this path-extension pass, define the routing data available for new seed ownership by
$
R_M=V_M\setminus\bigcup_j S_j^0.
$
Other immutable seed cores are absorbing terminals for paths but are excluded from source-owned routing sweeps. During a reconstruction in \cref{app:seed-pair-replay}, this set is defined using the original constituent seeds used in that reconstruction, with other unavailable commits and previously reserved cleanup data excluded. The physical measured patch $V_M$ is unchanged. A boundary crossing, or exposed measured half-edge, is an edge directed from the seed to adjacent routing data:

\begin{align}
\mathcal H_i&=\{(u,v):u\in S_i^0,\ v\in R_M,\ \{u,v\}\in E_M^{\rm walk}\}. \label{eq:A15a}
\end{align}
Group $\mathcal H_i$ by connected physical source face, outward direction, and outside measured component. Let $\mathcal G_i$ be these face/component groups. Propagate each group by \cref{eq:A12a}, retain its source-to-front block core, and fill and complete its measured enclosure according to \cref{eq:A12b,eq:A12c}. The routing-space extension is
\begin{align}
E_{i,f,U}
&=R_M\cap\mathcal E_{\ell_{\rm seed}}(S_i^0;f,U), \label{eq:A15}\\
E_i&=\bigcup_{(f,U)\in\mathcal G_i}E_{i,f,U}. \label{eq:A16}
\end{align}
A nonempty promoted enclosure must retain at least one source half-edge certified to reach length $\ell_{\rm seed}$. The endpoint, direction, and actual-hop support of its certified paths are retained separately as $\mathcal W_{i,h}^{(\ell_{\rm seed})}$, the union of the certified source-group families containing $h$. One diagonal half-edge can occur in two initial normal groups. This grouping permits the boundary paths of one physical face to close their common enclosure. It does not permit unrelated faces or disconnected routing components to donate endpoints to one another. A short exterior spur need not be retained merely because it lies at graph distance less than $\ell_{\rm seed}$; an active interior point is included when the enclosure rule admits it.

The promoted footprint can expose a horizontal or vertical face that stops a short distance from the measured exterior. A face qualifies when it is contiguous and lies wholly inside the authenticated wavefront extension; there is no minimum transverse width. Let $r_f$ be the outward active distance at which a strict majority of its lanes first reaches one common exterior plane through the same measured prefix. When every remaining lane can replay this prefix, and $0<r_f<d$, the complete intervening routing gap $K_f(S_i^0)$ is admitted. If the drawing contains an inactive coordinate gap, every promoted cross-section must replay the corresponding authenticated measured bridge. Otherwise $K_f(S_i^0)=\varnothing$. The distance threshold remains nominal $d$, not $\ell_{\rm seed}$. Only the qualifying face is extended towards its exterior plane; other faces are unchanged, and newly added cap data do not initiate another cap pass. Thus
\begin{align}
K_i&=\bigcup_fK_f(S_i^0), \label{eq:A17}\\
S_i^\star&=S_i^0\cup E_i\cup K_i. \label{eq:A18}
\end{align}
Here $S_i^\star$ is the complete length-$\ell_{\rm seed}$ proposal, including any admitted boundary cap, before conflicts with other seed proposals are resolved; it is not necessarily the final seed commit. Other immutable seed regions are absorbing: a path can enter such a region but cannot leave it again. Vertices lying in an absorbing foreign seed certify that the path reaches length $\ell_{\rm seed}$, but they are not assigned to the proposal of seed $i$. The source-owned enclosure and its certified paths satisfy
\begin{align}
\gamma&=(v_0,\ldots,v_{\ell_{\rm seed}}),\qquad
\gamma\in\mathcal W_{i,h}^{(\ell_{\rm seed})},\notag\\
&\Longrightarrow
\bigl(\{v_1,\ldots,v_{\ell_{\rm seed}}\}\cap R_M\bigr)
\subseteq E_i\subseteq S_i^\star. \label{eq:A18a}
\end{align}
All proposals $S_i^\star$ are computed from the same seed snapshot, so one proposed extension cannot block the wavefront or source certificates of another. The complete proposal, including its exterior cap, is protected throughout the separator search. No shortening of that seed extension is permitted, as specified in \cref{eq:A21,eq:A22}.

During conflict resolution, index the immutable seeds by $m$ and let $\mathfrak C$ be the current partition of these indices into seed classes. For $\alpha\in\mathfrak C$, let $I_\alpha$ be its immutable members, $S_\alpha^0=\bigcup_{m\in I_\alpha}S_m^0$, and $Q_\alpha=\bigcup_{m\in I_\alpha}Q_m$, where initially $Q_m=S_m^\star$ and every class contains one member. Define the stabilizer-incidence sets
\begin{align}
O_s(\mathbf Q)&=\{\alpha:\supp(s)\cap Q_\alpha\ne\varnothing\},\notag\\
O_s^0&=\{\alpha:\supp(s)\cap S_\alpha^0\ne\varnothing\}. \label{eq:A19a}
\end{align}
Two current classes conflict when their proposals acquire new overlap or new multi-class incidence on a measured support relative to their immutable seeds:
\begin{align}
\alpha\sim_{\mathbf Q}\beta\Longleftrightarrow\;&
\left(Q_\alpha\cap Q_\beta\right)
\setminus\left(S_\alpha^0\cap S_\beta^0\right)\ne\varnothing \notag\\
&\text{or there exists }s\text{ such that }\notag\\
&\hspace{1.5em}\{\alpha,\beta\}\subseteq O_s(\mathbf Q),\qquad
O_s(\mathbf Q)\ne O_s^0,\notag\\
&\hspace{1.5em}\{\alpha,\beta\}\nsubseteq O_s^0. \label{eq:A19b}
\end{align}
Let $L_m^0$ be the configured ownership-span limit of immutable seed $m$. The independently authorized ownership limits of the complete length-$\ell_{\rm seed}$ proposals and the separate buffer allowances are
\begin{align}
L_m^\star
&=\max\{L_m^0,\max\Span(S_m^\star)\},\notag\\
L_\alpha^{\rm own}
&=\max_{m\in I_\alpha}L_m^\star,\notag\\
L_m^{\rm buf}
&=L_m^0+\max\bigl\{
w_x(S_m^\star)-w_x(S_m^0),\notag\\
&\hspace{8em}w_y(S_m^\star)-w_y(S_m^0),0\bigr\}. \label{eq:A19c}
\end{align}
The class maximum applies before a required local relaxation. A seed can contain its full required extension even when that extension exceeds its initial bound; its ownership limit is the larger of the initial bound and the actual span needed by that extension. Buffer allowance is recorded separately and cannot make an oversized union of different commits admissible.
Let $A\subseteq\mathfrak C$ be a connected component of the current class-conflict relation, $U_A(\mathbf Q)=\bigcup_{\alpha\in A}Q_\alpha$, and $L_A(\mathbf Q)=\max_{\alpha\in A}L_\alpha^{\rm own}$. If the coordinate span of the current union fits this bound, the complete component can be admitted for coalescence:
\begin{align}
\max\Span(U_A(\mathbf Q))&\le L_A(\mathbf Q)\notag\\
&\Longrightarrow Q_A=U_A(\mathbf Q),\notag\\
&\hspace{4.4em}L_A^{\rm own}=L_A(\mathbf Q). \label{eq:A20}
\end{align}
Suppose instead that the proposed union is oversized. The separator must preserve the seed extension depth
\begin{align}
r_{\rm seed}&=\ell_{\rm seed}=\frac{d+1}{2}. \label{eq:A21}
\end{align}
The entire current proposal remains protected:
\begin{align}
\widehat P_i(\mathbf Q)&=Q_i. \label{eq:A22}
\end{align}
This includes the immutable seed, its complete wavefront enclosure, and every qualified exterior cap. A proposed separator cannot remove any of these data. The first eligible cross-section can lie immediately beyond the complete proposal, including its cap; this is not a requirement that the cut lie exactly $\ell_{\rm seed}$ steps from the original seed. A farther cross-section may be used if it satisfies the same geometric and ownership conditions. Connectivity and stabilizer ownership are checked on the resulting regions, rather than inferred from path length alone.

We next define the permitted geometric shapes of a cleanup separator. A horizontal or vertical arm is one coordinate wide and contains the complete active cross-section of the implicated local routing corridor in the direction perpendicular to the cut. An active side branch lying on the same coordinate line is not included merely because it is collinear; it belongs to the arm only when it crosses the cut direction. For a search target $T$, which can be the complete conflict component or one directly conflicting pair as defined below, let $\mathfrak A_x(T)$ and $\mathfrak A_y(T)$ denote the complete vertical and horizontal arm candidates. The geometric candidate family is
\begin{align}
\mathfrak X_T^{(1)}
&=\mathfrak A_x(T)\cup\mathfrak A_y(T),\notag\\
\mathfrak X_T^{(2)}
&=\{X_x\cup X_y:X_x\in\mathfrak A_x(T),\ X_y\in\mathfrak A_y(T)\},\notag\\
\mathfrak X_T
&=\mathfrak X_T^{(1)}\cup\mathfrak X_T^{(2)}. \label{eq:A23}
\end{align}
Before completing the ends of a candidate arm, let $X^0$ be its proposed cells and let $F$ be the protected requirements, reserved cleanup data, and foreign-owned data forbidden to that candidate. Completion can only add cells. Therefore
\begin{align}
X^0\cap F\ne\varnothing,\qquad X^0\subseteq\widehat X
&\Longrightarrow \widehat X\cap F\ne\varnothing .
\label{eq:A23a}
\end{align}
Such a candidate is rejected before reconstructing measured-stabilizer connectivity. A candidate that passes this first test is still completed to its full cross-section and checked afterward, since completion can introduce a new conflict. This changes neither the permitted cut shapes nor their preference order.

The complete arm geometry is determined from the original measured routing corridor, including its measured boundary connections. If completing an arm intersects a protected proposal, the complete candidate is rejected. Its endpoint is not removed to obtain a partial-width cut.

Thus a new separator contains either one complete straight arm or exactly one complete arm of each orientation, permitting clean L-, T-, and cross-shaped separators. A perpendicular arm pair must have a common grid intersection; diagonally offset arms do not define a clean T merely because a stabilizer connects them. Diagonal, jagged, staircase-shaped, partial-width, greedily thinned, and multiple-parallel-arm separators are excluded. Connectivity is checked as part of separator validity below. The new arm or arm pair can be connected through the structural cleanup graph, through a previously accepted forced-cleanup component serving as a fixed anchor, or through a measured stabilizer explicitly certified as a separator junction whose complete residual support is forced-cleanup-owned.

For $X\in\mathfrak X_T$, let $\Phi_m(X)$ contain any admitted completion of the retained face immediately adjacent to a straight cut; it is empty when no such completion is made. This completion is restricted to the tangential extent of that cut and must avoid other owners and reserved cleanup data. It can add data under the checks below, but cannot release any part of the protected proposal.

Let $\operatorname{CoreComp}_{S_m^0}(Y)$ denote the union of the active components of $Y$ connected to immutable core $S_m^0$, and let $\eta_\alpha=1$ when class $\alpha$ contains a previously unavoidable relaxed merge and $\eta_\alpha=0$ otherwise. The retained member and class proposals and their ownership limits are
\begin{align}
Q_m^+(X)
&=\operatorname{CoreComp}_{S_m^0}
\!\left(Q_m\setminus X\right)
\cup\Phi_m(X),\notag\\
L_m^+(X)
&=\max\{L_m^0,\max\Span(Q_m^+(X))\},\notag\\
Q_\alpha^+(X)
&=\bigcup_{m\in I_\alpha}Q_m^+(X),\notag\\
\widehat L_\alpha^+(X)
&=\max_{m\in I_\alpha}L_m^+(X),\notag\\
W_\alpha^+(X)&=\max\Span(Q_\alpha^+(X)),\notag\\
L_\alpha^+(X)
&=\begin{cases}
\widehat L_\alpha^+(X),
&\eta_\alpha=0,\\[1ex]
\max\{\widehat L_\alpha^+(X),W_\alpha^+(X)\},
&\eta_\alpha=1.
\end{cases} \label{eq:A24}
\end{align}
A retained class is admissible only when $\widehat P_m(\mathbf Q)\subseteq Q_m^+(X)$ for every member, every retained member is connected to its own immutable core, and $\max\Span(Q_\alpha^+(X))\le L_\alpha^+(X)$. Any loss of protected data rejects the candidate. A straight cut may complete its immediately adjacent retained row or column under the same connectivity, ownership, and span checks; no enclosing rectangular hull is filled. Unaffected faces retain their current data. The full-depth buffer allowance remains separate from the ownership bound and cannot authorize later commit growth.

Let $A\subseteq\mathfrak C$ be an oversized connected component of the current class-conflict graph. Let $\mathcal E(A)$ be its directly conflicting class pairs and let
\begin{align}
\mathcal E_{>}(A)
&=\bigl\{\{\alpha,\beta\}\in\mathcal E(A):\notag\\
&\hspace{2em}\max\Span(Q_\alpha\cup Q_\beta)
>\max\{L_\alpha^{\rm own},L_\beta^{\rm own}\}\bigr\}. \notag
\end{align}
A search target $T$ is either the complete component context $A$ or one pair in $\mathcal E_{>}(A)$. Write $\operatorname{Valid}_{A,T}(X)$ when the new separator cells $X\setminus X_{\rm forced}$ are disjoint from every immutable core, protected requirement, foreign proposal, and previously forced-cleanup cell; every arm is a complete local corridor cross-section; the new arms together with any qualified forced-cleanup anchor form one connected separator network in the structural graph augmented only by certified separator-junction stabilizers; the selected direct conflict is removed without a residual measured route around the network; every retained member and class satisfies \cref{eq:A24}; and every measured stabilizer has a unique residual cleanup owner. Define
\begin{align}
\mathfrak X_{A,T}
&=\{X\in\mathfrak X_T:\operatorname{Valid}_{A,T}(X)\}. \label{eq:A25}
\end{align}
For one target, $\operatorname{Best}_{A,T}$ denotes the deterministic result of the preliminary straight-separator search when that search succeeds, and otherwise the deterministic minimum from the general rectilinear search over \cref{eq:A23}. The general search considers both single-arm and perpendicular two-arm candidates with their associated local geometric completion. The complete component context is tried before any pair context:
\begin{align}
\mathcal Z_A
&=\bigl\{(e,\operatorname{Best}_{A,e}):
e\in\mathcal E_{>}(A),\notag\\
&\hspace{7em}\mathfrak X_{A,e}\ne\varnothing\bigr\},\notag\\
\mathcal Y_A
&=\begin{cases}
\{(A,\operatorname{Best}_{A,A})\},
&\mathfrak X_{A,A}\ne\varnothing,\\[1ex]
\mathcal Z_A,
&\mathfrak X_{A,A}=\varnothing,
\end{cases}\notag\\
(T_A,X_A)
&=\underset{(T,X)\in\mathcal Y_A}{\arg\min}\,\rho_A(T,X).
\label{eq:A25a}
\end{align}
The component-context search and, if needed, every oversized-pair search preserve the same complete proposals in \cref{eq:A22}. The ordering $\rho_A$ compares pair-context results by separator size, canonical pair geometry, and separator coordinates. Within one target, the preliminary straight search precedes the general rectilinear search, with deterministic geometric tie breakers based on seed-interface separation and balance, distance from the canonical midpoint, orientation, coordinate, and finally the complete separator geometry.

After accepting $(T_A,X_A)$, the retained member proposals and their ownership limits are updated, the separator is reserved for cleanup, and all affected classes are reconstructed:
\begin{align}
Q_m&\leftarrow Q_m^+(X_A)
\quad(m\text{ affected}),\notag\\
L_m^{\rm own}&\leftarrow L_m^+(X_A)
\quad(m\text{ affected}),\notag\\
X_{\rm forced}&\leftarrow X_{\rm forced}\cup X_A. \label{eq:A25b}
\end{align}
The complete conflict relation is then recomputed.

If $\mathcal Y_A=\varnothing$, choose one canonical directly conflicting pair
\begin{align}
e_A^\dagger
&=\begin{cases}
\min_{\kappa}\mathcal E_{>}(A),&\mathcal E_{>}(A)\ne\varnothing,\\
\min_{\kappa}\mathcal E(A),&\mathcal E_{>}(A)=\varnothing,
\end{cases}\notag\\
Q_{e_A^\dagger}
&=\bigcup_{\alpha\in e_A^\dagger}Q_\alpha,\notag\\
\overline L_{e_A^\dagger}
&=\max_{\alpha\in e_A^\dagger}L_\alpha^{\rm own},\notag\\
L_{e_A^\dagger}^{\rm own}
&=\begin{cases}
\displaystyle\max\!\left\{
\overline L_{e_A^\dagger},
\max\Span(Q_{e_A^\dagger})\right\},
&\mathcal E_{>}(A)\ne\varnothing,\\[1ex]
\overline L_{e_A^\dagger},
&\mathcal E_{>}(A)=\varnothing.
\end{cases} \label{eq:A25c}
\end{align}
The order $\kappa$ is the canonical immutable-core geometry order. The first case is the local unavoidable-merge exception and raises only the new class to the exact span required by its union. The second case handles a transitively oversized component for which every direct conflicting pair already fits an existing bound; one canonical pair is then combined without any span relaxation. In both cases, the new class is returned to the conflict calculation. No unrelated class or operation-wide span is enlarged.

After each accepted cut or required coalescence, the conflict relation in \cref{eq:A19b} is recomputed. Accepted separator reservations are preserved throughout this seed-conflict search. It continues until the remaining conflicts are resolved by combinations within the current ownership bound, by valid separators, or by the local unavoidable-merge exception in \cref{eq:A25c}.

After proposal coalescence, a measured check can have its committed support in one seed while its remaining support lies in two or more structural cleanup components. If that seed is the unique commit owner touching the check, and none of the residual data is reserved cleanup or unavailable foreign data, the complete residual support is assigned to that seed. The residual components are then recomputed, and this check is repeated until no further absorption is needed. The resulting seeds must still satisfy their ownership-span checks. This local completion gives the check one owner; it does not join unrelated cleanup components or remove a retained separator.

\subsubsection{Reconsidering oversized provisional seed pairs}
\label{app:seed-pair-replay}

The neighboring pairs introduced in \cref{eq:A13a} are provisional. They are treated as single sources while their extensions are checked against other seeds. A pair can nevertheless become oversized after extension, or can be contained in a larger aggregate that becomes oversized. The original constituent seed footprints and bounds are therefore retained.

Let $A$ denote a current aggregate containing an initial neighboring pair, let $\mathcal I_A$ index all of its recorded original constituent seeds, and let $Q_A$ be its current data. It is reconsidered only when
\begin{align}
L_A^{\rm pair}
&=\max_{m\in\mathcal I_A}L_m^0,\notag\\
\max\Span(Q_A)&>L_A^{\rm pair}. \label{eq:A25d}
\end{align}
The original constituent bound in \cref{eq:A25d}, rather than buffer allowance or a bound increased to fit the aggregate, determines whether this additional search is needed. A pair within that bound is not divided merely because another cut could fit.

When \cref{eq:A25d} holds, the wavefront-extension construction at the same $\ell_{\rm seed}$ and the separator search with complete proposal protection in \cref{eq:A21,eq:A22} are repeated from the separate original constituent seeds, still using nominal $d$ for physical-boundary completion. If a larger aggregate contains the pair, all of that aggregate's recorded constituents are included, so that none is omitted or assigned twice. Other current commits and accepted cleanup reservations remain unavailable. The complete measured patch is still used to identify physical exterior boundaries; an unavailable neighboring region is not treated as an exterior boundary.

The replacement is accepted only when it produces at least two commit owners, preserves all constituent logical patches exactly once, avoids other commits and reserved cleanup, and passes the ownership checks against every other region of the same multi-qubit Pauli measurement. No fully committed stabilizer may span distinct replacement commits, and each stabilizer's residual support must have one cleanup owner. An existing residual corridor may already separate the replacement owners, in which case no new separator is required. If no valid replacement is obtained, the original aggregate is retained with only its exact required local span admitted. The original operation assignment is preserved throughout this check.

\subsubsection{Deferred treatment of one-row or one-column strips}
\label{app:deferred-seed-strips}

Only after the preceding separator searches and oversized-pair checks do we consider an unassigned complete row or column between aligned seeds. The seeds must belong to the same multi-qubit Pauli measurement and connected measured component, have overlapping coordinate ranges along their facing boundaries, and be separated by exactly one measured coordinate layer. The intervening strip must have active width at least $d$ and touch both seeds through measured connections.

A strip already reserved as cleanup is excluded. Otherwise, assignment to the seed containing fewer active data qubits is considered, subject to that seed's ownership bound; geometry and then seed labels resolve a tie. Before this assignment is admitted, the transitive stabilizer coalescence that it would force is also checked. If that coalescence would exceed its bound, the strip is instead retained as cleanup when it independently satisfies the complete-cross-section and unique-residual-ownership conditions. A failed cleanup check does not authorize an invalid separation: any required stabilizer coalescence remains subject to the final ownership rule in \cref{eq:A38,eq:A39}.

Let $G_i^{(1)}$ be the strips admitted to current seed $Q_i$ by these checks, and let $X_{\rm strip}$ collect the additional complete strips retained as cleanup. Then
\begin{align}
S_i^{\rm adj}&=Q_i\cup G_i^{(1)},\notag\\
G_i^{(1)}\cap X_{\rm forced}&=\varnothing,\qquad
\max\Span(S_i^{\rm adj})\le L_i^{\rm own},\notag\\
X_{\rm forced}&\leftarrow X_{\rm forced}\cup X_{\rm strip}.
\label{eq:A25e}
\end{align}
Candidate geometry is taken from one seed snapshot. Candidates are processed in deterministic order with checks against previously accepted assignments, so strips cannot overlap or acquire conflicting owners. Assigning a strip changes data ownership at this stage, not the identities of the two seed records. Any fully committed stabilizer shared by their resulting regions requires their later combination.

Let $X_{\rm forced}^{XZ}$ include the separators accepted during the $X/Z$ conflict search and the provisional-pair checks. The horizontal and vertical $Y$-chain construction below contributes $X_Y$. Immediately before child construction, write $X_{\rm res}=X_{\rm forced}^{XZ}\cup X_{\rm strip}\cup X_Y$ for all reserved separator data. Ordinary child propagation and placement exclude these data. A full child at an unoccupied true junction may, however, claim the part of a separator inside its junction core, according to \cref{app:junction-children}. If $J_{\rm claim}$ is the union of these successfully claimed junction cores, the final forced cleanup set is $X_{\rm forced}=X_{\rm res}\setminus J_{\rm claim}$. Using the final sets in \cref{eq:A40,eq:A41},
\begin{align}
X_{\rm forced}\cap C_{\rm all}&=\varnothing,\qquad
X_{\rm forced}\subseteq L. \label{eq:A26}
\end{align}
Before any junction is claimed, $X_{\rm forced}$ denotes the current reservation. The exception changes ownership only inside an accepted junction child; the remaining separator cells are protected against later strip absorption and optional enlargement. The seed construction, including the additional pair and strip checks, is applied independently to every multi-qubit Pauli measurement and to its disconnected measured components.

\subsection{Cleanup separators within horizontal and vertical $Y$ chains}
\label{app:y-seed-separators}

The provisional assignment of overlapping $Y$-extension footprints in \cref{app:seed-regions} gives every active data qubit one seed owner. Without cleanup separators, the final stabilizer-ownership check can combine a long horizontal or vertical chain into one commit region. We test each still-uncoalesced provisional $Y$ seed that contains exactly two logical patches. This test is performed after the routed-lobe assignment, the $X/Z$ separator and oversized-pair searches, and the deferred strip treatment, but before buffers or child commits are constructed. All candidate geometry is taken from the same seed snapshot.

Let $i=(\lambda,\mu)$ index a provisional pair, let $D_\lambda,D_\mu$ be its original logical-patch data, and write $S_i^Y$ for its provisional seed footprint. The patch centers are the centers of their coordinate bounding boxes. The pair is treated as horizontal when
$|\bar x_\mu-\bar x_\lambda|\ge|\bar y_\mu-\bar y_\lambda|$,
and vertical otherwise. For a horizontal pair, order the patches from left to right, breaking ties by increasing $y$ and then by logical-qubit label. For a vertical pair, order them from top to bottom, breaking ties by increasing $x$ and then by label. Thus $\lambda$ is the left or upper patch. Physical $y$ coordinates increase upward.

For a horizontal pair, let $x_\lambda^+$ and $x_\mu^-$ be its facing right and left boundary coordinates, and require $x_\lambda^+<x_\mu^--1$. For a vertical pair, let $y_\lambda^-$ and $y_\mu^+$ be its facing lower and upper boundary coordinates, and require $y_\mu^+<y_\lambda^--1$. A pair failing the corresponding gap condition is not a separator candidate. Define the separation coordinate $a_i$ and the coordinate $b_i$ along the cut by
$$
(a_i(q),b_i(q))=
\begin{cases}
(x(q),y(q)),&i\text{ horizontal},\\
(y(q),x(q)),&i\text{ vertical}.
\end{cases}
$$
The midpoint coordinate and its active candidate cross-section are
\begin{align}
c_i&=
\begin{cases}
\displaystyle\left\lfloor\frac{x_\lambda^++x_\mu^-}{2}\right\rfloor,
&i\text{ horizontal},\\[1ex]
\displaystyle\left\lceil\frac{y_\lambda^-+y_\mu^+}{2}\right\rceil,
&i\text{ vertical},
\end{cases}\notag\\
\ell_i&=\min_{q\in S_i^Y}b_i(q),\qquad
h_i=\max_{q\in S_i^Y}b_i(q),\notag\\
P_i&=\{q\in V_M:a_i(q)=c_i\},\notag\\
X_i^Y&=S_i^Y\cap P_i. \label{eq:A26a}
\end{align}
The floor selects the middle column nearest the left patch; the ceiling selects the middle row nearest the upper patch. In both cases, only the specified midpoint is tested. The candidate consists of actual active data qubits, not every integer drawing coordinate on the line. An inactive coordinate can be crossed only through an existing measured connection.

Set $\epsilon_i=1$ for a horizontal pair and $\epsilon_i=-1$ for a vertical pair. The retained sides corresponding to patches $\lambda$ and $\mu$ are
\begin{align}
S_i^{Y,1}
&=\{q\in S_i^Y:\epsilon_i(a_i(q)-c_i)<0\},\notag\\
S_i^{Y,2}
&=\{q\in S_i^Y:\epsilon_i(a_i(q)-c_i)>0\}. \label{eq:A26b}
\end{align}
These are the left and right sides for a horizontal pair, and the upper and lower sides for a vertical pair.

Let $R_i$ be the measured routing data available to this pair within its multi-qubit Pauli measurement. Write $\operatorname{Conn}(A)$ when the induced active graph $G_M^{\rm str}[A]$ has exactly one connected component. To distinguish elongated checks that run across the proposed separation from those that run along the cut, define their coordinate extents using the complete measured support:
$$
\begin{aligned}
\Delta_{a_i}(s)
&=\max_{q\in\supp(s)}a_i(q)-\min_{q\in\supp(s)}a_i(q),\\
\Delta_{b_i}(s)
&=\max_{q\in\supp(s)}b_i(q)-\min_{q\in\supp(s)}b_i(q).
\end{aligned}
$$
Let $H_i=\{q\in V_M:\ell_i\le b_i(q)\le h_i\}$ and $U_{s,i}=\supp(s)\cap H_i$. Among the elongated measured stabilizers, only those with $\Delta_{a_i}(s)>\Delta_{b_i}(s)$ and $U_{s,i}\ne\varnothing$ enter the elongated-bridge exclusion below. Denote this set by $\mathcal E_i^\perp$, and for its members set
$a_{s,i}^-=\min_{q\in U_{s,i}}a_i(q)$ and
$a_{s,i}^+=\max_{q\in U_{s,i}}a_i(q)$.
In particular, the orientation test uses the complete support, not the support after restricting it to $H_i$.

A candidate is locally admissible only when
\begin{align}
&\varnothing\ne X_i^Y\subseteq R_i,\qquad
X_i^Y\cap(D_\lambda\cup D_\mu)=\varnothing,\notag\\
&\operatorname{Conn}(X_i^Y),\notag\\
&\min_{q\in X_i^Y}b_i(q)=\ell_i,\qquad
\max_{q\in X_i^Y}b_i(q)=h_i,\notag\\
&D_\lambda\subseteq S_i^{Y,1},\qquad
D_\mu\subseteq S_i^{Y,2},\notag\\
&\operatorname{Conn}(S_i^{Y,1}),\qquad
\operatorname{Conn}(S_i^{Y,2}),\notag\\
&\supp(s)\cap S_i^{Y,1}\ne\varnothing,\quad
\supp(s)\cap S_i^{Y,2}\ne\varnothing\notag\\
&\quad\Longrightarrow
\supp(s)\cap X_i^Y\ne\varnothing
\quad(s\in\mathcal S_M),\notag\\
&U_{s,i}\cap X_i^Y=\varnothing,\qquad
c_i\notin(a_{s,i}^-,a_{s,i}^+)
\quad(s\in\mathcal E_i^\perp). \label{eq:A26c}
\end{align}
The first three lines require a connected, full-width or full-height routing cut that avoids the logical-patch cores. The next two preserve each logical patch and its connected retained seed. The stabilizer condition prevents a measured check from connecting the retained commits without cleanup support. The final condition rejects an elongated connection directed across the separation. An elongated check parallel to the cut does not by itself prohibit the separator; it remains subject to the connectivity and stabilizer-ownership conditions.

Adjacent provisional pairs on the same midpoint line must be checked together. Construct this grouping from all candidates defined in \cref{eq:A26a}, including candidates that fail \cref{eq:A26c}; otherwise an invalid neighboring segment could be silently omitted. Candidates from different multi-qubit Pauli measurements or with different orientations or midpoint coordinates are never grouped.

Within one such common line, regard candidates $i,j$ as adjacent if any of the following holds: their intervals $[\ell_i,h_i]$ and $[\ell_j,h_j]$ overlap or are consecutive, meaning
$\max\{\ell_i,\ell_j\}\le\min\{h_i,h_j\}+1$;
an edge of $G_M^{\rm str}$ joins a qubit in $X_i^Y$ to a qubit in $X_j^Y$; or a measured stabilizer has support in both $S_i^Y$ and $S_j^Y$ and also has a support qubit on their common plane $P_i=P_j$. In this last condition, the support qubit on the plane need not belong to either candidate cut. Let $\mathfrak F$ be the connected components of this adjacency relation.

For $F\in\mathfrak F$, let $X_F^Y=\bigcup_{i\in F}X_i^Y$ and use the common coordinate $b$ along its cut. The family is accepted only if every member satisfies \cref{eq:A26c} and
$$
\begin{aligned}
&\operatorname{Conn}(X_F^Y),\\
&\min_{q\in X_F^Y}b(q)=\min_{i\in F}\ell_i,\\
&\max_{q\in X_F^Y}b(q)=\max_{i\in F}h_i.
\end{aligned}
$$
If any of these conditions fails, none of that family's partial segments is retained and all its provisional pairs remain unchanged. This ensures that the separator spans the complete active width or height rather than ending inside a neighboring overlap.

Let $\mathcal A$ contain the pair indices in accepted families. Replace each accepted provisional pair by its two retained sides and reserve the complete separator data:
\begin{align}
S_i^Y&\longmapsto\{S_i^{Y,1},S_i^{Y,2}\}
\quad(i\in\mathcal A),\notag\\
X_Y&=\bigcup_{i\in\mathcal A}X_i^Y,\notag\\
X_{\rm forced}&\leftarrow X_{\rm forced}\cup X_Y. \label{eq:A26d}
\end{align}
Inherited cleanup reservations are preserved. The available routing data for each retained side exclude those reservations and its newly accepted cut. Each side preserves the unused span allowance of its provisional pair, while the later stabilizer-ownership check in \cref{eq:A38,eq:A39} still determines the final commit regions.

The $X/Z$ separator search is not applied here: only the specified midpoint is tested, at each supported distance. A reserved row or column is not absorbed by optional thin-strip adjustments. The junction-child exception and the ownership invariant in \cref{eq:A26} apply to the final separator data. These are spatial ownership conditions; no assumption of statistically independent spatial and temporal noise is made.

Let $\mathcal C_{\rm seed}^{\rm fin}$ denote the seed family after all the preceding extension, separation, and strip operations.

\subsection{Buffers and directed child construction}
\label{app:child-construction}

Only after the retained seed family has been fixed do we construct its buffers. For every exposed face $f$ of a finalized seed or child commit $C$, use the same enclosure-first construction, separately on its outside measured components:
\begin{align}
B_f(C)&=\bigcup_{U}\mathcal E_d(C;f,U), \label{eq:A27}\\
B(C)&=\bigcup_{f\in\mathcal F(C)}B_f(C). \label{eq:A28}
\end{align}
Seed extensions and buffers use the same enclosure construction, but with different depths and assignments: the seed extension uses $\ell_{\rm seed}$, whereas every buffer uses nominal $d$. The seed extension is united with the seed, removing the old corridor-facing ownership interface. The buffer is constructed from the finalized, possibly combined seed boundary and does not enlarge the commit. Child fronts and pre-buffers are then constructed from these rebuilt boundaries using nominal $d$. Source-owned seed promotion excludes foreign absorbing data by \cref{eq:A15}; buffer context remains subject to its observation and terminal rules.

Buffers may overlap and can terminate on another commit. Every buffer cell outside the union of commit cores is cleanup-owned throughout the construction. Before the commit barrier it is additionally visible to the commit worker as likelihood context; after the barrier the cleanup worker acts on its residual syndrome. Thus a buffer is not a third correction-ownership class and does not change correction owner at the barrier. Let $\beta(C,f)$ be the outward terminal boundary of $B_f(C)$.

The finalized seed and buffer geometry determines whether a seed is an endpoint or a branch seed. An endpoint seed terminates or blocks its remaining routing corridor, whereas a branch seed is attached to the side of a corridor that continues past the attachment. This gives $\mathcal C_{\rm seed}^{\rm fin}=\mathcal C_{\rm end}\sqcup\mathcal C_{\rm branch}$. Contract the measured corridors to a skeleton $\Sigma_M$, and let $\mathcal E_{\rm branch}$ remove only the short branch-owned attachment while retaining the continuing corridor. The effective graph used for speculative child placement is
\begin{align}
G_M^{\rm eff}&=\mathcal E_{\rm branch}(G_M^{\rm walk},\mathcal C_{\rm branch}). \label{eq:A14}
\end{align}
This removal is local: it cannot suppress another independent corridor or a true junction.

Within each connected measured component, at most one endpoint-seed batch is active as a traversal root at a given time. The first root is selected deterministically. If it produces no routing progress, the next endpoint can be tried as an alternative root. Once progress has been made, another endpoint remains inactive until it is reached by an accepted walk. It can be activated without such a handoff only after no live face remains and only when it is required to reach a genuine junction that has not yet been explored. A branch seed is not an independent propagation root. Reserved seed separators are excluded from ordinary child propagation and placement. Complete unoccupied junction cores crossed by separators are restored before the true junctions are identified, as described in \cref{app:junction-children}. If $\mathcal R_{\rm active}$ denotes the set of active root batches, then
\begin{align}
|\mathcal R_{\rm active}|&\le 1. \label{eq:A28a}
\end{align}

\subsubsection{Candidate fronts and complete directed sweeps}
\label{app:directed-wavefront}

Candidate boundary finding and the subsequent complete-path sweep are different operations. Let $\Omega_t$ be the current admissible child-propagation universe after branch erasure and consumed-corridor exclusions, and let $p_\alpha(q)=\alpha\cdot q$ be progress along the original propagation direction. Candidate fronts are obtained by the layered walk recurrence
\begin{align}
F_0&=A,\notag\\
F_{r+1}
&=\{v\in\Omega_t:\exists u\in F_r,\ 
(u,v)\in E_M^{\rm eff},\
p_\alpha(v)\ge p_\alpha(u)\}.
\label{eq:A29a}
\end{align}
These are exact-hop walk layers, not minimum-distance layers. A vertex can appear at more than one depth. They propose boundaries but do not assign commit data.

Let $\operatorname{Front}_\alpha(F_d;A)$ collect the forward coordinate-extremal subset of $F_d$ and the perpendicular extremal subsets extending beyond the transverse span of $A$. Each connected candidate carries its outgoing direction, is completed to its measured cross-section, and is checked separately. At a complete terminal $d$-by-$d$ corner, the candidate can be moved onto the outgoing side of that same corner; this changes its face orientation without adding propagation depth.

The pre-buffer and child sweep require complete simple paths. They therefore use a second wavefront that retains precisely the history needed to exclude a revisit. Strict progress prevents a return to an earlier progress level. When each equal-progress graph is a forest, the state at one depth is $(v,\pi,b)$: the present vertex, the previous vertex on the same progress level (or $\bot$), and whether any required measured edge has been used. For an allowed measured edge $u\to v$ with $u\ne v$, $p_\alpha(v)\ge p_\alpha(u)$, and $v\ne\pi$, set
\begin{align}
\pi'&=
\begin{cases}
u,&p_\alpha(v)=p_\alpha(u),\\
\bot,&p_\alpha(v)>p_\alpha(u),
\end{cases}\notag\\
b'&=b\lor\mathbf1_{\{\{u,v\}\in E_{\rm req}\}},\qquad
(u,\pi,b)\longrightarrow(v,\pi',b').
\label{eq:A29b}
\end{align}
The next vertex must be outside the excluded source region and within $\Omega_t$. If the present vertex is absorbing, the next vertex must also be absorbing. Initialize $(a,\bot,0)$ for each admitted starting vertex $a$. Identical states at the same depth are combined, retaining all predecessor relations. No required-edge condition is imposed when $E_{\rm req}=\varnothing$.

Set $\mathcal D_0(A)=\varnothing$, since a sweep excludes its source. For $m\ge1$, let $\mathscr L_r$ be the states at depth $r$. Retain the terminal states at depth $m$ satisfying the required-edge condition, and propagate success backward:
\begin{align}
\mathscr R_m
&=\{s\in\mathscr L_m:s\text{ satisfies the terminal condition}\},\notag\\
\mathscr R_r
&=\{s\in\mathscr L_r:\exists s'\in\mathscr R_{r+1},\ s\to s'\},\notag\\
\mathcal D_m(A)
&=\bigcup_{r=1}^{m}\{v:(v,\ldots)\in\mathscr R_r\}.
\label{eq:A29c}
\end{align}
The source, excluded region, direction, absorbing sets and current component are arguments of $\mathcal D_m$ whenever they are not clear from the construction. A reached vertex that cannot be continued to a successful depth-$m$ state is not in this sweep.

The forest rule is exact: within a forest, a non-backtracking walk cannot revisit a vertex, and strict progress prevents a return to any previous level. When an equal-progress graph has cycles, decompose its underlying undirected graph into biconnected blocks. Replace $\pi$ by the current block $j$ and its visited subset $H$. A step within $j$ requires $v\notin H$ and adds $v$ to $H$. Upon entering a different block, retain only its entry and new vertex; upon strict progress, clear the transverse history. A return to a departed block would repeat an articulation vertex, so its old history cannot affect a successful simple continuation.

This cyclic-block rule remains exponential in the largest retained block, not in the total number of vertices traversed. The implementation limits a block to 12 vertices and a query to 250,000 states and 2,000,000 transitions. A limit violation is reported before a region is accepted. It is not interpreted as an empty sweep.

\subsubsection{The frontier, pre-buffer, and child commit}

A traversal state $(t,\alpha)$ leaves an exposed face of a source commit $C$. Starting from the outward boundary $\beta(C,f)$ of its buffer, apply \cref{eq:A29a}. The face-specific candidates are
\begin{align}
\{(\partial_{t,j},\alpha_{t,j})\}_{j\in J_t}
&=\operatorname{Front}_{\alpha}
\bigl(F_d;\beta(C,f)\bigr).
\label{eq:A29}
\end{align}
The index $j$ distinguishes candidates even when they share the same outgoing direction. Select one candidate $j$ and abbreviate its front and direction as $\partial_{t,\alpha}$ and $\alpha$. This direction can differ from the incoming direction when the corridor turns.

Move one active step beyond $\partial_{t,\alpha}$ to its complete next measured face $A_{t,\alpha}=\operatorname{Adv}_\alpha(\partial_{t,\alpha})$. The first face and a further $d-1$ measured steps define the raw pre-buffer. With $F_{t,\alpha}=\beta(P_{t,\alpha},\alpha)$, the raw child is constructed by $d$ further measured steps outside the pre-buffer:
\begin{align}
P_{t,\alpha}
&=\operatorname{cl}^{\square}_M
\left(A_{t,\alpha}\cup\mathcal D_{d-1}(A_{t,\alpha})\right),
\label{eq:A30}\\
\widetilde C_{t,\alpha}
&=\operatorname{cl}^{\square}_M
\left(\mathcal D_d(F_{t,\alpha};\,\text{outside }P_{t,\alpha})\right).
\label{eq:A31}
\end{align}
The pre-buffer thus contains $d$ active sites along an ordinary forward rail, rather than one outward step followed by an additional $d$ steps. Each outgoing component of the pre-buffer is treated separately. The raw child is clipped to the connected measured component anchored to its complete first forward face. A separate side-branch component can be discarded; within the retained component, outline pruning preserves its required directed path support. A measured trapezoidal edge can justify an additional lateral boundary band only when a complete directed path is certified to use that edge; it does not authorize an extra straight forward layer.

An ordinary child is accepted only when it remains in $V_M$, is connected, contains the required active depth, fits its applicable ownership span, avoids existing commits and $X_{\rm forced}$, and has a valid complete forward exit. Rectilinear closure alone does not replace these checks.

\subsubsection{Endpoint and branch interception without exhaustive search}
\label{app:wavefront-certificates}

Endpoint seeds remain hard terminals during speculative placement. Before accepting a frontier, pre-buffer, or child, we test whether its defining propagation reaches an endpoint seed at the required depth. A completed reachability relaxation that excludes the target proves a negative. To prove a positive, join a shortest legal first-entry prefix of length $\ell$ to a path of length $d-\ell$ in a breadth-first tree entirely inside the absorbing endpoint. The concatenation is accepted only after checking its exact length, distinct vertices, measured edges and absorption. If this first witness is too short, a bounded native search tries to construct one exact-length simple path in the full depth-$d$ reachable query graph. It can enter the endpoint later, after a longer outside prefix, rather than searching only for a longer suffix within the seed. The graph retains the original measured edges, excludes re-entry into the pre-buffer, and enforces absorption across the full endpoint-seed union. Reverse distances to the target prune paths that cannot finish in the remaining steps. Every returned path is checked against the original query. The relaxation, graph construction, search and witness checks share a fixed work budget; an empty or budget-exhausted search remains unresolved, not a negative proof.

Branch interception uses its original heading rule rather than this endpoint predicate. A branch is intercepted only when the first admissible entry has depth $d$ and a depth-$d$ entry arrives along the main local heading. The local heading changes only when the remaining straight rail cannot supply the required continuation. A voluntary sideways step stays off-main until a genuine main-rail advance; it cannot subsequently be relabeled as a forced turn.

The branch relaxation retains current vertex, local heading, off-main status and depth. It proves rejection if no depth-$d$ main-heading entry exists even in that relaxation. A verified earlier-entry path also proves rejection. Acceptance requires both a verified simple depth-$d$ main-heading witness and the absence of any earlier entry in the completed relaxation. Remaining cases are unresolved, not positive or negative by default. Straight-rail continuation is evaluated by a longest-path calculation on a graph of strictly increasing heading coordinate; a rotated-heading witness also excludes its actual earlier prefix.

An unresolved required endpoint or branch decision is reported as unsupported geometry and stops the construction explicitly. An exhaustive compatibility calculation can be requested separately, but is not part of the default wavefront procedure or its reported scaling.

If an endpoint is reached before a child can be placed, the candidate is rejected and the traversal hands off to that seed. Its incoming face is consumed and its other unexplored faces are considered independently. Branch seeds are omitted from this endpoint rejection test. An unoccupied true junction reachable within active distance $d$ is considered before an ordinary child. After a child is accepted, its buffer is rebuilt and only its unconsumed complete $d$-wide face-specific terminal fronts are considered. A rejected hypothetical pre-buffer never becomes walked geometry and cannot suppress another face.

\subsubsection{Child commits at routing junctions}
\label{app:junction-children}

A junction is identified from the complete measured routing geometry before reserved separator data are removed. Let $\mathcal J_{\rm true}$ contain the canonical junction cores remaining after short branch-owned attachments have been removed according to \cref{eq:A14}. A true junction joins at least three independently continuing routing corridors. A short side attachment alone does not define one.

Let $C_{\rm seed,all}$ be the union of finalized seed commits. The separator-crossed junctions requiring a child are
\begin{align}
\mathcal J_X
&=\left\{J\in\mathcal J_{\rm true}:
w_x(J)=w_y(J)=d,\right.\notag\\
&\hspace{3em}\left.
J\cap X_{\rm res}\ne\varnothing,\quad
J\cap C_{\rm seed,all}=\varnothing
\right\}. \label{eq:A31a}
\end{align}
Their complete measured cores are retained in the propagation and placement domains. A child containing each such core must be placed:
\begin{align}
J&\subseteq C_J\qquad(J\in\mathcal J_X),\notag\\
J_{\rm claim}&=\bigcup_{J\in\mathcal J_X}J. \label{eq:A31b}
\end{align}
A construction with an unclaimed required junction is rejected. A junction already touched by a seed does not receive a second overlapping child under this rule. If removing a short branch attachment eliminates the third independent arm, the provisional junction is rejected and its separator cells remain excluded.

Only after the child is accepted are the reserved cells inside its junction core reassigned to commit. The remaining separator cells stay cleanup-owned. Let $\mathcal S_X^{\rm res}$ be the stabilizers previously certified to connect arms of a reserved separator. The final reservation and surviving certificates are
\begin{align}
X_{\rm forced}
&=X_{\rm res}\setminus J_{\rm claim},\notag\\
\mathcal S_X
&=\left\{s\in\mathcal S_X^{\rm res}:
\supp(s)\cap(X_{\rm res}\cap J_{\rm claim})
=\varnothing\right\}. \label{eq:A31c}
\end{align}
Thus a child can divide a separator into cleanup portions on its sides. A former separator certificate touching newly committed separator data cannot reconnect those cleanup portions through the child. The buffer is constructed from the accepted child boundary and propagation continues independently through its unexplored outgoing faces.

Each directed state has a corridor key $\kappa(C,f)$ recording the source, face, outgoing component, and initial direction. If $\mathcal T$ is the consumed set, then
\begin{align}
\kappa(C,f)\in\mathcal T&\Longrightarrow \kappa(C,f)\text{ is not re-enqueued}. \label{eq:A32}
\end{align}
Two faces with the same compass direction can still have different keys when they belong to different corridors.

\subsection{Geometric normalization and buffer recomputation}
\label{app:geometric-normalization}

The separator set used below is updated whenever a required junction child is accepted, as in \cref{eq:A31c}. That reassignment is distinct from optional strip absorption or commit enlargement.

Let $G(C,S)$ be a complete local one- or two-coordinate-wide routing strip between aligned faces of a child commit $C$ and a seed commit $S$ in the same measurement context, and let $w(G(C,S))$ denote the number of coordinate rows or columns across that strip. Let $L^{\rm own}(C,S)$ be their applicable combined ownership-admission span. The three regions are combined only when the strip is not a forced cleanup separator and the result remains within that span:
\begin{align}
C^+&=C\cup G(C,S)\cup S,\notag\\
w(G(C,S))&\in\{1,2\},\qquad
G(C,S)\cap X_{\rm forced}=\varnothing,\notag\\
\max\Span(C^+)&\le L^{\rm own}(C,S)\notag\\
&\Longrightarrow C_{\rm merge}=C^+. \label{eq:A33}
\end{align}
If any condition in \cref{eq:A33} fails, the seed, child, and intervening cleanup data remain separate. A junction child can be enlarged by the minimum row or column needed so that no measured stabilizer couples two intended residual branches, but only when the enlarged child remains within its effective ownership-admission span. A local span relaxation inherited from an unavoidable seed coalescence cannot be increased by either optional normalization.

Every operation $\mathcal N_{\rm geom}$ that changes commit geometry invalidates its old buffer:
\begin{align}
C\leftarrow\mathcal N_{\rm geom}(C)
&\Longrightarrow B_{\rm old}(C)\text{ is invalid}, \label{eq:A34}\\
B_{\rm new}(C)&=\bigcup_{f\in\mathcal F(C)}B_f(C). \label{eq:A35}
\end{align}
No later child can be derived from a buffer constructed before the child was moved to its junction footprint, enlarged, or combined with another commit.

\subsection{Final coalescence and cleanup regions}
\label{app:cleanup-ownership}

Let $\widehat{\mathcal C}=\{\widehat C_1,\ldots,\widehat C_{\widehat N_C}\}$ be the provisional family after the directed traversal. Its combined data and residual data are
\begin{align}
\widehat C_{\rm all}&=\bigcup_i\widehat C_i, \label{eq:A36}\\
\widehat L&=V_M\setminus\widehat C_{\rm all}. \label{eq:A37}
\end{align}
For a measured stabilizer $s$, let $I_C(s)=\{i:\supp(s)\cap\widehat C_i\ne\varnothing\}$. A stabilizer is sandwiched when
\begin{align}
|I_C(s)|&\ge 2,\qquad \supp(s)\cap\widehat L=\varnothing. \label{eq:A38}
\end{align}
Construct an undirected graph on the provisional commits, joining two vertices whenever one sandwiched stabilizer touches both. If $A_1,\ldots,A_{N_C}$ are its connected components, the final commits are
\begin{align}
C_a&=\bigcup_{i\in A_a}\widehat C_i, \qquad
\mathcal C=\{C_1,\ldots,C_{N_C}\}. \label{eq:A39}
\end{align}
This coalescence also resolves the provisional interfaces produced by \cref{eq:A25e}: a fully commit-owned stabilizer crossing such an interface places the two provisional regions in the same final commit region. Further, this transitive coalescence changes ownership interfaces but does not add committed data. In particular, it combines provisional pair and singleton blocks within a continuous $Y$-extension chain only across interfaces having no residual cleanup support. For the portions of a $Y$ separator remaining outside junction children, $X_Y\cap X_{\rm forced}\subseteq\widehat L$. A stabilizer with support in these remaining cleanup data does not satisfy \cref{eq:A38} and therefore cannot itself require the commits on its two sides to be combined. Where separator data have instead been assigned to a junction child, the final ownership test uses the updated commit and cleanup regions and the surviving separator certificates in \cref{eq:A31c}. Uncut interfaces, including the required elongated-stabilizer connections, are still combined transitively when their complete supports are committed. Because this coalescence is required to give a fully committed stabilizer one decoder, the resulting commit can exceed a previous ownership span. In that case, only its exact resulting ownership-admission span is retained:
\begin{align}
L_a^{\rm own}&=\max\Span(C_a). \label{eq:A39a}
\end{align}
The buffer-span metadata can preserve unused headroom from the constituent commits when the final buffer is constructed, but this headroom is not part of $L_a^{\rm own}$ and cannot authorize a later commit-owned enlargement. Every affected buffer is recomputed from the final composite perimeter. Traversal references to constituent commits are updated to their final composite commit, while preserving the multi-qubit Pauli measurement to which the traversal belongs.

The provisional residual components are then examined for a one-row or one-column component, including a singleton. Such a component is absorbed only when its complete measured boundary and every incident stabilizer support outside the component already belong to one commit in the same operation context, the enlarged commit remains within its effective ownership-admission span, and no second commit is cardinally adjacent. A forced seed separator is never eligible. These absorptions are decided from one immutable residual snapshot. After they are applied simultaneously, every affected buffer and every cleanup component is recomputed.

Let
\begin{align}
C_{\rm all}&=\bigcup_{i=1}^{N_C}C_i, \label{eq:A40}\\
L&=V_M\setminus C_{\rm all}, \label{eq:A41}\\
\{\widetilde{\mathrm{CL}}_1,\ldots,
\widetilde{\mathrm{CL}}_{\widetilde N_{\rm CL}}\}
&=\Comp(G_M^{\rm str}[L]). \label{eq:A42}
\end{align}
These are the preliminary cleanup components. Two arms of an accepted multi-arm separator can meet through a measured stabilizer without being adjacent in $G_M^{\rm str}$. Use the surviving separator-junction certificates $\mathcal S_X$ from \cref{eq:A31c}; certificates touching separator cells claimed by a junction child have already been removed. Let $g(j)$ identify the multi-qubit Pauli measurement owning $\widetilde{\mathrm{CL}}_j$. Define
\begin{align}
j\approx_X k\Longleftrightarrow\;&g(j)=g(k),\notag\\
&\text{there exists }s\in\mathcal S_X\text{ such that}\notag\\
&\supp(s)\cap\widetilde{\mathrm{CL}}_j\ne\varnothing,\notag\\
&\supp(s)\cap\widetilde{\mathrm{CL}}_k\ne\varnothing,\notag\\
&\supp(s)\cap L\subseteq X_{\rm forced}. \label{eq:A42a}
\end{align}
Let $\equiv_X$ be the equivalence closure of $\approx_X$. If $J_1,\ldots,J_{N_{\rm CL}}$ are its equivalence classes, the final cleanup regions are
\begin{align}
\mathrm{CL}_a&=\bigcup_{j\in J_a}\widetilde{\mathrm{CL}}_j,\qquad
\mathcal C_{\rm cleanup}=\{\mathrm{CL}_1,\ldots,\mathrm{CL}_{N_{\rm CL}}\}. \label{eq:A42b}
\end{align}
Ordinary disconnected components are not combined by this rule. If a measured stabilizer intersects more than one resulting cleanup region without satisfying the certified separator-junction condition, the proposed construction is rejected.
The active data therefore admit the disjoint partition
\begin{align}
V_M&=C_{\rm all}\sqcup\biggl(\bigsqcup_j\mathrm{CL}_j\biggr). \label{eq:A43}
\end{align}

A stabilizer is commit-owned only when its complete active support lies in one final commit:
\begin{align}
\mathcal S(C_i)&=\{s\in\mathcal S_M:\supp(s)\subseteq C_i\}, \label{eq:A44}\\
\mathcal S(\mathrm{CL}_j)&=\{s\in\mathcal S_M:\supp(s)\cap\mathrm{CL}_j\ne\varnothing\}. \label{eq:A45}
\end{align}
The final normalization verifies that one stabilizer intersects at most one final cleanup region. Boundary-crossing stabilizers can be observed through $B(C_i)$ but remain cleanup-owned whenever they have residual support. Define $\mathcal S_{\rm commit}=\bigcup_i\mathcal S(C_i)$ and $\mathcal S_{\rm cleanup}=\bigcup_j\mathcal S(\mathrm{CL}_j)$.

\paragraph{Proposition 1.}
The commit- and cleanup-owned stabilizer sets are disjoint and exhaustive:
\begin{align}
\mathcal S_{\rm commit}\cap\mathcal S_{\rm cleanup}&=\varnothing, \label{eq:A46}\\
\mathcal S_M&=\mathcal S_{\rm commit}\sqcup\mathcal S_{\rm cleanup}. \label{eq:A47}
\end{align}
If $\supp(s)\cap L\ne\varnothing$, \cref{eq:A42,eq:A42a,eq:A42b} and the final ownership verification place its complete residual support in one cleanup region, so $s$ has one cleanup owner and cannot be contained in a commit. If $\supp(s)\cap L=\varnothing$, its support lies in the provisional committed union. A stabilizer touching more than one provisional commit is sandwiched and joins those commits in the coalescence graph. It is therefore contained in one final composite commit. This proves both claims.

\subsection{Space-time separation of cleanup regions}
\label{app:cleanup-separation}

Let $\mathcal R_\ell$ be the operation rounds for one measurement layer and let $V_r$ be the active data-qubit set in round $r$. The final spatial commit, buffer, and cleanup volumes are
\begin{align}
\mathcal C_{i,\ell}&=\{(q,r):q\in C_i\cap V_r,\ r\in\mathcal R_\ell\}, \label{eq:A48}\\
\mathcal B_{i,\ell}&=\{(q,r):q\in B_i\cap V_r,\ r\in\mathcal R_\ell\}, \label{eq:A49}\\
\mathcal L_{j,\ell}&=\{(q,r):q\in\mathrm{CL}_j\cap V_r,\ r\in\mathcal R_\ell\}. \label{eq:A50}
\end{align}
Let $\mathcal I_{\lambda,\ell}^-$ and $\mathcal I_{\lambda,\ell}^+$ denote the contiguous idle-round intervals immediately before and after the operation on logical-patch worldline $\lambda$, and let $V_{\lambda,r}$ be the active data associated with that logical patch in round $r$. The corresponding temporal volumes are
\begin{align}
\mathcal T_{\lambda,\ell}^{\pm}
&=\{(q,r):q\in V_{\lambda,r},\ r\in\mathcal I_{\lambda,\ell}^{\pm}\},
\qquad |\mathcal I_{\lambda,\ell}^{\pm}|=d. \label{eq:A50a}
\end{align}
When two consecutive measurement layers $\ell$ and $\ell'$ use the same intervening idle-round interval on worldline $\lambda$, the future volume of the first operation and the past volume of the second operation are one and the same temporal block:
\begin{align}
\mathcal I_{\lambda,\ell}^{+}=\mathcal I_{\lambda,\ell'}^{-}
&\Longrightarrow
\mathcal T_{\lambda,\ell}^{+}=\mathcal T_{\lambda,\ell'}^{-}.
\label{eq:A50b}
\end{align}
This shared volume is created once and records its dependence on both neighboring commit decoders. The temporal volumes are indexed by logical-patch worldline rather than by final composite commit, since one commit can contain several logical patches. Before the commit corrections are applied, an adjacent idle-round volume supplies temporal buffer syndrome to the commit decoder or decoders that border it. After the relevant commit barriers, that same volume is acted upon as one temporal cleanup block.

For an $X/Z$ seed $i$, let $a(i)$ denote its final composite commit. For each originating boundary crossing $h$, let $K_{i,h}$ be the qualified exterior cap and let $F_{i,h}^{\rm cap}$ be the face that generated that cap. The complete source-owned extension remains committed:
\begin{align}
R_{i,h}^{\rm seed}
&=\left(\bigcup_{\gamma\in\mathcal W_{i,h}^{(\ell_{\rm seed})}}
\Sweep^\circ(\{\gamma\};S_i^0)\right)\cap R_M\notag\\
&\hspace{1em}\cup K_{i,h}\cup F_{i,h}^{\rm cap},\notag\\
R_{i,h}^{\rm seed}\times\mathcal R_\ell
&\subseteq\mathcal C_{a(i),\ell}. \label{eq:A51}
\end{align}
The enclosure and its qualified cap remain protected by \cref{eq:A22}; no shorter path prefix is substituted. A foreign seed terminal is excluded from the source-owned routing part and retains its own owner. When a provisional pair is reconstructed, these definitions refer to the constituent seeds used by its accepted reconstruction, with the same depth $\ell_{\rm seed}$ and nominal-$d$ boundary-completion rule.

The retained routing data provide commit ownership between the logical-patch temporal region and residual spatial cleanup at that interface. This is a geometric condition. Whether the resulting cleanup corrections can be applied independently is certified by \cref{eq:A51a} for the actual circuit and DEM. If their authorized detector effects overlap, the corresponding cleanup regions are decoded together. A short dead end that does not contribute to an enclosure certified at $\ell_{\rm seed}$ need not be promoted.

This is a local separation statement. It does not require the raw detector-error-model factor projections used by the temporal and spatial cleanup decoders to be disjoint. To state the execution rule precisely, let $\mathfrak B$ be the cleanup observation blocks, let $z(b)$ be the authenticated cleanup correction zone that owns block $b$, and let $\mathcal A_z$ be the corrections authorized to zone $z$. If $\Delta(c)$ is the set of detector outcomes changed by correction $c$, define
\begin{align}
\mathfrak B_z&=\{b\in\mathfrak B:z(b)=z\},\notag\\
\Delta_z&=\bigcup_{c\in\mathcal A_z}\Delta(c),\notag\\
z\sim_{\rm exec}z'&\Longleftrightarrow
z=z'\ \text{or}\ \Delta_z\cap\Delta_{z'}\ne\varnothing.
\label{eq:A51a}
\end{align}
All observation blocks in the same authenticated correction zone are therefore assigned to one owner first. The execution components are the equivalence classes generated by $\sim_{\rm exec}$: distinct zones are combined only when their authorized corrections have intersecting detector effects. The same raw DEM factor can appear in more than one likelihood context without combining otherwise distinct execution components. An accepted $X/Z$ cleanup separator resolves oversized competing path extensions while preserving the retained commit-owned collar. An accepted $Y$-chain separator instead prevents overlapping fixed $Y$-extension footprints from producing an unbounded chain-wide commit. Their remaining separator portions outside junction children are cleanup-owned, and the resulting regions are subject to the same execution test in \cref{eq:A51a}. Distinct execution components have disjoint authorized correction effects on the detectors. This is a statement about independent correction execution, not statistical independence of their likelihood models; the same physical DEM factor can contribute to more than one local likelihood model.

\subsection{Termination and buffer sufficiency}
\label{app:buffer-sufficiency}

The merged patch has finitely many active data qubits and finitely many directed corridor keys. Since each key is processed at most once,
\begin{align}
N_{\rm iterations}&\le |\mathcal K_M|, \label{eq:A52}
\end{align}
where $\mathcal K_M$ is the finite key set. Moving a child to its junction footprint, merging, and expansion retain the consumed history and cannot restart a backward traversal. Final coalescence is one connected-component computation on a finite graph and is idempotent.

\subsubsection{Construction cost}
\label{app:wavefront-scaling}

For a fixed logical-patch layout and measurement pattern, our numerical studies are consistent with approximately $O(d^3)$ scaling of the wavefront commit/cleanup construction over the distances tested. This is an empirical scaling observation, not a proven worst-case bound for every distance, Pauli measurement, or routing geometry.

The reason for the improvement is that repeated paths no longer cause repeated exploration of their shared geometric state. The ordinary distance layers, block-cut pruning and exterior flood are linear in the local graph and raster sizes. In a directed forest query with $N$ reachable vertices and maximum measured degree $\Delta$, the state in \cref{eq:A29b} gives at most $O(dN\Delta)$ states and $O(dN\Delta^2)$ transitions. For bounded $\Delta$ and $N=O(d^2)$, this particular kernel therefore has $O(d^3)$ work. A bounded transverse block introduces a factor depending on its size, which is fixed by the certificate limit. The sparse finite-pocket and endpoint-witness searches have separate fixed work limits and can report an unresolved query; they do not supply a general polynomial-time guarantee.

This kernel bound is not a proof of the same bound for the complete planner. The number of source faces and repeated queries, the source-specific path certificates, the size of their intermediate records, cleanup-separator candidates, and geometric adjustments also contribute. The number of logical patches and the routing topology are separate parameters from $d$. The spatial construction cost must also be distinguished from materializing $d$ rounds of circuits and DEMs, whose data volume already grows with the space-time volume.

The old and new constructions can be compared by their final commit data, buffers, cleanup data, and measured-stabilizer ownership. Agreement on tested golden examples and focused circuit/DEM checks is numerical evidence for those examples. It is not a claim that every intermediate endpoint or source-certificate set equals the set obtained by enumerating all simple paths on an arbitrary graph.

Let
\begin{align}
t_{\rm corr}&=\left\lfloor\frac{d-1}{2}\right\rfloor. \label{eq:A53}
\end{align}
Every recomputed buffer is the full measured enclosure obtained from the final commit boundary by \cref{eq:A12a,eq:A12b,eq:A12c,eq:A27,eq:A28}. It includes its admitted boundary and interior data, together with the measured terminal completions. This defines the geometric decoding context; it does not require an equality between the new source certificates and the intermediate sweep of every simple path in the old construction. To infer correction of all errors caused by at most $t_{\rm corr}$ circuit faults, the circuit and noise model must additionally ensure that every such fault pattern capable of changing the commit decision is represented in the commit-plus-buffer detector model. The geometric inequality
\begin{align}
t_{\rm corr}&<d \label{eq:A54}
\end{align}
alone does not establish this correspondence: one circuit fault can affect several data qubits or detector outcomes. The schedule, noise model, allowed corrections, and detector relations at the region boundaries must be included when verifying fault tolerance. This circuit-level condition is separate from the choice of the shorter seed extension; buffers continue to use the full depth-$d$ enclosure construction. Corrections selected by a commit decoder remain restricted to its commit region, and unresolved boundary effects enter the residual syndrome supplied to cleanup.

The complete construction therefore produces final composite commits, cleanup-owned cells that are additionally exposed as commit-buffer context, and residual cleanup blocks whose size is controlled by $d$ and the local routing topology except when an unavoidable ownership coalescence requires a larger commit. Commit tasks can run in parallel on the original syndrome. Their effects form the post-commit residual supplied to cleanup. Cleanup observation blocks belonging to the same authenticated correction zone are first treated as one owner. Distinct cleanup zones are then grouped into maximal execution components only when their authorized corrections have intersecting detector effects, after which distinct cleanup execution components can run in parallel.

\section{Specifying and decoding \texttt{ExpD} in Patchwork}
\label{app:expd-spec}

Patchwork (to be released) takes a declarative description of a lattice-surgery program and produces everything a decoding experiment needs: the noisy syndrome-extraction circuit, its DEM, the space-time partition of that DEM into commit, buffer, and cleanup regions (\cref{sec:spacetime_partition,sec:spacetime_corrections}), and the spacelike and timelike logical observables (\cref{sec:Logic_Auto_Gen,sec:Timelike_log_gen}). The user never writes circuits, stabilizer schedules, or detector definitions; all of these follow from the rules of \cref{sec:Layout,sec:MeasureXZ,sec:FaultTolerantY,sec:Gate_Schedule}. \texttt{ExpD}, as shown in \cref{sec:numerics}, is specified in a single TOML file, the essential parts of which we reproduce here.

The header fixes the code distance, the routing-space grid, and the noise model. Every syndrome-round count below is a multiple of the code distance, so changing \texttt{distance} re-derives the entire experiment.

\begin{lstlisting}
name = "ExpD"
distance = 7

[layout]
# 4 by 4 cells of logical patches with
# routing corridors between them
grid = [4, 4]

[noise]
# uniform circuit-level depolarizing noise
kind = "single_p"
p = 0.001
\end{lstlisting}

Each logical qubit is declared with its preparation state and final readout basis, as in \cref{tab:exp4_logical_qubits}; patch names refer to positions in the grid layout.

\begin{lstlisting}
[patches]
Q2  = { prepare = "0",  readout = "Z" }
Q6  = { prepare = "+",  readout = "X" }
Q15 = { prepare = "+",  readout = "Z" }
Q19 = { prepare = "Y+", readout = "Y" }
Q22 = { prepare = "Y+", readout = "Y" }
# ... Q10, Q13, Q33, Q36, Q54 likewise
\end{lstlisting}

The program itself is a sequence of layers, each lasting a stated number of syndrome rounds. A layer either idles a set of patches (\texttt{memory}) or performs one or more multi-qubit Pauli measurements. A measurement is written as a labelled Pauli product over patch names. From this string alone Patchwork chooses the merged-patch region and its checkpoint lines (\cref{sec:Layout}), solves the multi-commodity flow problem for the routing paths (\cref{subsec:MCFAlgo}), places the elongated, boundary, and $Y$-extension stabilizers and twist defects (\cref{sec:MeasureXZ,sec:FaultTolerantY}), and assigns the gate schedules of \cref{sec:Gate_Schedule}. The three measurements of \cref{eq:exp4PsM1,eq:exp4PsM2,eq:exp4PsM3}, separated by idle layers of $d$ rounds, read

\begin{lstlisting}
[[layers]]
id = "pre_idle"          # d idle rounds
rounds = 7
prepare = ["Q2", "Q6", "Q10", "Q13", "Q15",
           "Q19", "Q22", "Q33", "Q36", "Q54"]

[[layers]]
id = "first_ppm"         # measures P1
rounds = 7
measure = ["P1: Y(Q22), Z(Q10), Z(Q2),
                X(Q6), X(Q13), X(Q15)"]
memory  = ["Q19", "Q33", "Q36", "Q54"]

[[layers]]
id = "first_middle_idle" # d idle rounds
rounds = 7
memory = ["Q2", "Q6", "Q10", "Q13", "Q15",
          "Q19", "Q22", "Q33", "Q36", "Q54"]

[[layers]]
id = "second_ppm"        # measures P2
rounds = 7
measure = ["P2: Z(Q2), Z(Q15), X(Q54)"]
memory  = ["Q6", "Q10", "Q13", "Q19",
           "Q22", "Q33", "Q36"]

# ... second idle layer; third_ppm measuring
#     P3: Y(Q22), Y(Q36), Y(Q19), Y(Q33);
#     final idle layer with readout of all
#     ten patches
\end{lstlisting}

Finally, the logical observables whose failures define the total LER of \cref{fig:LERresult} are declared as parities of measurement outcomes. A timelike observable is the joint outcome of a multi-qubit Pauli measurement (\cref{sec:Timelike_log_gen}); a spacelike observable is the final readout of a logical qubit (\cref{sec:Logic_Auto_Gen}). The two combine when the measured product does not commute with the readout basis, as for $Q_{15}$, which is prepared in $|+\rangle$, participates in the $Z$-type measurement $P_2$, and is read out in $Z$.

\begin{lstlisting}
[[observables]]
id  = "T_first_ppm"      # timelike
xor = ["joint:P1"]

[[observables]]
id  = "T_second_ppm_xor_final_q15_z"
xor = ["joint:P2", "final:Q15"]

[[observables]]
id  = "final_q19_y"      # spacelike
xor = ["final:Q19"]

# ... one observable per measurement and per
#     logical qubit, twelve in total
\end{lstlisting}

Compiling this file yields the circuit, the DEM, the partition into commit, buffer, and cleanup regions, and the observable definitions, written together as one self-describing compiled program. Alongside the DEM the compiled program carries the correction contracts of \cref{sec:spacetime_corrections}: for every region, the corrections it is permitted to apply, the detectors it may observe through its buffer, and the cleanup likelihood models obtained by replaying the commit correction rule on each modeled error. These contracts are what make the staged procedure of \cref{subsec:spacetime_decoding} well defined; a decoder that observes an edge of the DEM is not thereby authorized to apply every correction associated with it, and decoding the regions independently without the contracts does not reproduce the results of \cref{fig:LERresult}.

Decoding is performed by companion libraries, which reads the compiled program, builds and deploys one decoder per commit region and per cleanup region under the correction contracts. The commit decoders act in parallel on the original syndrome and apply only their permitted corrections; the cleanup decoders then act on the residual syndrome $\boldsymbol{s}_{\rm res}$, with all selected correction effects entering the final logical-frame update. A decoding experiment samples detector outcomes and observable values from the DEM, decodes them with the staged decoder, and counts a shot as a failure when any of the twelve predicted observables disagrees with its sampled value. The monolithic reference of \cref{fig:LERresult} uses the same compiled program: its DEM is decoded whole by correlated minimum-weight perfect matching~\cite{HiggottPyMatch}, so the two curves differ only in the decoder, not in the circuit, noise model, or observables.
 
\bibliography{q}
 
\end{document}